\documentclass[11pt,a4paper]{article}
\pdfoutput=1

\usepackage{jheppub}
\usepackage{latexsym}
\usepackage{multirow}
\usepackage{color}
\usepackage[usenames,dvipsnames,table]{xcolor}
\usepackage{graphicx}
\usepackage{epsfig} 
\usepackage{epsf}   
\usepackage{dcolumn}
\usepackage{bm}
\usepackage{dcolumn}
\usepackage{textcomp}
\usepackage{float}
\usepackage{subfig}
\usepackage{hypcap}
\usepackage[]{hyperref}
\usepackage{makecell}
\usepackage{color}
\usepackage{pifont}
\usepackage{appendix}
\usepackage{amsmath}
\usepackage{multirow,bigdelim}  
\usepackage{lineno}

\usepackage[compat=1.0.0]{tikz-feynman}

\usepackage{amsmath}
\usepackage{amssymb}
\usepackage{amssymb, amsmath, amsthm}
\usepackage{eucal}

\usepackage[normalem]{ulem}
\graphicspath{{fig/}}

\definecolor{usergreen}{RGB}{100,255,200}

\newcommand{\comment}[1]{}
\usepackage{amssymb}
\hypersetup{
	bookmarks=true,         
	unicode=false,          
	pdftoolbar=true,        
	pdfmenubar=true,        
	pdffitwindow=true,   
	pdfstartview={FitH},    
	pdfsubject={scalar NSI},   
	pdfnewwindow=true,      
	pdfcreator={RevTeX},
	colorlinks=true,     
	linkcolor=red,         
	citecolor=blue,        
	filecolor=black,     
	urlcolor=blue,           
}

\usepackage{tikz,xcolor,hyperref}

\definecolor{lime}{HTML}{A6CE39}
\DeclareRobustCommand{\orcidicon}{\hspace{-4pt}
	\begin{tikzpicture}
		\draw[lime, fill=lime] (0,0) 
		circle [radius=0.16] 
		node[white] {\hspace{0.1mm}{\fontfamily{qag}\selectfont \tiny ID}};
		\draw[white, fill=white] (-0.07,0.1) 
		circle [radius=0.01];
	\end{tikzpicture}
	\hspace{-3.2mm}
}

\foreach \x in {A, ..., Z}{\expandafter\xdef\csname 
	orcid\x\endcsname{\noexpand\href{https://orcid.org/\csname orcidauthor\x\endcsname}
		{\noexpand\orcidicon}}
}

\preprint{}

\title{Neutrino Oscillations in Presence of Diagonal Elements of Scalar NSI: An Analytic Approach}

\author[a]{Dharitree~Bezboruah\orcidA{},}
\author[b]{Dibya~S.~Chattopadhyay\orcidB{},}
\author[a, c]{Abinash~Medhi\orcidC{},}
\author[a]{Arnab~Sarker\orcidD{},}
\author[a]{Moon~Moon~Devi\orcidE{}}

\affiliation[a]{Department of Physics, Tezpur University, Napaam, Sonitpur, Assam 784028, India}
\affiliation[b]{Tata Institute of Fundamental Research, Homi Bhabha Road, Colaba, Mumbai 400005, India}
\affiliation[c]{Indian Institute of Technology, Guwahati, Assam 781039, India}

\emailAdd{dbbphy1@tezu.ernet.in}
\emailAdd{d.s.chattopadhyay@theory.tifr.res.in}
\emailAdd{amedhi0@rnd.iitg.ac.in}
\emailAdd{arnabs@tezu.ernet.in}
\emailAdd{devimm@tezu.ac.in}

\date{\today}

	\abstract { Scalar Non-Standard Interactions (SNSI) in neutrinos can arise when a scalar mediator couples to both neutrinos and standard model fermions. This beyond the Standard Model (BSM) scenario is particularly interesting as the SNSI contribution appears as a density-dependent perturbation to the neutrino mass, rather than
appearing as a matter-induced potential, and the neutrino oscillation probabilities uniquely depend on the absolute neutrino masses. In this work, we show the complex dependence of the SNSI contributions on the neutrino masses and discuss how the mass of the lightest neutrino would regulate any possible SNSI contribution in both mass ordering scenarios.  We derive the analytic expressions for neutrino oscillation probabilities, employing the Cayley-Hamilton theorem, in the presence of diagonal
elements of SNSI. The expressions are compact and shows explicit dependence on matter effects and the absolute neutrino masses.
The analytic expressions calculated here allow us to obtain the dependence of the SNSI contribution on mass terms of the form $m_1 + m_2$,  $m_2 - m_1$, $m_1c_{12}^2 + m_2s_{12}^2,$ $ m_1s_{12}^2 + m_2c_{12}^2$, and $m_3$. 
We then explore the non-trivial impact of neutrino mass ordering on the SNSI contribution.
The dependence of the SNSI contribution on the 3$\nu$ parameters is then thoroughly explored using our analytic expressions.}

\keywords{Scalar NSI, Non-standard interactions, BSM, Neutrino Oscillations}
\arxivnumber{}

\begin{document}
	\maketitle
\section{Introduction} \label{sec:introduction}
The formalism of neutrino oscillations and the subsequent discovery, jointly by Super--Kamiokande (SK)~\cite{Super-Kamiokande:1998kpq} and Sudbury Neutrino Observatory (SNO)~\cite{SNO:2002tuh}, has led to further probes of new--physics beyond the Standard Model (BSM). Neutrino oscillations essentially confirm that neutrinos are massive and provide the first clear experimental hint of BSM--physics. The parameters associated with the neutrino oscillations are being widely probed in different neutrino experiments \cite{Super-Kamiokande:2004orf,KamLAND:2004mhv,MINOS:2008kxu,MINOS:2011neo}. The neutrinos are one of the most promising portals for exploring new physics in the leptonic sector. One such new physics scenario is that of the non--standard interactions (NSIs), where neutrinos couple to fermions via some BSM mediator. NSIs have been extensively explored in both theoretical and experimental settings~\cite{Liao:2016orc,Friedland:2012tq,Coelho:2012bp,Rahman:2015vqa,Coloma:2015kiu,deGouvea:2015ndi,Liao:2016hsa,Forero:2016cmb,Huitu:2016bmb,Bakhti:2016prn,Kumar:2021lrn,Agarwalla:2015cta,Agarwalla:2014bsa,Agarwalla:2012wf,Blennow:2016etl,Blennow:2015nxa,Deepthi:2016erc,Masud:2021ves,Soumya:2019kto,Masud:2018pig,Masud:2017kdi,Masud:2015xva,Ge:2016dlx,Fukasawa:2016lew,Chatterjee:2021wac,Medhi:2023ebi,Chaves:2021kxe,Brahma:2023wlf,Davoudiasl:2023uiq,Chatterjee:2020kkm,Choubey:2014iia,Singha:2021jkn,Denton:2018xmq,Denton:2020uda,Farzan:2015hkd,Ge:2018uhz, Medhi:2021wxj}. Looking at the unprecedented accuracy and precision provided by the current and upcoming neutrino experiments, these subdominant effects on neutrino oscillations may have a significant impact on their physics reach. 

Current and upcoming neutrino experiments have entered the era of precision neutrino physics. The primary objectives of these experiments are to resolve three major unknowns in the neutrino sector: the neutrino mass hierarchy~\cite{Capozzi:2017ipn}, the octant of the mixing angle $\theta_{23}$~\cite{Agarwalla:2013ju}, and the determination of the CP-violating phase ($\delta_{CP}$) in the leptonic sector~\cite{Kobayashi:1973fv}. The robustness of these neutrino experiments makes them sensitive to possible subdominant effects in neutrino interactions, arising from BSM scenarios, one such scenario is Non-Standard Interactions. NSIs, if present, may also significantly influence the measurement of oscillation parameters across various neutrino experiments.
NSIs, initially proposed by Wolfenstein~\cite{Wolfenstein:1977ue}, involve the non-standard coupling of neutrinos with the environmental fermions through a vector mediator. This vector-mediated interaction introduces an additional matter potential term in the neutrino oscillation Hamiltonian. The vector NSI framework has been extensively studied~\cite{Miranda:2015dra, Farzan:2017xzy, Biggio:2009nt, Babu:2019mfe, Ohlsson:2012kf}, positioning it as a promising candidate for exploring physics beyond the Standard Model.
Such NSI interactions can also considerably impact the physics reach of various neutrino experiments~\cite{Liao:2016orc,Friedland:2012tq,Coelho:2012bp,Rahman:2015vqa,Coloma:2015kiu,deGouvea:2015ndi,Liao:2016hsa,Forero:2016cmb,Huitu:2016bmb,Bakhti:2016prn,Kumar:2021lrn,Agarwalla:2015cta,Agarwalla:2014bsa,Agarwalla:2012wf,Blennow:2016etl,Blennow:2015nxa,Deepthi:2016erc,Masud:2021ves,Soumya:2019kto,Masud:2018pig,Masud:2017kdi,Masud:2015xva,Ge:2016dlx,Fukasawa:2016lew,Chatterjee:2021wac}. These effects are actively being investigated~\cite{Khatun:2019tad,Chatterjee:2014gxa,Super-Kamiokande:2011dam, Davidson:2003ha,Choubey:2014iia,Denton:2018xmq,Farzan:2015hkd,Farzan:2015doa,Esmaili:2013fva,Khan:2021wzy,Liu:2020emq,Chatterjee:2020kkm,Denton:2020uda,Babu:2020nna,Flores:2020lji,Farzan:2019xor,Pandey:2019apj}, and global constraints on vector NSI parameters can be found in~\cite{Esteban:2019lfo,Coloma:2019mbs}.

In this study, we investigate the non-standard coupling of neutrinos via a scalar mediator~\cite{Ge:2018uhz, Yang:2018yvk, Khan:2019jvr, Medhi:2021wxj}. The scalar-mediated NSIs influence the neutrino mass term in the oscillation Hamiltonian and offer unique phenomenological signatures in neutrino oscillations. Scalar field-mediated non-standard interactions, as discussed in \cite{Ge:2018uhz}, directly affect the neutrino masses through Yukawa-like terms in the Lagrangian.
Furthermore, the dependence on absolute neutrino masses becomes particularly significant in the context of scalar-mediated NSIs. In this case, the effects of new physics are influenced by all three neutrino masses and the scalar NSI SNSI parameters. The SNSI contributions scale linearly with the environmental matter density, making long-baseline neutrino experiments particularly well-suited for probing such interactions. The phenomenological aspects of SNSI in the context of various experiments are explored in recent studies \cite{Ge:2018uhz,Medhi:2021wxj,Medhi:2022qmu,Denton:2022pxt,Singha:2023set,ESSnuSB:2023lbg,Sarker:2023qzp,Medhi:2023ebi,Chaves:2021kxe,Gupta:2023wct,Sarker:2024ytu,Dutta:2022fdt,Venzor:2020ova,Smirnov:2019cae,Arguelles:2019xgp}. 

Although there are currently no stringent bounds on the SNSI parameters, a few studies have tried putting some constraints through astrophysical and cosmological limits \cite{Babu:2019iml,Venzor:2020ova}. In a recent paper \cite{Denton:2024upc}, the authors put constraints on the SNSI parameters as well as the absolute neutrino mass scale by combining solar and reactor data.
Additionally, \cite{Gupta:2023wct} constrains SNSI parameters considering the JUNO (Jiangmen Underground Neutrino Observatory) \cite{JUNO:2021vlw} experiment, while \cite{Dutta:2024hqq} explores the sensitivities of DUNE (Deep Underground Neutrino Experiment) and compares DUNE's constraints with other non-oscillatory probes. In reference \cite{Medhi:2023ebi}, the authors discuss constraining SNSI at DUNE, further pointing out that the absolute neutrino mass can also be constrained in the presence of SNSI.
Reference \cite{Denton:2022pxt} explores distinguishing different new physics scenarios, including sterile neutrinos and vector and scalar NSI at NO$\nu$A (NuMI Off-axis $\nu_e$ Appearance) \cite{NOvA:2004blv,NOvA:2016kwd}, T2K (Tokai to Kamioka) \cite{T2K:2014xyt}, and DUNE.
SNSI using $\nu$($\bar{\nu}$)-disappearance in oscillation data across various experiments is analyzed in \cite{Chaves:2021kxe}.

The works in \cite{Medhi:2021wxj,Medhi:2022qmu} investigate CP violation (CPV) sensitivities at DUNE \cite{DUNE:2016hlj, DUNE:2015lol, DUNE:2016rla}, focusing on the effects of diagonal SNSI elements, and conducting a synergy study of DUNE, T2HK (Tokai-to-Hyper-Kamiokande) \cite{Hyper-KamiokandeProto-:2015xww}, and T2HKK (Tokai-to-Hyper-Kamiokande to Korea)  \cite{Hyper-Kamiokande:2016srs}.
A similar analysis for DUNE and P2SO (Protvino to Super-ORCA) \cite{Akindinov:2019flp} concerning CPV and octant sensitivities in the presence of diagonal SNSI elements is presented in \cite{Singha:2023set}, which also explores the constraining capabilities of DUNE and P2SO towards these SNSI parameters.
In reference~\cite{Sarker:2023qzp} the impact of SNSI on neutrino mass ordering in the context of long-baseline (LBL) experiments like DUNE, HK, and HK+KNO are examined. In \cite{ESSnuSB:2023lbg}, the effect of SNSI on $\delta_{CP}$ measurement is studied within the ESS$\nu$SB (European Spallation Source neutrino Super Beam) \cite{ESSnuSB:2021azq} experiment framework.

 Analytic approximations of neutrino oscillation probabilities often offer crucial insights into the physics of neutrino oscillations. By offering explicit functional forms, analytic expressions reveal how oscillation probabilities depend on experimental parameters such as baseline and neutrino energy. This aids in identifying optimal conditions for the precise measurement of relevant parameters and thus guides the design and optimization of various experiments. These expressions also assist in data analysis and interpretation, support phenomenological studies, and improve computational efficiency.  Furthermore, analytic expressions provide a clear and direct understanding of how neutrino oscillations depend on fundamental parameters like mixing angles, mass-squared differences, and the CP-violating phase.  These expressions serve as benchmarks for validating numerical simulations and approximations, ensuring the accuracy and reliability of computational methods. They also enable straightforward exploration of new physics scenarios, including non-standard interactions, sterile neutrinos, etc., providing a flexible and powerful tool for investigating physics beyond the Standard Model.

In this study, we present the analytic approximation for $\nu$-oscillation probabilities when the effects of SNSI are included.
The analytic expressions for neutrino mixing probabilities in the presence of standard matter effects have been widely explored under different approximations~\cite{Akhmedov:2004ny,TorrenteLujan:1995qi,Kimura:2002wd,Arafune:1997hd,Cervera:2000kp,Asano:2011nj,Freund:2001pn,Denton:2016wmg,Denton:2018hal,Agarwalla:2013tza}, similar works have also been done concerning different possible new physics effects~\cite{Meloni:2009ia,Friedland:2006pi,Chaves:2018sih,Kopp:2007ne,Gronroos:2024jbs,Chattopadhyay:2022ftv,Chattopadhyay:2022hkw,Kostelecky:2003cr}.
Some analytic expressions for SNSI have been studied under certain approximations in references \cite{Denton:2022pxt, Singha:2023set,Gupta:2023wct}. The analytic approximations derived in these studies are geared towards the specific experimental scenarios considered. We compute and present all leading contributions to the conversion and survival probability ($P_{\mu e}$ and $P_{\mu\mu}$) to a precision that will suffice for all upcoming long-baseline neutrino experiments. Furthermore, our analytic approximations have an explicit dependence on matter effects. 
Our concise expressions exhibit substantial accuracy and enable a precise and physical interpretation of the effects of SNSI parameters. To obtain the analytic expressions, we perturbatively expand in terms of the power of a book-keeping parameter $\lambda \equiv 0.2$, by associating the small parameters $s_{13} \equiv \sin(\theta_{13})$, $\alpha$, and the SNSI element $\eta_{\alpha\alpha}$ with powers of $\lambda$. These expressions are presented in a format very similar to the existing expressions for standard matter effects \cite{Akhmedov:2004ny}. Note that unlike the vector NSI scenario, where the new physics effects grow as we increase the energy, the SNSI contribution leads to an energy-independent correction to the mass term, which makes it harder to disentangle standard oscillation terms from SNSI. Our analysis enables the probe of SNSI via different oscillation channels and allows us to point out the most potent ways to distinguish 3-$\nu$ matter mixing from SNSI.

Our analysis uncovers interesting terms in the probability expressions, suggesting a promising avenue for probing and potentially constraining the absolute neutrino masses through $\nu$-oscillation experiments. The analytic expressions reveal patterns that can help explain various behaviors in SNSI phenomenology. We calculate the oscillation probabilities for $\nu_e$ appearance channel $P_{\mu e}$ and $\nu_\mu$ disappearance channel $P_{\mu \mu}$; relevant mainly for the LBL and the atmospheric neutrino sectors, as long as the constant density approximation holds.  We note interesting dependence of the probability expressions on the absolute neutrino masses through terms of the form  $m_1+m_2$, $m_1 c_{12}^2 + m_2 s_{12}^2$, $m_1 s_{12}^2 + m_2 c_{12}^2$, $m_2 - m_1$ and $m_3$. These individual terms would show distinct behavior for the two neutrino mass ordering scenarios, indicating that the SNSI would contribute differently in normal ordering (NO) and inverted ordering (IO) scenarios. In addition, we have also looked into the dependence of the oscillation probabilities on the leptonic CP phase ($\delta_{CP}$) in the presence of SNSI.

This paper is organized as follows. In section \ref{sec:framework}, we discuss the detailed framework of SNSI. In section \ref{sec:ana_prob}, we derive the analytic expressions of the appearance as well as disappearance probabilities. In section \ref{sec:applicability}, we discuss the accuracy and applicability of the analytic probability expressions. In section~\ref{sec:SNSI_mixing_param}, we explore the complex behaviour of the SNSI contribution to the neutrino probabilities and point out their dependence on the absolute neutrino mass and the leptonic CP phase ($\delta_{CP}$). Finally, in section \ref{sec:summary}, we summarize our findings.

\section{Scalar NSI Formalism} \label{sec:framework}

The non-standard coupling of neutrinos with standard model fermions mediated by a scalar~\cite{Ge:2018uhz,Babu:2019iml} is an interesting sector to probe new--physics beyond SM. We discuss here, the formalism used to incorporate SNSI effects. The effective Lagrangian for SNSI via a scalar mediator, $\phi$ may be framed as, 
\begin{align}
{\cal L}_{\rm eff}^{\rm S} = \frac{y_f y_{\alpha\beta}}{m_\phi^2}(\bar{\nu}_\alpha(p_3) \nu_\beta(p_2))(\bar{f}(p_1)f(p_4)) \,, 
\label{eq:nsi_L}
\end{align}
where, 
\begin{itemize}
    \item 
\noindent $\alpha$, $\beta$ refer to the neutrino flavours e, $\mu$, $\tau$,
\item  \noindent $f = e, u, d$ indicate the matter fermions, ($e$: electron, $u$:
up-quark, $d$: down-quark), while, $\bar{f}$ indicates the corresponding anti fermions, 
 \item  \noindent $y_{\alpha\beta}$ is the Yukawa couplings of the neutrinos with $\phi$, 
  \item  \noindent$y_f$ is the Yukawa coupling of $\phi$ with $f$, and, 
  \item \noindent $m_\phi$ is the mass of $\phi$.
  \end{itemize}

 \begin{figure}[t]
    \centering
    \includegraphics[width=0.3\linewidth, ]{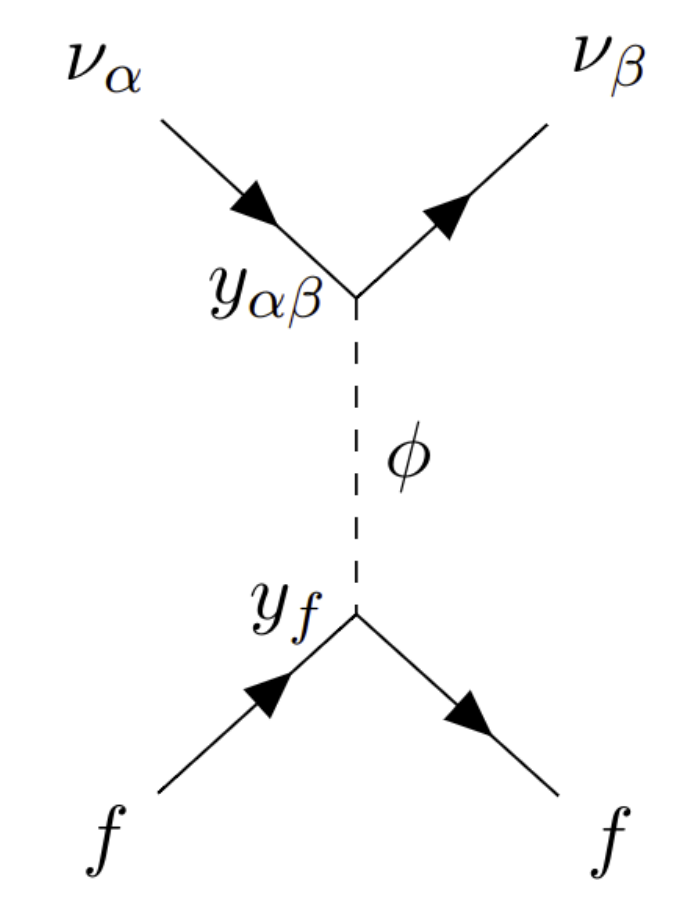}
    \hspace{2cm}
    \includegraphics[width=0.3\linewidth, ]{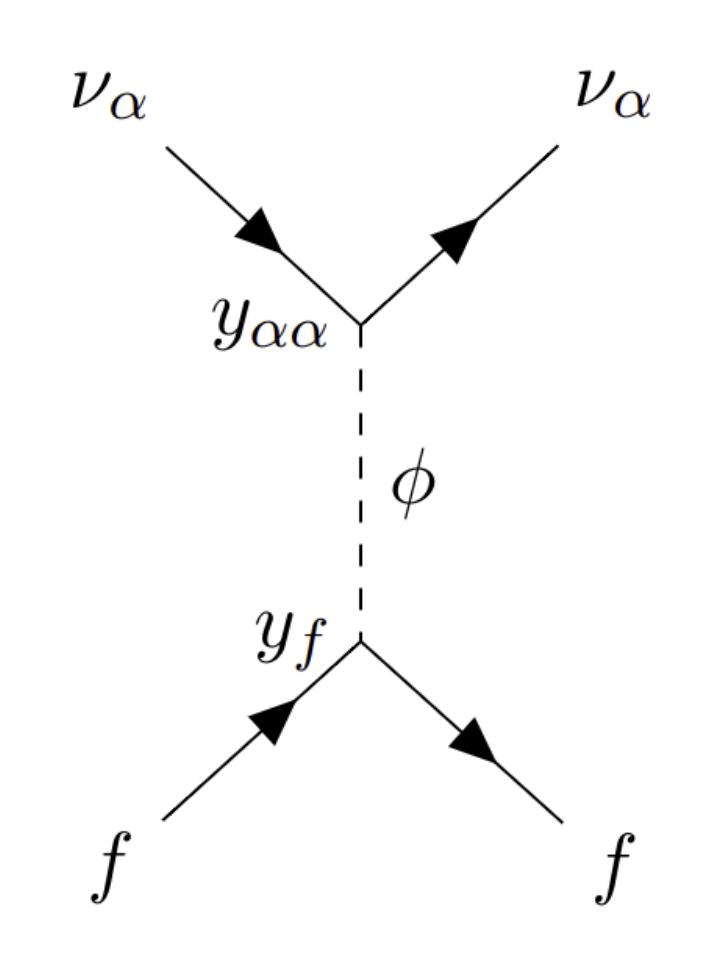}
    \caption{(Left) A general Feynman diagram for scalar NSI in neutrinos, (Right: A Feynman diagram for one non-zero scalar NSI element, $\eta_{\alpha\alpha}$.}
    \label{fig:feynman_snsi}
\end{figure}

A typical Feynman diagram for such neutrino-fermion scattering, mediated by the scalar $\phi$ is shown in figure \ref{fig:feynman_snsi}. Here the left panel represents a general interaction and the right panel shows flavor-conserving SNSI. Note that, the SNSI contribution does not appear as a matter potential since the lagrangian is composed of Yukuwa terms which constrain us from converting it to vector currents. Rather, it appears as a density-dependent correction to the neutrino mass. The corresponding Dirac equation taking into account the effect of SNSI and spin summation over the fermions inducing SNSI may take the following form,
\begin{equation}
  \bar \nu_\beta
\left[
  i \partial_\mu \gamma^\mu
+
\left(
  M_{\beta \alpha}
+ \frac {\sum_f n_f y_f y_{\alpha \beta}}{m^2_\phi}
\right)
\right] \nu_\alpha
=
  0 \,,
\end{equation}
where $n_f$ is the number density of the environmental fermions.

We, therefore, see that the effect of SNSI appears as a perturbation to the neutrino mass term. So, the effective Hamiltonian in the presence of  SNSI takes the form,
\begin{equation}
\mathcal H_{\rm SNSI}
\approx
E_\nu
+ \frac { M_{\rm eff}  M_{\rm eff}^\dagger}{2 E_\nu}
\pm V_{\rm SI} \,,
\label{eq:Hs}
\end{equation}
where, $M_{\rm eff}$ = $M + \delta M$, is the effective mass matrix that includes both the regular mass matrix $M$ and the contribution from the SNSI, $\delta M  \equiv \sum_f n_f y_f y_{\alpha\beta} / m^2_\phi$. We have parameterized the scalar NSI contribution \cite{Ge:2018uhz,Medhi:2021wxj,Medhi:2022qmu,Denton:2022pxt,Singha:2023set} as,
\begin{equation}
\delta M
\equiv
S_m
\begin{bmatrix}
\eta_{ee}     & \eta_{e \mu}    & \eta_{e \tau}   \\
\eta_{e \mu}^* & \eta_{\mu \mu}  & \eta_{\mu \tau} \\
\eta_{e \tau}^* & \eta_{\mu \tau}^* & \eta_{\tau \tau}
\end{bmatrix}.
\label{eq:dM}
\end{equation}
Here, we introduce the constant scaling term $S_m$, with the dimension as of mass, in order to make the $\eta_{\alpha \beta}$ parameters dimensionless. In this work, we have used
\begin{equation}
   S_m = \sqrt{2.5 \times 10^{-3}~{\rm eV^2}} \approx 0.05~{\rm eV} , 
\end{equation}
which corresponds to a typical atmospheric mass squared difference. 

The dimensionless elements $\eta_{\alpha \beta}$ quantify the size of SNSI. The Hermicity of the Hamiltonian allows the diagonal elements to be real and off-diagonal elements to be complex. In this work, we have calculated the oscillation probabilities for the diagonal elements of the SNSI matrix, taking one at a time. For the three cases that we have used, the effective modified Hamiltonian takes the following forms:
\begin{subequations}
\begin{align}
~~~~~& {\rm {In~presence~of ~\eta_{ee}}:}~ M_{\rm eff} = \mathcal{U} \cdot {\rm diag}\left(m_1, m_2, m_3
 \right) \cdot \mathcal{U}^\dag + S_m  ~ \rm diag \left( \eta_{ee}, 0, 0
 \right) \label{MeffCase1}\\
~~~~~& {\rm {In~presence~of ~\eta_{\mu \mu}}:}~ M_{\rm eff} = \mathcal{U} \cdot {\rm diag}\left(m_1, m_2, m_3
 \right) \cdot \mathcal{U}^\dag + S_m  ~ \rm diag \left( 0, \eta_{\mu\mu}, 0
 \right) \label{MeffCase2}\\
~~~~~& {\rm {In~presence~of ~\eta_{\tau \tau}}:}~ M_{\rm eff} = \mathcal{U} \cdot {\rm diag}\left(m_1, m_2, m_3
 \right) \cdot \mathcal{U}^\dag + S_m  ~ \rm diag \left( 0, 0, \eta_{\tau\tau}
 \right) \label{MeffCase3}
 \end{align}
  \label{eq:Meff}
\end{subequations}

The unique and most interesting aspect of SNSI is its dependence on the absolute masses of neutrinos. It brings the possibility of probing absolute neutrino mass and constraining the ($\eta_{\alpha\alpha} - m_{\ell}$) parameter space in future oscillation experiments \cite{Medhi:2023ebi}. 

Note that, the presence of SNSI would contaminate the measurements of all the neutrino mixing parameters. Thus, the ``true'' values of the neutrino mixing parameters (in vacuum), and SNSI contribution (due to the presence of a scalar field interacting with standard model fermions) would be hard to disentangle.
However, since the effect of SNSI scales linearly with matter density, upcoming experiments like DUNE with longer baselines ($\sim 1300$ km) would observe a more dominant contribution from SNSI.
This is because, for a longer baseline, the average matter density experienced by neutrinos as they propagate would also be higher~\cite{Dziewonski:1981xy}.
We define the $\eta_{\alpha\beta}$ terms in equation~\ref{eq:dM} as,
\begin{equation}
    \eta_{\alpha\beta} = \eta_{\alpha\beta}^{\rm (true)} \left(\frac{\rho_{\text  {\tiny DUNE}} - \rho_0}{\rho_0}\right)\; ,
\end{equation}
where $\eta_{\alpha\beta}^{\rm (true)}$ is the true value of the SNSI parameter, $\rho_{\text{\tiny DUNE}}$ is the average matter density experienced by neutrinos in DUNE, and $\rho_0$ is the average matter density for reactor and long-baseline experiments from which the neutrino mixing parameters are currently determined.

Analytic approximations for SNSI contributions with exact dependence on the matter potential, and dependence on the absolute mass of neutrinos would allow us to explore the effects of SNSI in a more systematic and detailed manner.
The dependence SNSI contributions on the absolute mass of the lightest neutrino $m_{\ell}$, mass ordering, $\nu$/$\bar \nu$ channel, as well as survival/ conversion channel may allow us to disentangle further the contributions from SNSI and from the standard oscillation terms. 

In this work, we take some benchmark values of $\eta_{\alpha\beta}$ and $m_{\ell}$ which can lead to observable but sub-leading changes in the neutrino oscillation probabilities observed in DUNE. we have considered the specifications of DUNE \cite{DUNE:2016hlj, DUNE:2015lol, DUNE:2016rla} for the purpose of illustration in the long baseline sector, and we have fixed the density at $\rho = 2.8 \text{ g/cm}^3$.  In table \ref{tab:3nu_params}, we also list the Values of the $3\nu$ oscillation parameters used in our analysis.

\begin{table}[!h]
\centering
 \begin{tabular}{|c|c|c|c|c|c|} 
 \hline
 $\theta_{12} [^\circ]$ & $\theta_{13} [^\circ]$ & $\theta_{23} [^\circ]$ & $\delta_{CP} [^\circ]$& $\Delta m_{21}^2 [eV^2]$ & $|\Delta m_{31}^2| [eV^2]$\\ [0.5ex] 
 \hline
 33.5 & 8.5 & 45 & 45 & $7.5 \times 10^{-5}$ & $ 2.5 \times 10^{-3}$\\ 
 \hline
 \end{tabular}
 \caption{Values of the $3\nu$ oscillation parameters used in our analysis. }
 \label{tab:3nu_params}
\end{table}

\subsection{Relative scalar NSI contribution}
The SNSI, if present in nature, is expected to be subdominant, especially for current and upcoming long-baseline neutrino experiments. To constrain SNSI scenarios, we first need to understand how a non-zero value of an SNSI element would affect the neutrino oscillation Hamiltonian and thus the neutrino probabilities. We here look into the relative contribution of SNSI to the standard interaction Hamiltonian for various choices of $\eta_{\alpha\alpha}$. This may help us to obtain a ballpark approximation of the values of SNSI parameters. To quantify the relative SNSI contribution, we define a ratio
\begin{equation}\label{eq:ratio}
  \mathcal{R}_{SNSI} = \Bigg | \frac{H_{SNSI} \big[i,j\big]}{H_{SI}\big[i,j\big]} \Bigg |_{\max},
\end{equation}
where,
\begin{itemize}
\item $H_{SNSI} [i,j]$ is the matrix element corresponding to $i^\text{th}$ row and $j^\text{th}$ column of $H_{SNSI}$, the contribution of SNSI in the effective Hamiltonian, and, 
\item $H_{SI}[i,j]$ is the matrix element corresponding to $i^\text{th}$ row and $j^\text{th}$ column of  $H_{SI}$, the standard matter Hamiltonian.
\item Note that, since the overall phase in neutrino oscillations has no physical significance, we subtract the [1,1] matrix element of the ratio from all the diagonal elements which leaves us with only eight ratios. We then consider the ratio with the highest value.
\end{itemize}

We have used  $\mathcal{R}_{SNSI}$ to naively explore the strength of SNSI contributions at the Hamiltonian level. However, it is intended to be used only to get a rough view, since contributions from different elements can add up or cancel out while calculating the probabilities, potentially resulting in slightly different contributions. Furthermore, different neutrino channels will also be more sensitive to different SNSI elements, this is also not captured by $\mathcal{R}_{SNSI}$.

\begin{figure}[!t]
    \centering
    \includegraphics[width=0.45\textwidth]{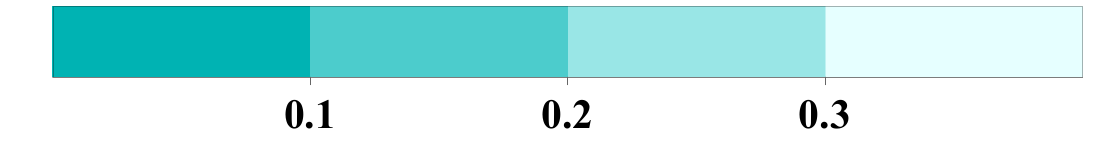}\\[2ex]
    \includegraphics[width=0.325\textwidth]{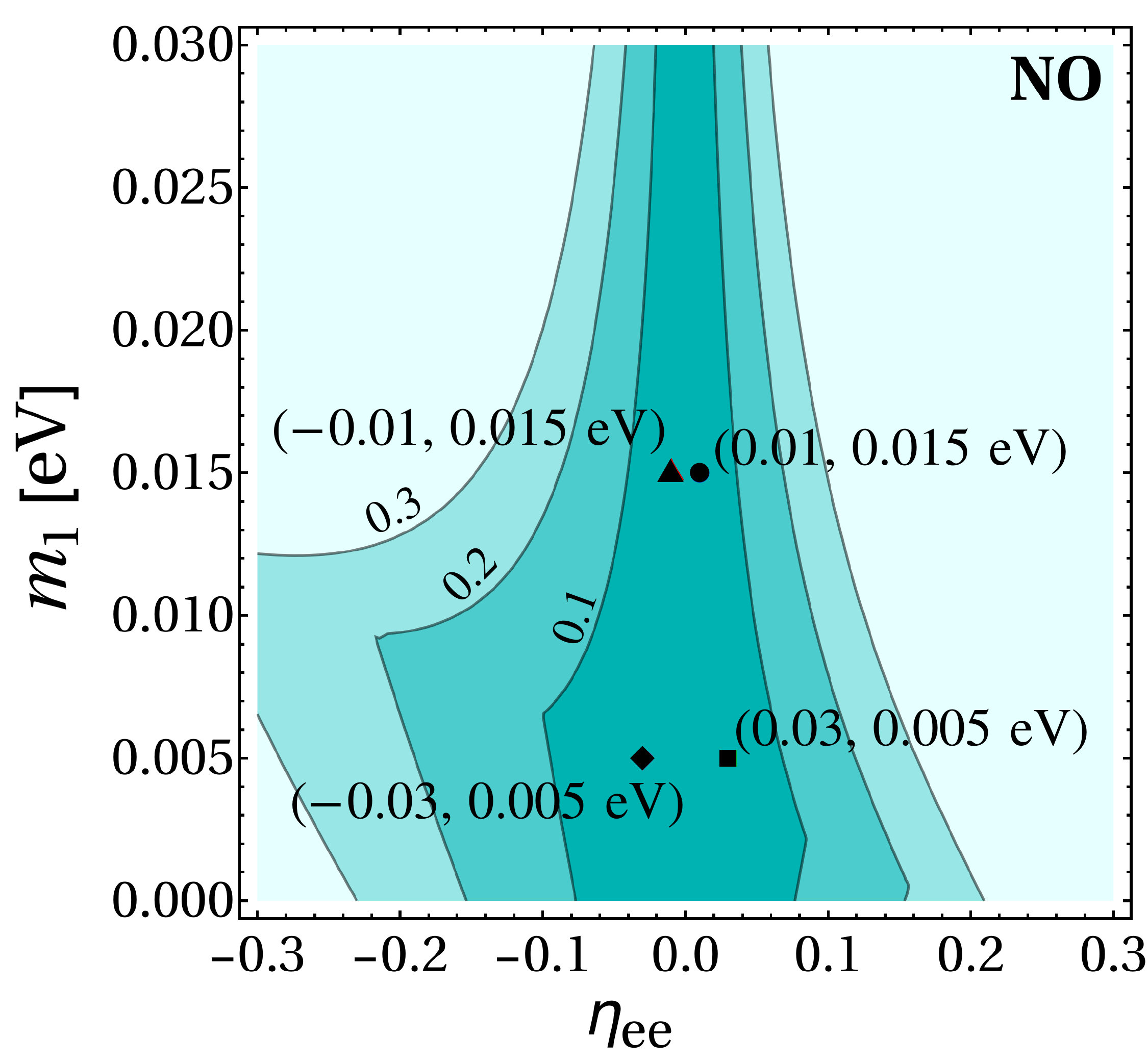}
    \includegraphics[width=0.325\textwidth]{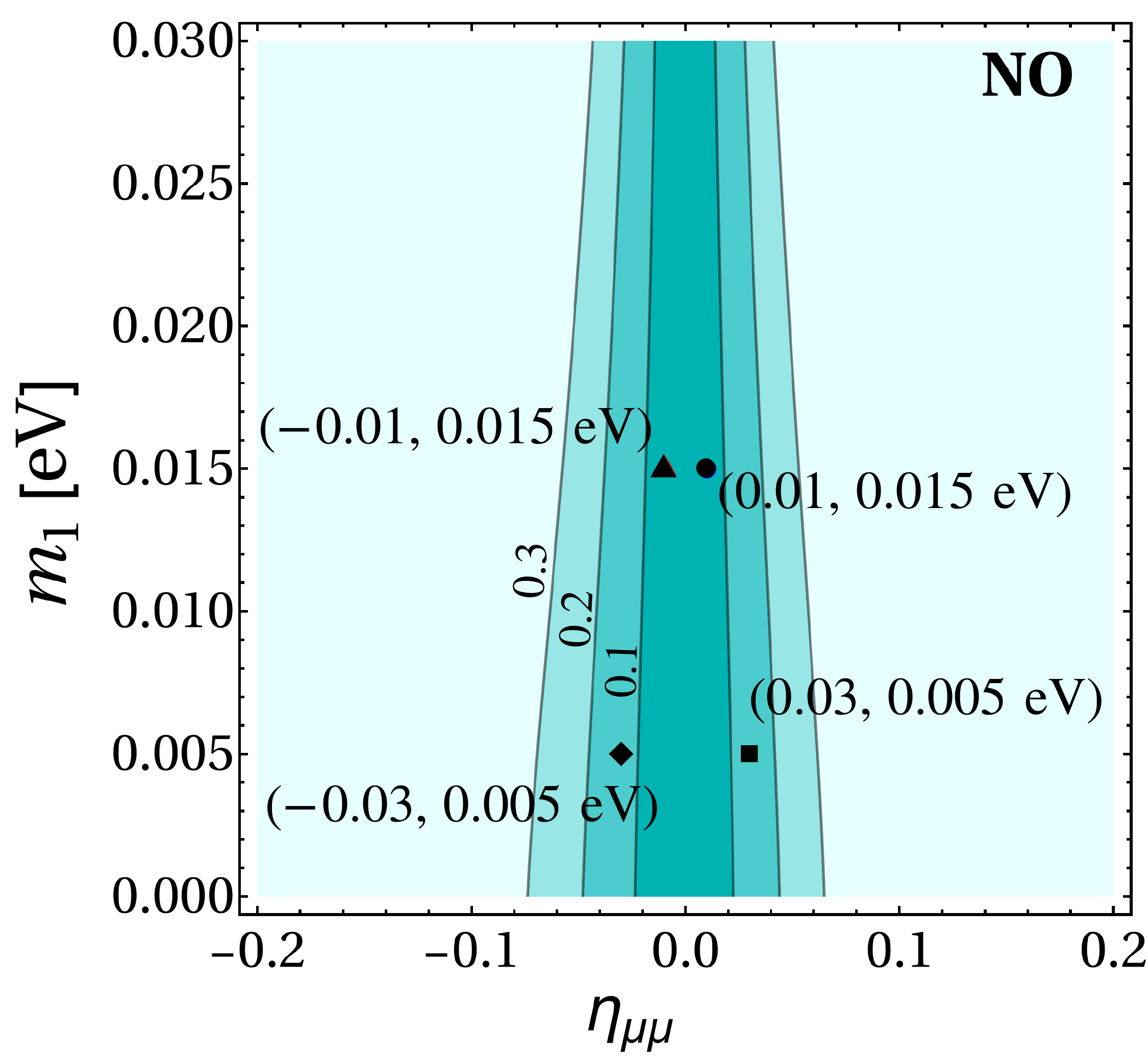}
    \includegraphics[width=0.325\textwidth]{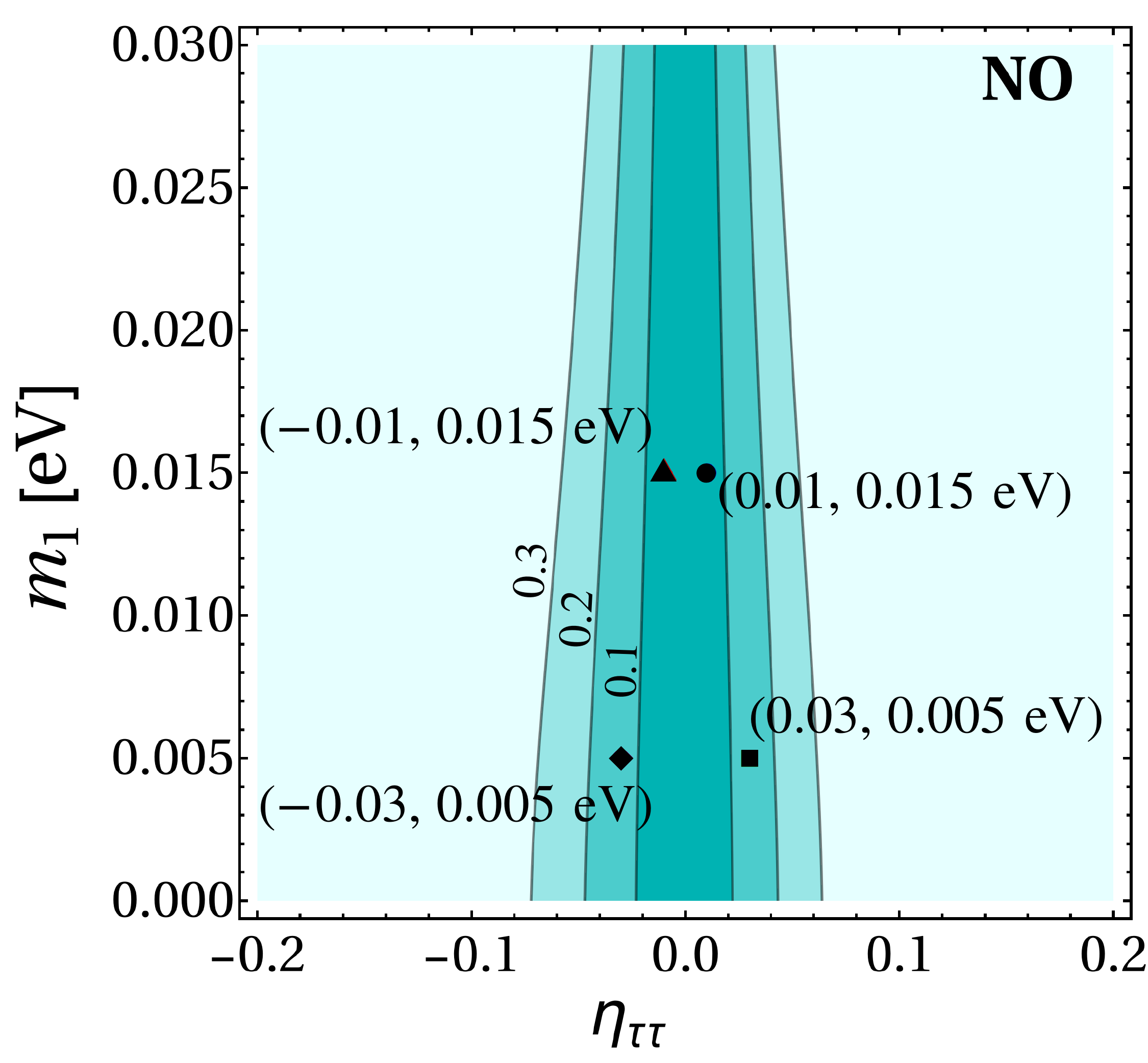}\\[2ex]
    \includegraphics[width=0.325\textwidth]{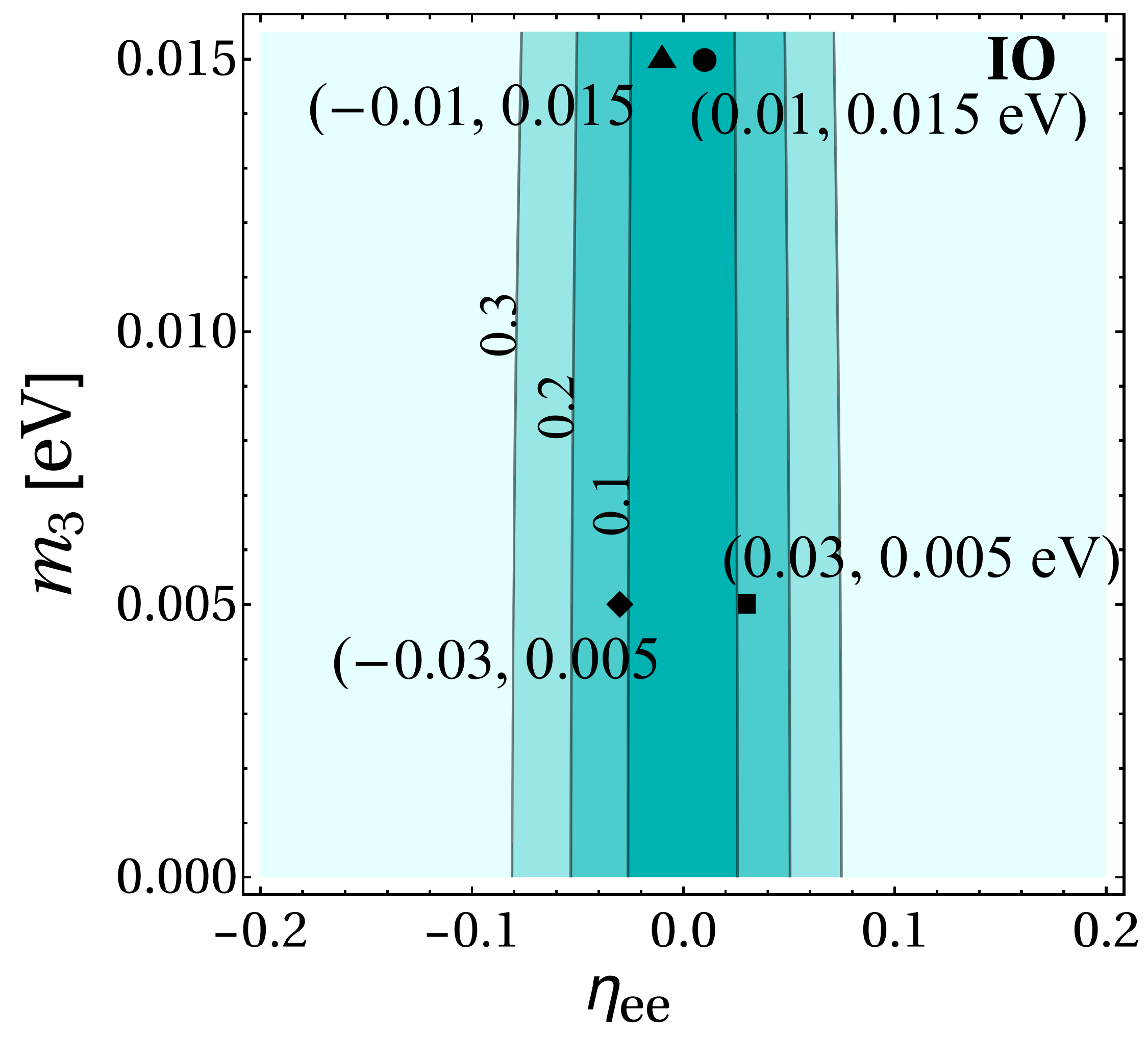}
    \includegraphics[width=0.325\textwidth]{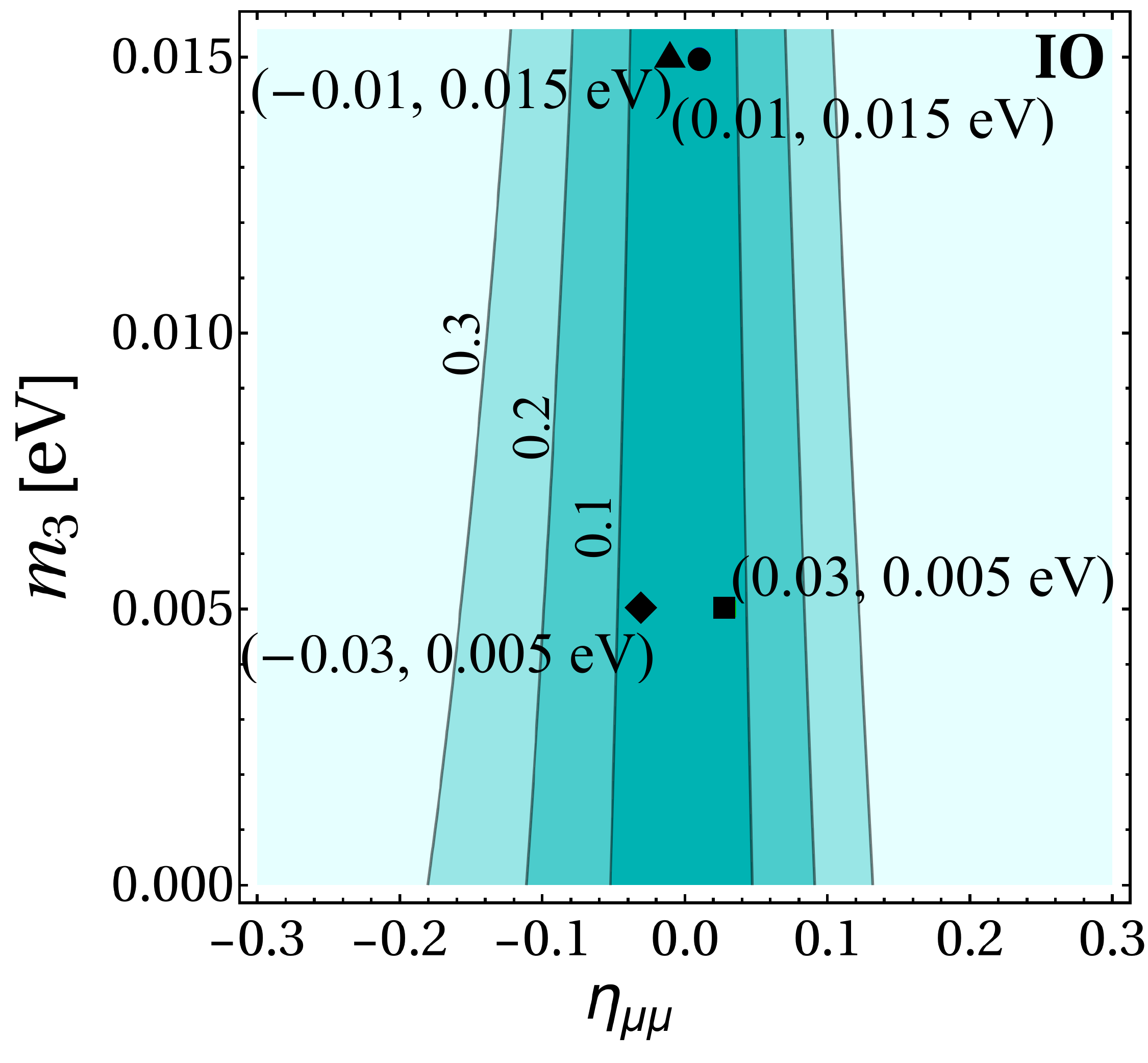}
     \includegraphics[width=0.325\textwidth]{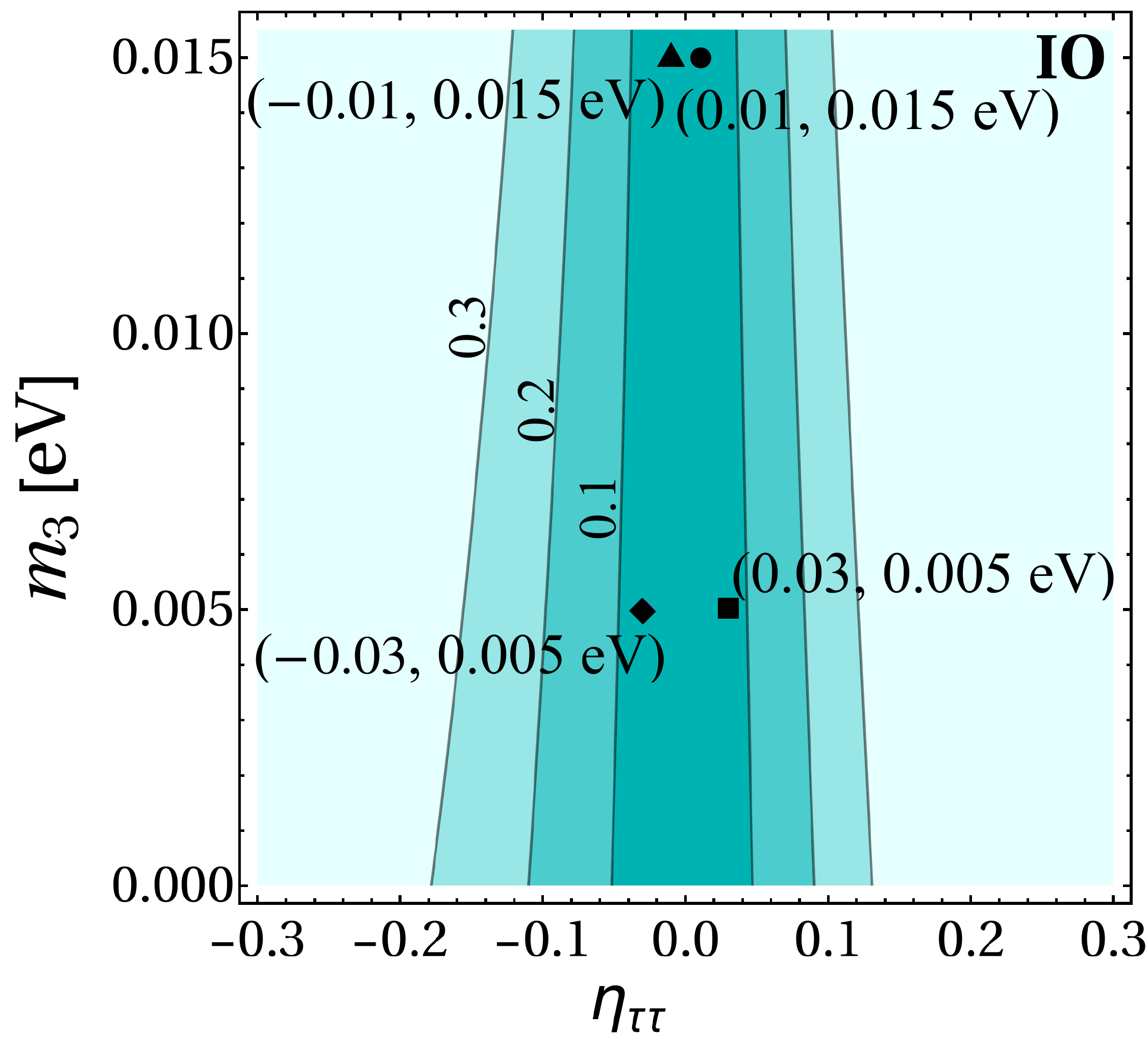}
    \caption{Contour plots for the ratio $\mathcal{R}_{SNSI}$ in the ($\eta_{\alpha\alpha} - m_\ell$) plane. The top (bottom) panels represent the case of NO (IO). The left, middle, and right panels correspond to the case of $\eta_{ee}$, $\eta_{\mu\mu}$ and $\eta_{\tau\tau}$.}
    \label{fig:ratio_H}
\end{figure}

In figure \ref{fig:ratio_H}, we have shown $\mathcal{R}_{SNSI}$ on the $(\eta_{\alpha\alpha}$ -- $m_\ell)$ plane, where $m_\ell$ is lightest neutrino mass ( i.e., $m_1$ for NO and $m_3$ for IO) for both of the mass orderings. The values of the $3 \nu$ oscillation parameters used are listed in table \ref{tab:3nu_params}. From the figure, we note the following.
\begin{itemize}
    \item We observe that different values of $\eta_{\alpha \alpha}$ can result in a similar level of relative contribution due to the non-trivial dependence of $H_{SNSI}$ on $\nu$-masses.
    \item The contours for $\eta_{\mu \mu}$ and $\eta_{\tau \tau}$ are nearly similar with marginal changes for both NO and IO. The range of $\eta_{ee}$ is relatively wider as compared to $\eta_{\mu \mu}$ and $\eta_{\tau \tau}$ for NO. Whereas for IO, a much narrower range of $\eta_{ee}$ can contribute to similar values of $\mathcal{R}_{SNSI}$.
    \item  We mostly note minimal dependence on the lightest mass $m_\ell$, for (NO,  $\eta_{ee}$) case the dependence on $m_\ell$ being most prominent. Note that, as we will show later, the neutrino probabilities depend crucially on the lightest neutrino mass $m_\ell$ in most cases.
\end{itemize}

\begin{table}[h]
    \centering
\begin{tabular}{|c|c|}
    \hline 
    Parameters & Choices\tabularnewline
    \hline 
    {($\eta_{\alpha\alpha},m_{\ell}$ [eV])} & (0.03, 0.005), (-0.03, 0.005), (0.01, 0.015),  (-0.01, 0.015) \tabularnewline
    \hline 
    \end{tabular}
    \caption{Different choices of SNSI parameter $\eta_{\alpha \alpha}$ and $m_\ell$ used in our analysis. We denote all three diagonal SNSI elements with $\eta_{\alpha \alpha}$ (i.e., $\eta_{ee}$, $\eta_{\mu \mu}$, and $\eta_{\tau \tau}$), and $m_\ell$ corresponds to the lightest neutrino mass ( i.e., $m_1$ for NO and $m_3$ for IO).}
    \label{tab:value_eta_m}
\end{table}

The values of $\eta_{\alpha \alpha}$ and $m_1$ (NO) or $m_3$ (IO) commonly used in our work are listed in table \ref{tab:value_eta_m}. These points are also highlighted on the $\eta_{\alpha \alpha}$ -- $m_\ell$ plane in figure \ref{fig:ratio_H}. For this work, for demonstration purposes, we have chosen $\eta_{\alpha \alpha}$ and $m_\ell$ such that the corresponding value of $\mathcal{R}_{SNSI}$ falls below 0.2.

In the next section, we present the approximate analytic expressions for the SNSI contributions to the neutrino oscillation probabilities $P_{\mu e}$ and $P_{\mu\mu}$, with matter effects explicitly included.

\section{The Analytic Probability Expressions in Presence of Scalar NSI} \label{sec:ana_prob} 
In this section, we calculate the analytic probability expressions in the presence of SNSI parameters. We have perturbatively calculated the probabilities expanding them in the small parameters $\alpha$, $s_{13} \equiv \sin \theta_{13}$ and the diagonal SNSI parameters $\eta_{\alpha\alpha}$. As in references \cite{Chattopadhyay:2022ftv,Chattopadhyay:2022hkw}, an accounting parameter $\lambda \equiv 0.2$ is used and the variables used for expansion are expressed as,
\begin{equation}
    \alpha \simeq 0.03 \approx O(\lambda^2), \qquad s_{13} \simeq 0.14 \approx O(\lambda), \qquad \eta_{\alpha \alpha} \approx O(\lambda^2),
\end{equation}
in powers of $\lambda$. The following dimensionless quantities are defined as they will frequently appear in our expressions:
\begin{equation}
    \alpha = \frac{\Delta m_{21}^2}{\Delta m_{31}^2}, \qquad A_e =\frac{2 E_\nu V_{SI}}{\Delta m_{31}^2}, \qquad \Delta =\frac{\Delta m_{31}^2 L}{4 E_\nu}.
\end{equation}
We have obtained the transition amplitude using the effective SNSI Hamiltonian from equation \ref{eq:Hs} which is written in flavor basis, 
\begin{equation} \label{eq:time_evol}
    \mathcal{A} \equiv \text{exp}(-i \, \mathcal{H}_{SNSI} \, L).
\end{equation}
Then the probability of transition from $\alpha$ flavor to $\beta$ flavor can be calculated as
\begin{equation}
    P_{\alpha \beta} = |\mathcal{A}_{\beta \alpha}|^2,
\end{equation}
where $\mathcal{A}_{\beta \alpha}$ is the amplitude of transition from $\alpha$ to $\beta$ flavor. For the expansion of the transition amplitude, Cayley-Hamilton theorem \cite{Ohlsson:1999xb} is implemented; according to which we can express any function f($\mathbb{M}$) of a matrix $\mathbb{M}$ as,
\begin{equation}
    f(\mathbb{M}) = \sum_{i=1}^k M_i f(\Lambda_i), \qquad \text{with} \quad M_i = \prod_{j=1, j \neq i}^k \frac{1}{\Lambda_i -\Lambda_j} (\mathbb{M} - \Lambda_j \mathbb{I}),
\end{equation}
where $\Lambda_j$ are the eigenvalues of the matrix $\mathbb{M}$. To employ the Cayley-Hamilton theorem, we have used $\mathbb{M} \equiv -i\mathcal{H}_{SNSI} L$ and have calculated the eigenvalues in the presence of SNSI terms, $\eta_{ee}$, $\eta_{\mu\mu}$, and  $\eta_{\tau\tau}$, respectively. The linear term in $\eta_{\alpha\alpha}$ will have the dominant effect on the SNSI contribution. To obtain the SNSI contribution, we must calculate the eigenvalues up to the first order in $\eta_{\alpha\alpha}$. Therefore, let us define the eigenvalues as
\begin{equation}
    E = E ^ {(0)} + E ^{(\eta_{\alpha\alpha})} .
\end{equation} 
Where, $E ^ {(0)}$ is the eigenvalue for standard 3$\nu$ Hamiltonian and $E ^{(\eta_{\alpha\alpha})}$ represents the contribution due to the diagonal SNSI terms. The eigenvalues for the $3\nu$ contribution are given by,
\begin{subequations}
    \begin{align}
        E_1^{(0)} = \frac{  \Delta m^2_{31} }{2 E_\nu } & \left( A_e + \alpha \,s_{12}^2 + s_{13}^2 \, \frac{A_e }{A_e - 1} -s_{13}^4 \frac{A_e^2}{\left(A_e-1\right)^3} + \alpha ^2 \sin ^2\left(2 \theta _{12}\right)\frac{1}{4\, A_e} \right. \nonumber\\
        & \left. \qquad \qquad - \alpha \, s_{13}^2 \, s_{12}^2 \frac{A_e^2}{\left(A_e-1\right)^2} \right)\\
        E_2^{(0)} = \frac{  \Delta m^2_{31} }{2 E_\nu } & \left( \alpha \, c_{12}^2 -\alpha ^2 \sin ^2\left(2 \theta _{12}\right)\frac{1}{4\, A_e} \right)\\
        E_3^{(0)} = \frac{  \Delta m^2_{31} }{2 E_\nu } & \left( 1 - s_{13}^2 \, \frac{A_e }{A_e - 1} + s_{13}^4 \frac{A_e^2}{\left(A_e-1\right)^3} + \alpha \, s_{13}^2 \, s_{12}^2 \frac{A_e^2}{\left(A_e-1\right)^2} \right)
    \end{align}
    \label{eq:SI_eigen}
\end{subequations}

The complete probability expression in the presence of SNSI can be defined as,
\begin{equation}
    P_{\alpha \beta} = P_{\alpha \beta}^{(0)}  + P_{\alpha \beta}^{(\eta_{\alpha\alpha})}
\end{equation}
where $P_{\alpha \beta}^{(0)}$ is the standard $3\nu$ probability and $P_{\alpha \beta} ^{(\eta_{\alpha\alpha})}$ is the SNSI contribution to the probabilities.  Furthermore, for the SNSI contribution $P_{\alpha \beta}^{(\eta_{\alpha\alpha})}$, if the analytic expression is complicated, then we express it as
\begin{equation}
    P_{\alpha \beta}^{(\eta_{\alpha\alpha})} \equiv \sum_i \left[P_{\mu e}^{(\eta_{ee})} \right]^{(i)} \; ,
\end{equation}
where the number `$(i)$' in the superscript denotes the approximate power of $\lambda$ at which this term contributes. For example, $\left[P_{\mu e}^{(\eta_{ee})} \right]^{(i)}$ would only contain $O(\lambda^i)$ terms, where we count the powers of $\lambda$ by associating $s_{13} \sim \lambda$, $\alpha \sim \lambda^2$, and $\eta_{\alpha\alpha} \sim \lambda^2$.

The standard $3\nu$ probability contribution for the appearance and disappearance channel calculated up to $O(\lambda^2)$ for $P_{\mu e}$ and $P_{\mu \mu}$, are given by,
\begin{align}\label{eq:pmm_si}
    P_{\mu e}^{(0)} = & \; 4 s_{13}^2 s_{23}^2 \frac{\sin ^2\left[\left(A_e-1\right) \Delta \right]}{\left(A_e-1\right)^2} \nonumber\\
    & + 2 \alpha  s_{13} \sin \left(2 \theta _{12}\right) \sin \left(2 \theta _{23}\right) \cos \left(\delta _{\text{CP}}+\Delta \right) \frac{\sin \left[ \left(A_e-1\right) \Delta \right]}{A_e-1}\frac{\sin \left[ A_e \Delta \right]}{A_e} \\
    P_{\mu\mu}^{(0)}= & \; 1 -\sin ^2\left(2 \theta _{23}\right) \sin ^2[\Delta ]  +4 \alpha  \Delta s_{23}^2 c_{12}^2 c_{23}^2 \sin [2 \Delta ] \nonumber\\
    &- \frac{2 s_{13}^2 \sin ^2\left(2 \theta _{23}\right)}{A_e-1} \left( \sin [\Delta ] \cos \left[\Delta  A_e\right] \frac{\sin \left[ \left(A_e-1\right) \Delta \right]}{A_e-1} -\frac{A_e}{2} \Delta  \sin [2 \Delta ] \right) \nonumber\\
    &-4 s_{13}^2 s_{23}^2 \frac{\sin ^2\left[ \left(A_e-1\right) \Delta \right]}{\left(A_e-1\right){}^2} \; .
\end{align}
We present these standard 3$\nu$ oscillation probabilities where we expand in the first order of $\alpha$ and $s_{13}$ \cite{Akhmedov:2004ny}. However, several expressions for standard 3$\nu$ oscillation probabilities exist in literature \cite{Arafune:1997hd,Cervera:2000kp,Asano:2011nj,Freund:2001pn,Akhmedov:2004ny,Friedland:2006pi,Denton:2016wmg,Denton:2018hal,Agarwalla:2013tza}. These analytic expressions are mostly obtained by using perturbation theory and expanding around small parameters including ratios of $\Delta m^2$'s, $s_{13}$ and associated terms. Some expressions are obtained by first performing two flavor rotations, followed by perturbation theory. A detailed comparison of the accuracy and computation speed of various expressions can be found in reference \cite{Barenboim:2019pfp}. One may prefer to use 3$\nu$ probability contribution from expressions like DMP (Denton, Minakata, and Parke) \cite{Denton:2016wmg,Denton:2018hal} for better precision.

In this work, we have calculated the probabilities for the three cases mentioned in equation \ref{eq:Meff} considering one $\eta_{\alpha\alpha}$ at a time. We calculate and show all the dominant contributions to the conversion probability up-to an accuracy of $O(\lambda^5)$, while for survival probability, we calculate $P_{\mu\mu}^{(\eta_{ee})}$ up-to an accuracy of $O(\lambda^4)$, since the effect of the $\eta_{ee}$ term on the survival probability is small. We calculate $P_{\mu\mu}^{(\eta_{\mu\mu})}$ and $P_{\mu\mu}^{(\eta_{\tau\tau})}$ up-to $O(\lambda^2)$, as SNSI significantly affects the survival channel in these cases. For the sake of simplicity, we only show linear $\eta_{\alpha\alpha}$ terms in our analysis, this has negligible effects on the accuracy of the SNSI contributions for our purpose.

The SNSI contributions are obtained in terms of mass terms of the form $m_1 + m_2$,  $m_2 - m_1$, $m_1c_{12}^2 + m_2s_{12}^2,$ $ m_1s_{12}^2 + m_2c_{12}^2$, and $m_3$.
The $m_2-m_1$ term is especially interesting as its maximum value is obtained for $m_1 =0$ (unlike the other terms), this is because we can express
\begin{equation}
    m_2-m_1 = \frac{\Delta m^2_{21}}{m_1+m_2} \quad \Rightarrow \quad  \left[m_2-m_1 \right]_{\max} = \sqrt{\Delta m^2_{21}}\; ,
\end{equation}
where the maximum value is achievable only in the Normal Ordering scenario.
Thus, a contribution of the form $\left(m_2-m_1\right) \left(S_m/\Delta m^2_{31}\right)$ can have a maximum value of 
\begin{equation}
   \left[ \left(m_2-m_1\right) \left(\frac{S_m}{\Delta m^2_{31}}\right) \right]_{\max} = \sqrt{\alpha} \approx O(\lambda) \;.
\end{equation}
Whereas, a contribution of the form $\left(m_1+m_2\right) \left(S_m/\Delta m^2_{31}\right)$, or $m_3 \left(S_m/\Delta m^2_{31}\right)$ can have a maximum value of $\sim O(1)$, in I.O. and N.O., respectively.
This implicit suppression of at least $O(\lambda)$ in the $m_2-m_1$ terms is important, and we consider this effect in our calculations, removing all $m_2-m_1$ terms which lead to negligible contributions, and also considering only terms of the form $m_1+m_2$ or $m_3$ for higher order corrections to the SNSI contribution.

We show the SNSI contribution to the relevant neutrino probabilities for each case in the following subsections. Note that our primary focus is on long-baseline neutrino experiments.

\subsection{In presence of $\eta_{ee}$}
The contribution for the presence of SNSI element $\eta_{ee}$, i.e. Case I (as in equation \ref{MeffCase1}), is shown below,
\begin{subequations}
    \begin{align}
          E_1^{(\eta_{ee})} = & \,\frac{ S_m}{E_\nu } \left( \left(m_1 c_{12}^2 +  m_2 s_{12}^2\right) + \alpha\, \sin ^2\left(2 \theta _{12}\right) \frac{1}{4 \, A_e} \left(m_2-m_1\right)\right. \nonumber \\
          & \left. \qquad -s_{13}^2 \left[ \left(\frac{A_e}{A_e-1}+\frac{1}{\left(A_e-1\right){}^2}\right) \left(m_1 c_{12}^2 + m_2 s_{12}^2 \right)-\frac{A_e}{A_e-1} m_3 \right] \right) \eta_{ee} \\
          E_2^{(\eta_{ee})} = & \,\frac{ S_m} {E_\nu } \left( - \alpha \sin ^2\left(2 \theta _{12}\right) \frac{1}{4 A_e} \left(m_2-m_1\right) \right) \eta_{ee} \\
          E_3^{(\eta_{ee})} = & \,\frac{ S_m} {E_\nu } \left( s_{13}^2 \frac{A_e}{\left(A_e-1\right)^2} \left(m_1 c_{12}^2+ m_2 s_{12}^2 \right)- s_{13}^2 \frac{1}{A_e-1} m_3  \right) \eta_{ee}
    \end{align}
\end{subequations}

We have perturbatively calculated the probabilities, expanding them in the small parameters $\alpha$, $s_{13}$ and $\eta_{\alpha\alpha}$. The probability contribution for $\eta_{ee}$ for the appearance and disappearance channels is shown below. We express the $\eta_{ee}$ contribution to $P_{\mu e}$ as $P_{\mu e}^{(\eta_{ee})} \equiv \left[P_{\mu e}^{(\eta_{ee})} \right]^{(3)} +  \left[P_{\mu e}^{(\eta_{ee})} \right]^{(4)} +  \left[P_{\mu e}^{(\eta_{ee})} \right]^{(5)}$, where the number in the superscript denotes the approximate power of $\lambda$ at which this term contributes, as discussed previously. The SNSI ($\eta_{ee}$) contribution to $P_{\mu e}$ can thus be expressed as:

\begin{subequations}
\allowdisplaybreaks
\begin{align}
    \left[P_{\mu e}^{(\eta_{ee})} \right]^{(3)} = & \, 2 s_{13} \sin \left(2 \theta _{12}\right) \sin \left(2 \theta _{23}\right) \cos \left(\delta _{\text{CP}}+\Delta \right) \frac{\sin \left[ \left(A_e-1\right) \Delta\right]}{A_e-1} \frac{\sin \left[ A_e \Delta \right]}{A_e} (m_2-m_1) \nonumber\\
    &  \quad \times \left( \frac{S_m}{\Delta m^2_{31}} \right) \eta_{ee} \\
    \left[P_{\mu e}^{(\eta_{ee})} \right]^{(4)} = & \, \Bigg[ 2  \alpha \, c_{23}^2 \sin^2 \left(2 \theta _{12}\right) \frac{\sin^2 \left[A_e \Delta \right]}{A_e^2} \Big( m_2-m_1 \Big) \Bigg. \nonumber \\
    & \quad + 8 s_{13}^2 s_{23}^2 \frac{\sin \left[ \left(A_e-1\right) \Delta \right]}{A_e-1} \Bigg( \frac{\sin \left[  \left(A_e-1\right) \Delta \right]}{A_e-1} \, m_3 \Bigg.  \nonumber\\
    &\quad + \frac{1}{A_e-1} \left[ 2 \, \Delta  \cos \left[ \left(A_e-1\right)\Delta \right]- \left(A_e+1\right) \frac{ \sin \left[ \left(A_e-1\right) \Delta \right]}{A_e-1} \right] \nonumber \\
    & \Bigg. \Bigg. \qquad \qquad \quad \times \Big( m_1 c_{12}^2 +m_2 s_{12}^2 \Big) \Bigg) \Bigg]  \left( \frac{S_m}{\Delta m^2_{31}} \right) \eta_{ee} \\
    \left[P_{\mu e}^{(\eta_{ee})} \right]^{(5)} = & \, \alpha s_{13} \sin \left(2 \theta _{23}\right) \sin \left(2 \theta _{12}\right) \cos \left(\delta _{\text{CP}}+\Delta \right) \Bigg[ \Bigg( 2 \,\frac{\sin \left[ A_e \Delta\right]}{A_e} \frac{\sin \left[ \left(A_e-1\right) \Delta \right]}{A_e-1}\, m_3\Bigg.\nonumber\\
    & \Bigg. - \frac{1}{A_e-1} \left[ \frac{2}{A_e} \bigg( \sin \left[\left(A_e-1\right) \Delta  \right] \frac{ \sin \left[ A_e \Delta \right]}{A_e} -\Delta  \sin \left[(2 A_e-1) \Delta \right]  \bigg)  \right. \nonumber\\
    & \Bigg. \left. + \left(A_e+1\right) \frac{\sin \left[ A_e \Delta \right]}{A_e} \frac{\sin \left[ \left(A_e-1\right) \Delta \right]}{A_e-1} \right] \left( m_1 + m_2 \right)\Bigg) \Bigg] \left( \frac{S_m}{\Delta m^2_{31}} \right) \eta_{ee} \; .
\end{align}
\label{eq:pme_etaee}
\end{subequations}
Similarly, the $\eta_{ee}$ contribution to $P_{\mu \mu}$ can be expressed as $P_{\mu \mu}^{(\eta_{ee})} =\left[P_{\mu \mu}^{(\eta_{ee})} \right]^{(3)} +\left[P_{\mu \mu}^{(\eta_{ee})} \right]^{(4)} $, which leads to
\begin{subequations}
    \begin{align}
    \left[P_{\mu \mu}^{(\eta_{ee})} \right]^{(3)} = & \,  2 s_{13} \sin \left(2 \theta _{12}\right) \sin \left(2 \theta _{23}\right) \cos \left(\delta _{\text{CP}}\right) \frac{1}{A_e-1} \times \nonumber\\
    & \Bigg[ \left( \cos \left(2 \theta _{23}\right) \sin ^2(\Delta ) - \frac{\sin \left[A_e\Delta \right]}{A_e} \Big[s_{23}^2 \sin \left[\left(A_e-2\right) \Delta  \right]+ c_{23}^2 \sin \left[A_e \Delta  \right] \Big]  \right) \Bigg. \nonumber\\
    & \Bigg. \quad \times \Big( m_2-m_1 \Big) \Bigg] \left(\frac{S_m}{\Delta m^2_{31}}\right) \eta_{\text{ee}} \\  
    \left[P_{\mu \mu}^{(\eta_{ee})} \right]^{(4)} = & \,  8 s_{13}^2 s_{23}^2 \frac{1}{A_e-1} \Biggr[ \bigg( \Delta c_{23}^2 \sin [2 \Delta] \bigg. \Biggr. \nonumber \\
    & \bigg. \qquad \qquad - \frac{\sin \left[ \left(A_e-1\right) \Delta \right]}{A_e-1} \Big[ 2 \sin (\Delta ) c_{23}^2 \cos \left[ A_e \Delta \right]+\sin \left[ \left(A_e-1\right) \Delta \right] \Big] \bigg) m_3 \nonumber\\
    & \, - \frac{1}{A_e-1} \bigg( \Delta \Big[ \cos \left(2 \theta _{23}\right) \sin (\Delta )  \cos \left[\left(2 A_e-1\right) \Delta  \right]+\cos (\Delta ) \sin \left[\left(2 A_e-1\right) \Delta  \right] \Big. \bigg. \nonumber \\
    & \Big. \quad \qquad \qquad + A_e c_{23}^2 \sin (2 \Delta )   \Big] - \big( A_e+1 \big)  \frac{\sin \left[ \left(A_e-1\right) \Delta \right]}{A_e-1} \times  \nonumber\\
    & \Biggr. \bigg. \quad \qquad \qquad \Big[ 2 c_{23}^2 \sin (\Delta ) \cos \left[ A_e \Delta \right]+\sin \left[ \left(A_e-1\right) \Delta \right] \Big] \bigg) \left(  m_1 c_{12}^2 + m_2 s_{12}^2 \right) \Biggr] \nonumber\\
    &  \qquad \qquad \times  \left(\frac{S_m}{\Delta m^2_{31}}\right) \eta_{\text{ee}} 
    \end{align}
    \label{eq:pmm_etaee}
\end{subequations}
The contribution to the probabilities in the presence of the diagonal SNSI element $\eta_{ee}$ can be expressed as factors of $(m_2 - m_1)$, $(m_1 + m_2)$, $(m_1 c_{12}^2 + m_2 s_{12}^2)$, and $m_3$ for both appearance and disappearance channel. 

\subsection{In presence of $\eta_{\mu \mu}$}

The contribution of $\eta_{\mu\mu}$, i.e. Case II (as in equation \ref{MeffCase2}), is given by
\begin{subequations}
    \begin{align}
          E_1^{(\eta_{\mu\mu})} = & \,\frac{ S_m} {E_\nu } \Bigg( \frac{1}{4} \sin \left(2 \theta _{12}\right) \left[ s_{13}  \sin \left(2 \theta _{23}\right) \cos \left(\delta _{\text{CP}}\right) \frac{1}{A_e-1} + \alpha \, c_{23}^2 \sin \left(2 \theta _{12}\right) \frac{1}{A_e}  \right] \Big( m_2-m_1 \Big) \Bigg. \nonumber\\
          &\Bigg. \qquad +s_{13}^2 s_{23}^2 \left[ -\frac{1}{A_e-1} \Big(m_1 c_{12}^2 + m_2 s_{12}^2 \Big) + \frac{A_e}{\left(A_e-1\right)^2} \, m_3 \right] \Bigg) \eta_{\mu\mu} \\
          E_2^{(\eta_{\mu\mu})} = & \,\frac{ S_m} {E_\nu } \Bigg( c_{23}^2 \Big(m_1 s_{12}^2 + m_2 c_{12}^2 \Big) \Bigg. \nonumber \\
          & \Bigg. \qquad -\frac{1}{4} \sin \left(2 \theta _{12}\right) \left[ s_{13}  \sin \left(2 \theta _{23}\right) \cos \left(\delta _{\text{CP}}\right)+ \alpha \, c_{23}^2 \sin \left(2 \theta _{12}\right) \frac{1}{A_e}  \right]  \Big(m_2-m_1\Big) \Bigg) \eta_{\mu\mu}\\
          E_3^{(\eta_{\mu\mu})} = & \,\frac{ S_m} {E_\nu } \Bigg(s_{23}^2 \, m_3  -\frac{1}{4} s_{13} \sin \left(2 \theta _{12}\right) \sin \left(2 \theta _{23}\right) \cos \left(\delta _{\text{CP}}\right)  \frac{ A_e }{A_e-1} \Big(m_2-m_1\Big) \Bigg. \nonumber\\
          & \Bigg. \qquad + s_{13}^2 s_{23}^2 \left[ \frac{A_e}{A_e-1} \Big(m_1 c_{12}^2 + m_2 s_{12}^2 \Big) - \left(1 + \frac{A_e}{A_e-1}\right) m_3 \right] \Bigg) \eta_{\mu\mu}
    \end{align}
\end{subequations}

We calculate the probability contribution for $\eta_{\mu\mu}$ for the appearance and disappearance channels. Expressing $ P_{\mu e}^{(\eta_{\mu\mu})} \equiv \left[P_{\mu e}^{(\eta_{\mu\mu})} \right]^{(3)}+\left[P_{\mu e}^{(\eta_{\mu\mu})} \right]^{(4)}+\left[P_{\mu e}^{(\eta_{\mu\mu})} \right]^{(5)}$, we obtain
\begingroup
\allowdisplaybreaks
\begin{subequations}
    \begin{align}
        \left[P_{\mu e}^{(\eta_{\mu\mu})} \right]^{(3)} = & \, s_{13} \sin \left(2 \theta _{12}\right) \sin \left(2 \theta _{23}\right) \frac{\sin \left[ \left(A_e-1\right) \Delta \right]}{A_e-1} \Biggr[  \Bigg( \cos \left(\delta _{\text{CP}}\right) \frac{\sin \left[ \left(A_e-1\right) \Delta \right]}{A_e-1} \Bigg.\Biggr. \nonumber \\
        & \; + \frac{1}{A_e-1} \bigg[ \sin (\Delta ) \cos \left(2 \theta _{23}\right) \cos \left(\Delta  A_e+\delta _{\text{CP}}\right)+\sin \left(\Delta  A_e\right) \cos \left(\delta _{\text{CP}}+\Delta \right) \bigg. \nonumber\\
        & \Biggr. \Bigg. \bigg. \qquad \quad -2 c_{23}^2 \cos \left(\delta _{\text{CP}}+\Delta \right) \frac{\sin \left[ A_e \Delta \right]}{A_e} \bigg] \Bigg) \big(m_2 -m_1 \big) \Biggr] \left(\frac{S_m}{\Delta m^2_{31}}\right) \eta _{\mu \mu } \\
        \left[P_{\mu e}^{(\eta_{\mu\mu})} \right]^{(4)} = & \Biggr[ 4 \, s_{13}^2 \, s_{23}^2 \frac{\sin \left[ \left(A_e-1\right) \Delta \right]}{A_e-1} \Bigg( \cos \left(2 \theta _{12}\right) \frac{\sin \left[\left(A_e-1\right) \Delta  \right]}{A_e-1} \Big( m_2-m_1 \Big) \Bigg. \Biggr.  \nonumber\\
        & +\frac{1}{A_e-1} \bigg[ 2 c_{23}^2 \cos (\Delta )\frac{\sin \left[ A_e \Delta \right]}{A_e} \nonumber \bigg. \\
        & \bigg. \qquad \qquad - \Big( \cos \left(2 \theta _{23}\right) \sin (\Delta ) \cos \left[ A_e \Delta \right]+\cos (\Delta ) \sin \left[A_e \Delta  \right] \Big)\bigg] \Big( m_1+m_2 \Big) \nonumber \\
        & \Bigg.  + \frac{2}{A_e-1} \bigg[ \Big(A_e-\cos \left(2 \theta _{23}\right)\Big) \frac{\sin \left[ \left(A_e-1\right) \Delta \right]}{A_e-1} - 2 \Delta  s_{23}^2 \cos \left[ \left(A_e-1\right) \Delta \right]\bigg] m_3 \! \Bigg) \nonumber \\
        & + 2 \,\alpha \, c_{23}^2 \sin ^2\left(2 \theta _{12}\right) \frac{\sin \left[ A_e \Delta \right]}{A_e} \frac{1}{A_e-1} \Bigg( -c_{23}^2 \frac{\sin \left[ A_e \Delta \right]}{A_e}  \Bigg. \nonumber\\
        & \Biggr. \Bigg. \quad + \Big[ \sin \left[ A_e \Delta \right]-s_{23}^2 \sin (\Delta ) \cos \left[\left(A_e-1\right) \Delta  \right] \Big]\Bigg)\Big( m_2-m_1 \Big)\Biggr] \left(\frac{S_m}{\Delta m^2_{31}}\right) \eta _{\mu \mu } \\
        \left[P_{\mu e}^{(\eta_{\mu\mu})} \right]^{(5)} = & \, \alpha s_{13} \sin \left(2 \theta _{12}\right) \sin \left(2 \theta _{23}\right) \Biggr[ \Bigg(  s_{23}^2 \frac{\sin \left[\left(A_e-1\right) \Delta \right]}{A_e-1} \frac{1}{A_e-1} \times \Bigg. \Biggr. \nonumber\\
        & \bigg[ \sin \left(\delta _{\text{CP}}\right) \cos \left[\left(A_e-1\right) \Delta  \right] -\sin \left[\left(A_e+1\right) \Delta +\delta _{\text{CP}}\right] \bigg. \nonumber\\
        &\bigg. \qquad + 2 \cos \left(\delta_{\text{CP}}+\Delta \right)\frac{\sin \left[ A_e \Delta \right]}{A_e} \bigg] \nonumber\\
        & -2 \Delta  c_{23}^2 \cos \left[\left(A_e+1\right) \Delta  +\delta _{\text{CP}}\right] \frac{1}{A_e} \frac{\sin \left[ \left(A_e-1\right) \Delta \right]}{A_e-1} -c_{23}^2 \frac{\sin \left[ A_e \Delta \right]}{A_e} \times\nonumber\\
        & \bigg[ \frac{1}{A_e-1} \Big( \sin \left(\Delta  A_e-\delta _{\text{CP}}\right)+ \sin \left(\delta _{\text{CP}}+\Delta \right) \cos \left[\left(A_e-1\right) \Delta  \right] \Big) \bigg. \nonumber\\
        &\Bigg. \bigg. -2 \cos \left(\delta _{\text{CP}}\right) \frac{1}{A_e-1} \frac{\sin \left[ A_e \Delta \right]}{ A_e} -2 \cos \left(\delta _{\text{CP}}+\Delta \right) \frac{1}{A_e} \frac{\sin \left[ \left(A_e-1\right) \Delta \right]}{A_e-1} \bigg] \Bigg) \nonumber\\
        & \times \Big(m_1+m_2\Big) \nonumber\\
        &+ \frac{1}{A_e-1} \Bigg( 2 \frac{\sin \left[\left(A_e-1\right) \Delta  \right]}{A_e-1} \Big[ \sin \left(\Delta  A_e\right) \cos \left(\delta _{\text{CP}}+\Delta \right) \Big. \Bigg. \nonumber\\
        & \Big. \; - s_{23}^2 \sin (\Delta ) \cos \left(\Delta  A_e+\delta _{\text{CP}}\right) \Big] - \frac{\sin \left[ A_e \Delta \right]}{A_e} \bigg[ 4 \Delta s_{23}^2 \cos \left[\left(A_e-2\right)\Delta  -\delta _{\text{CP}}\right] \bigg.\nonumber\\
        & \Biggr. \Bigg. \bigg. \; - \left(1-3 \cos \left(2 \theta _{23}\right)\right) \cos \left(\Delta + \delta _{\text{CP}}\right) \frac{\sin \left[ \left(A_e-1\right) \Delta \right]}{A_e-1} \bigg]\Bigg) m_3 \Biggr] \left(\frac{S_m}{\Delta m^2_{31}}\right) \eta _{\mu \mu }
        \end{align}
        \label{eq:pme_etamm}
\end{subequations}
\endgroup
The $\eta_{\mu \mu}$ contribution to the survival channel is given by,
\begin{align}\label{eq:pmm_etamm}
    P_{\mu \mu}^{(\eta_{\mu \mu})} = & \; 2 \sin ^2\left(2 \theta _{23}\right) \sin [\Delta ]\nonumber \\
    & \times \Bigg[ \biggl( 2 \Delta c_{23}^2 \cos [\Delta ]- \cos \left(2 \theta _{23}\right) \sin [\Delta ]  \biggr) \Big( m_1 s_{12}^2 +m_2 c_{12}^2 \Big)  \Big. \nonumber\\
    & \, \Big. \quad -\biggl(2\, \Delta s_{23}^2 \cos [\Delta] + \cos \left(2 \theta _{23}\right) \sin [\Delta ] \biggr) m_3 \Bigg]  \left(\frac{S_m}{\Delta m^2_{31}}\right) \eta _{\mu \mu }
\end{align}
The contribution to $P_{\mu e}$ in the presence of the diagonal SNSI element $\eta_{\mu\mu}$ can be expressed as factors of $(m_2 - m_1)$, ($m_1 + m_2$), $(m_1 c_{12}^2 + m_2 s_{12}^2)$, and $m_3$. However, the contribution to $P_{\mu\mu}$ can be expressed as factors of $(m_1 s_{12}^2 + m_2 c_{12}^2)$ and $m_3$ only.

\subsection{In presence of $\eta_{\tau \tau}$}

The contribution of $\eta_{\tau \tau}$ i.e. Case III (as in equation \ref{MeffCase3}), is given by the modifications to the eigenvalues expressed as
\begin{subequations}
\begin{align}
    E_1^{(\eta_{\tau \tau})} = &\frac{  S_m} {E_\nu } \Bigg(\frac{1}{4} \sin \left(2 \theta _{12}\right) \left[ - s_{13} \sin \left(2 \theta _{23}\right) \cos \left(\delta _{\text{CP}}\right) \frac{1}{A_e-1} + \alpha \, s_{23}^2 \sin \left(2 \theta _{12}\right) \frac{1}{A_e} \right] \Big( m_2 -m_1 \Big) \Bigg. \nonumber\\
    & \Bigg. + s_{13}^2 \, c_{23}^2 \left[ -\frac{1}{A_e-1} \Big( m_1 c_{12}^2 + m_2 s_{12}^2 \Big) + \frac{A_e}{\left(A_e-1\right)^2} \, m_3 \right] \Bigg) \, \eta_{\tau\tau} \,\\
    E_2^{(\eta_{\tau \tau})} = & \frac{  S_m }{E_\nu } \Bigg( s_{23}^2 \left(m_1 s_{12}^2 + m_2 c_{12}^2 \right)\Bigg. \nonumber\\
    &\Bigg. \qquad +\frac{1}{4} \sin \left(2 \theta _{12}\right) \left[s_{13} \sin \left(2 \theta _{23}\right) \cos \left(\delta _{\text{CP}}\right) - \alpha \, s_{23}^2 \sin \left(2 \theta _{12}\right) \frac{1}{A_e} \right] \Big(m_2-m_1\Big) \Bigg) \, \eta_{\tau\tau}\, \\
    E_3^{(\eta_{\tau \tau})} = & \frac{ S_m }{E_\nu } \Bigg(  c_{23}^2 m_3 + \frac{1}{4} s_{13} \sin \left(2 \theta _{12}\right) \sin \left(2 \theta _{23}\right) \cos \left(\delta _{\text{CP}}\right)  \frac{A_e}{A_e-1} \Big(m_2 -m_1\Big) \Bigg.\nonumber\\
    &\Bigg. \qquad +s_{13}^2 \, c_{23}^2 \left[ \frac{A_e}{A_e-1} \Big( m_1 c_{12}^2 + m_2 s_{12}^2 \Big) - \left(1+\frac{A_e}{A_e-1}\right) \, m_3 \right] \Bigg) \, \eta_{\tau\tau}\, .
\end{align}
\end{subequations}

Similar to the previous subsections, the contribution for $\eta_{\tau\tau}$ to the neutrino survival and conversion probabilities can also be calculated. Defining the SNSI contribution to the conversion probability as $ P_{\mu e}^{(\eta_{\tau\tau})} = \left[P_{\mu e}^{(\eta_{\tau\tau})} \right]^{(3)} + \left[P_{\mu e}^{(\eta_{\tau\tau})} \right]^{(4)} + \left[P_{\mu e}^{(\eta_{\tau\tau})} \right]^{(5)}$, we calculate $P_{\mu e}^{(\eta_{\tau\tau})}$ and obtain

\begin{subequations}
\allowdisplaybreaks
    \begin{align}
        \left[P_{\mu e}^{(\eta_{\tau\tau})} \right]^{(3)} = & \, 2 s_{13} s_{23}^2 \sin \left(2 \theta _{12}\right) \sin \left(2 \theta _{23}\right) \frac{\sin \left[\left(A_e-1\right) \Delta  \right]}{A_e-1} \frac{1}{A_e-1} \nonumber\\
        & \; \times \Biggr[\Bigg(\cos \left(\Delta  A_e+\delta _{\text{CP}}\right) \sin (\Delta ) - \cos \left(\delta _{\text{CP}}+\Delta \right) \frac{\sin \left[ A_e \Delta \right]}{A_e} \Bigg) \Big(m_2-m_1\Big)\Biggr] \nonumber\\
        &\quad \times \left(\frac{S_m}{\Delta m^2_{31}}\right) \eta _{\tau \tau } \\
        \left[P_{\mu e}^{(\eta_{\tau\tau})} \right]^{(4)} = & \, \sin ^2\left(2 \theta _{23}\right) \frac{1}{A_e-1} \Biggr[ \alpha  \sin ^2\left(2 \theta _{12}\right) \frac{\sin \left[ A_e \Delta \right]}{4 A_e}  \Biggr.  \nonumber\\
        & \; \times \Bigg( \Big[\sin \left[ A_e \Delta \right]-\sin \left[ \left(A_e-2\right) \Delta \right] \Big] -2\frac{\sin \left[ A_e \Delta \right]}{A_e} \Bigg) \Big(m_2 -m_1\Big)\nonumber\\
        & \; -2 s_{13}^2 \frac{\sin \left[\left(A_e-1\right) \Delta  \right]}{A_e-1} \Bigg( \bigg[\cos (\Delta )  \frac{\sin \left[ A_e \Delta \right]}{A_e}-\sin (\Delta ) \cos \left[ A_e \Delta \right] \bigg] \Big( m_1 + m_2 \Big) \Bigg. \nonumber\\
        & \Biggr. \Bigg. \quad \quad + 2 \bigg[ \Delta  \cos \left[ \left(A_e-1\right) \Delta \right]-\frac{\sin \left[\left(A_e-1\right) \Delta \right]}{A_e-1} \bigg] m_3 \Bigg) \Biggr] \left(\frac{S_m}{\Delta m^2_{31}}\right) \eta _{\tau \tau } \\
        & \nonumber\\ 
        \left[P_{\mu e}^{(\eta_{\tau\tau})} \right]^{(5)} = & \, \alpha s_{13} \sin \left(2 \theta _{12}\right) \sin \left(2 \theta _{23}\right) \times \nonumber\\
        &\! \Biggl[ \Bigg( \! 2 c_{23}^2 \frac{\sin \left[ A_e \Delta \right]}{A_e} \frac{1}{A_e-1} \bigg[ \sin (\Delta ) \cos \left[ \left(A_e-1\right) \Delta -\delta _{\text{CP}}\right]- \cos \left(\delta _{\text{CP}}\right) \frac{\sin \left[ A_e \Delta \right]}{A_e} \bigg]\Bigg. \Biggr. \nonumber\\
        & + s_{23}^2 \frac{\sin \left[ \left(A_e-1\right) \Delta \right]}{A_e-1} \frac{1}{A_e} \bigg[ \cos \left(\delta _{\text{CP}}\right) \sin (\Delta )  \cos \left[A_e \Delta  \right]\frac{A_e}{A_e-1}  \bigg. \nonumber\\
        & \quad + \frac{\sin \left[ A_e \Delta \right]}{A_e} \left(  \cos \left(\delta _{\text{CP}}\right)\cos (\Delta )  \left(1-\frac{1}{A_e-1}\right) -\left(A_e+2\right)  \sin \left(\delta _{\text{CP}}\right) \sin (\Delta ) \right) \nonumber\\
        &\Bigg. \bigg. \quad - 2 \Delta  \cos \left[\left(A_e+1\right) \Delta  +\delta _{\text{CP}}\right] \bigg] \Bigg) \Big( m_1+m_2 \Big) \nonumber\\
        & -\frac{1}{A_e-1} \Bigg( 4 \Delta  c_{23}^2 \cos \left[ \left(A_e-2\right) \Delta - \delta _{\text{CP}}\right] \frac{\sin \left[ A_e \Delta \right]}{A_e} \Bigg. \nonumber\\
        & \quad - \frac{\sin \left[ \left(A_e-1\right) \Delta \right]}{A_e-1} \bigg[ \Big(1+ 3 \cos \left(2 \theta _{23}\right)\Big) \cos \left(\delta _{\text{CP}}+\Delta \right) \frac{\sin \left(\Delta  A_e\right)}{A_e} \bigg. \nonumber\\
        & \Bigg. \Biggl. \bigg. \qquad \qquad \qquad + 2 s_{23}^2 \sin (\Delta ) \cos \left[ A_e \Delta  +\delta _{\text{CP}}\right] \bigg] \Bigg) m_3 \Biggr] \left(\frac{S_m}{\Delta m^2_{31}}\right) \eta _{\tau \tau } 
    \end{align}
    \label{eq:pme_etatt}
\end{subequations}
The SNSI contribution to the survival probability is given by
\begin{align}\label{eq:pmm_etatt}
    P_{\mu \mu}^{(\eta_{\tau\tau})} = & \;  2 \sin ^2\left(2 \theta _{23}\right) \sin [\Delta ] \nonumber\\
    & \, \times \Bigg[ \biggl( 2 \Delta s_{23}^2 \cos [\Delta ] +\cos \left(2 \theta _{23}\right) \sin [\Delta]  \biggr)
    \Big( m_1 s_{12}^2+m_2 c_{12}^2 \Big) \nonumber \Bigg.\\
    & \Bigg. \qquad -\Big( 2 \Delta c_{23}^2 \cos [\Delta ] -\cos \left(2 \theta _{23}\right)\sin [\Delta ] \Big) m_3 \Bigg] \left(\frac{S_m}{\Delta m^2_{31}}\right) \eta _{\tau \tau }
\end{align}
The contribution to $P_{\mu e}$ in the presence of the diagonal SNSI element $\eta_{\tau\tau}$ can be expressed as factors of $(m_2 - m_1)$, ($m_1 + m_2$), and $m_3$. However, the contribution to $P_{\mu\mu}$ can be expressed as factors of $(m_1 s_{12}^2 + m_2 c_{12}^2)$ and $m_3$ only.

In the next section, we highlight the key features of the SNSI contribution for the $P_{\mu e}$ and $P_{\mu \mu}$ channels. We also probe the accuracy of our analytic approximations obtained in this section and show that the analytic expressions would be applicable for future long-baseline experiments.

\section{The Applicability of the Analytic Probability Expressions} \label{sec:applicability}

Analytic expressions for neutrino oscillation probabilities offer crucial insights into how neutrino oscillations depend on parameters like the PMNS mixing angles, mass-squared differences, and the CP-violating phase. These expressions reveal the functional dependence of oscillation probabilities on factors like baseline distance and neutrino energy, allowing us to identify optimal conditions for precise measurement of specific parameters.

While numerical methods are commonly used to calculate exact oscillation probabilities by diagonalizing the Hamiltonian, these do not provide intuitive insight into the role of mixing parameters. We have calculated exact probabilities numerically using the time evolution operator in equation \ref{eq:time_evol}, primarily for validation and comparison with the analytic approximations. The main purpose of the analytic approximations,  which omit higher-order terms, is to highlight the interesting features and dependence which will be discussed further in the following sections.

We quantify the  accuracy of our analytic expressions for the SNSI contributions by defining the quantity $\Delta P_{\alpha \beta}^{(\eta_{\alpha\alpha})}$, the absolute error for the SNSI contributions, explicitly as given below:
\begin{equation}\label{eq:accuracy_snsi}
\Delta P_{\alpha \beta}^{(\eta_{\alpha\alpha})}    =|P_{\alpha \beta}^{(\eta_{\alpha\alpha})} (\text{Numerical})-P_{\alpha \beta}^{(\eta_{\alpha\alpha})} (\text{Analytic})| \; ,
\end{equation}
where,
\begin{itemize}
    \item $\alpha, \beta$ = the neutrino flavors (e, $\mu$, $\tau$) ,
    \item $P_{\alpha \beta}^{(\eta_{\alpha\alpha})} \rm(Numerical)$ are the SNSI contributions to the exact oscillation probabilities calculated using the numerical recipe,
    \item $P_{\alpha \beta}^{(\eta_{\alpha\alpha})} \rm(Analytic)$ are the SNSI contributions to the oscillation probabilities calculated using the analytic formulae given in section \ref{sec:ana_prob}.
\end{itemize}

We first compare the numerical and analytic expressions for $P_{\mu e}^{(\eta_{\alpha \alpha})}$ and $P_{\mu \mu}^{(\eta_{\alpha \alpha})}$ for both NO and IO.  We take $E_\nu=(0.5-5)$ GeV and L $=$ 1300 km, which correspond to DUNE. The values of the standard oscillation parameters used are as shown in table \ref{tab:3nu_params}. All other parameters except the sign of $\Delta m_{31}^2$ are kept fixed for both NO and IO. These values of the oscillation parameters are consistent with the 3$\sigma$ bound for the current best-fit values \cite{Esteban:2020cvm,ParticleDataGroup:2024cfk, NuFIT_5.3,deSalas:2020pgw,Capozzi:2021fjo}.

\begin{figure}[!t]
        \hspace{1.1cm}
        \includegraphics[width=0.9\linewidth]{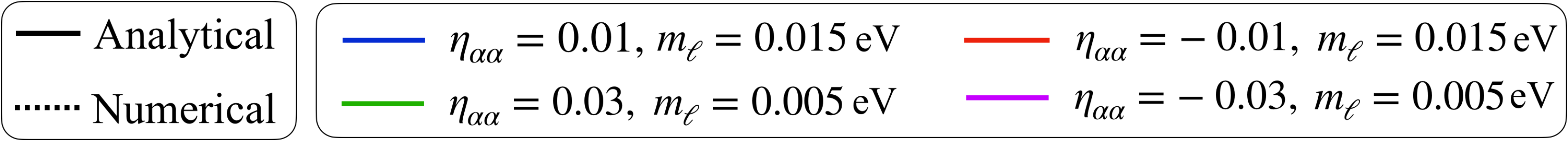} \\[6pt]
       \includegraphics[width=0.325\linewidth]{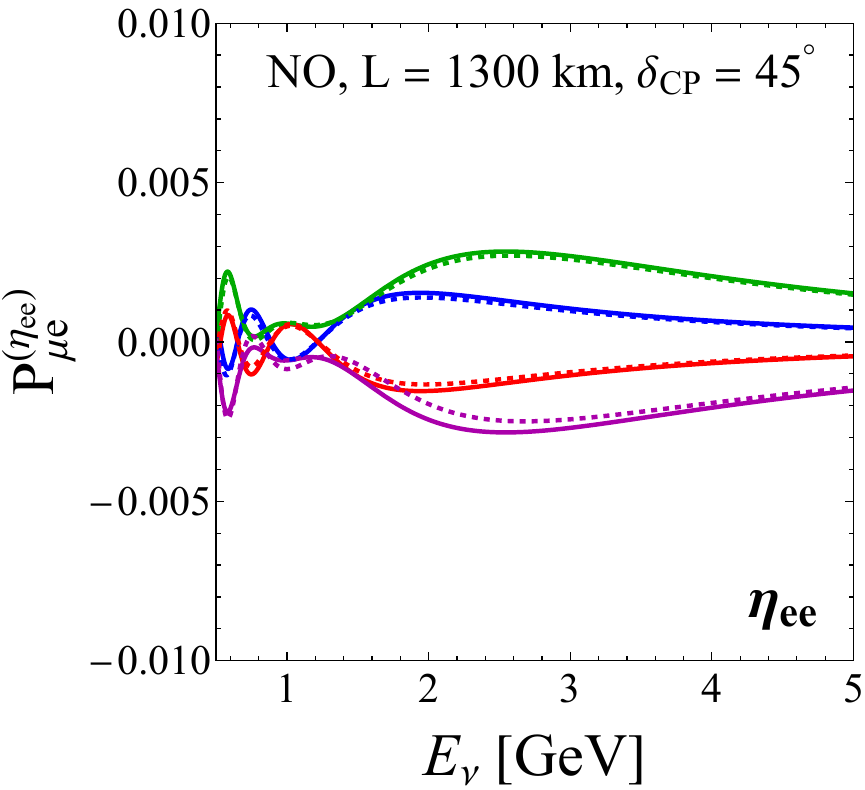} 
 	\includegraphics[width=0.325\linewidth]{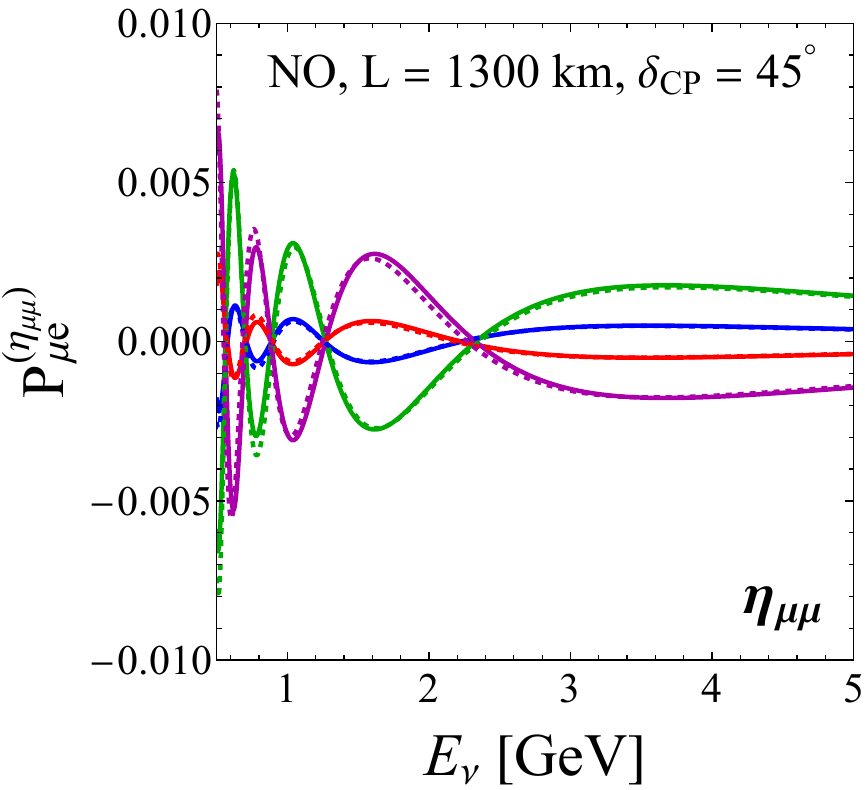} 
        \includegraphics[width=0.325\linewidth]{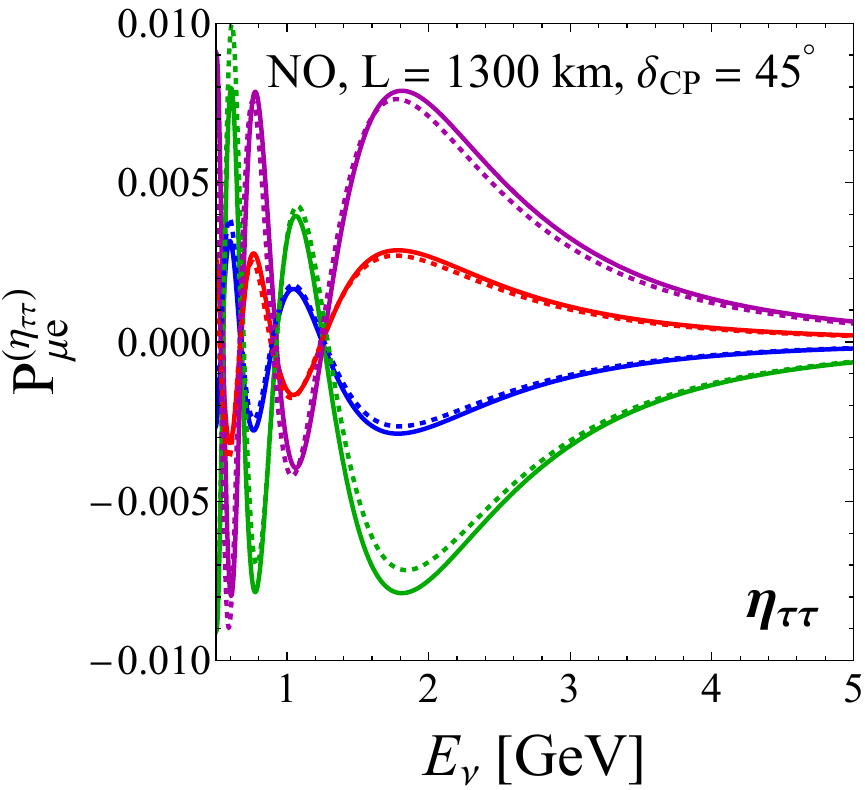} \\
       \includegraphics[width=0.325\linewidth]{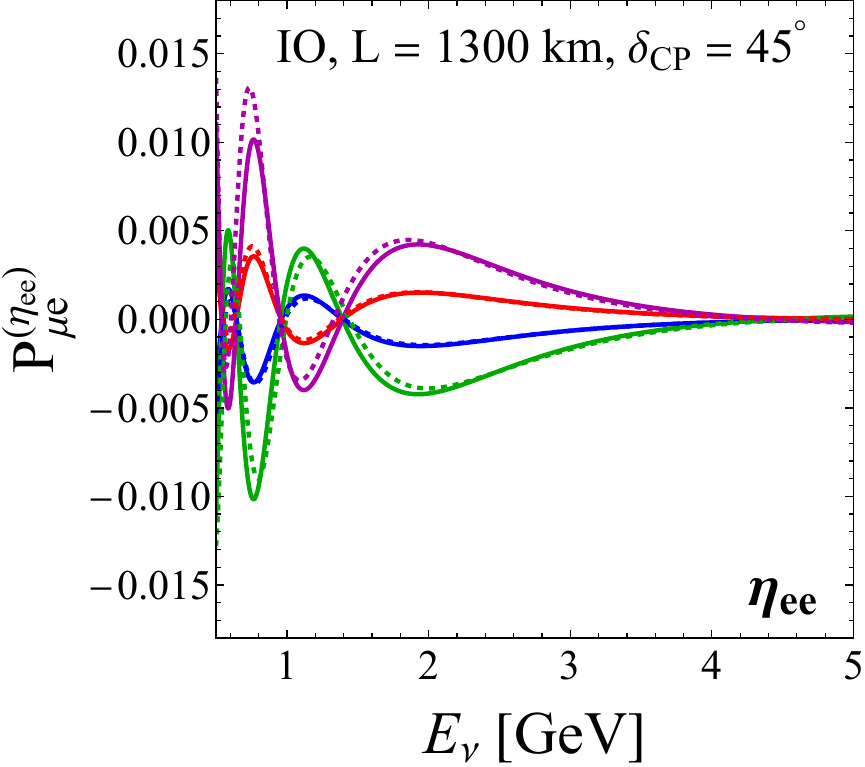} 
 	\includegraphics[width=0.325\linewidth]{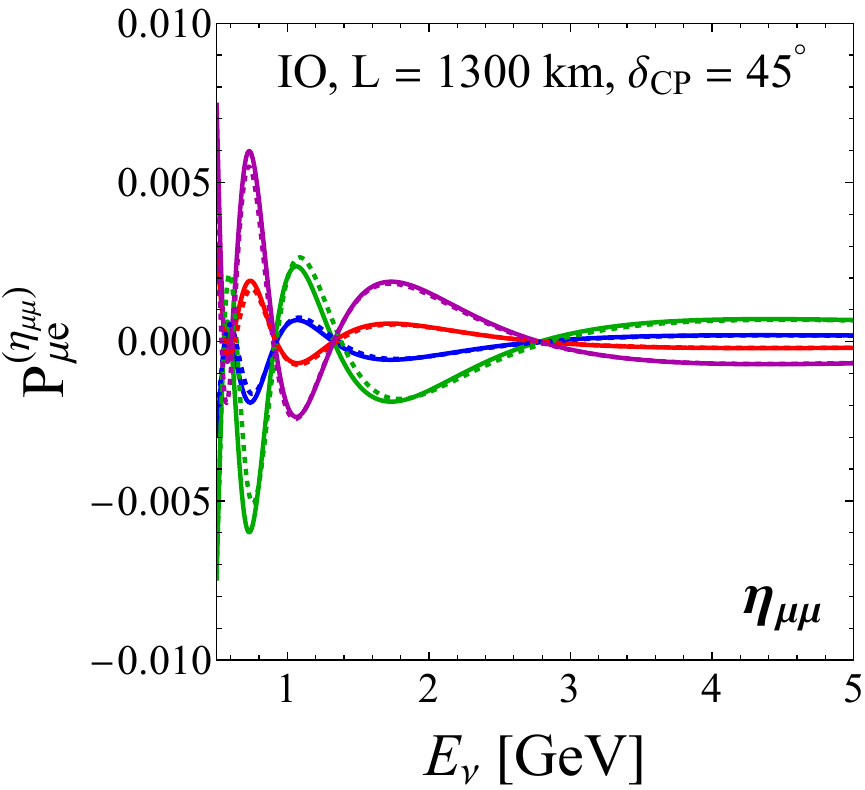} 
        \includegraphics[width=0.325\linewidth]{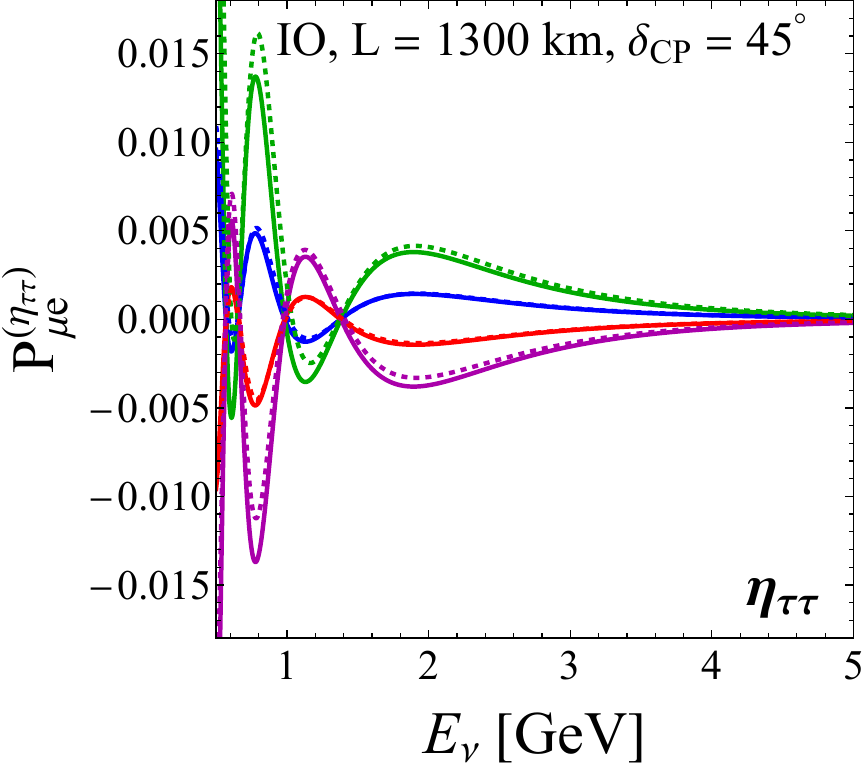} 

 	\caption{ $P_{\mu e}^{\eta_{\alpha\alpha}}$ vs $E_\nu$ for $\eta_{ee}$ (left), $\eta_{\mu \mu}$ (middle) and $\eta_{\tau \tau}$ (right), with  NO (top panel) and IO (bottom panel). The solid (dashed) lines correspond to analytically (numerically) calculated values.}
 	\label{fig:pme_snsi}
 \end{figure}

 In figure \ref{fig:pme_snsi}, we show the contribution of the SNSI term $(\eta_{\alpha \alpha})$ to appearance channel $P_{\mu e}^{(\eta_{\alpha \alpha})}$. The dashed lines correspond to numerically calculated probabilities and the solid lines are obtained from our analytic expressions shown in equations \ref{eq:pme_etaee}, \ref{eq:pme_etamm} and \ref{eq:pme_etatt}. The green and blue (red and magenta) lines correspond to positive (negative) values of $\eta_{\alpha \alpha}$ parameters, We plot for two different choices ($0.005$ eV and $0.015$ eV) for the mass of the lightest neutrino $m_\ell$ (i.e. $m_1$ in NO, and $m_3$ in IO). From the figure, we note the following:
\begin{itemize}
    \item For both $\eta_{\mu \mu}$ and $\eta_{\tau \tau}$, in both NO and IO, the SNSI contribution to $P_{\mu e}$ is observed to scale proportionally with the value of $\eta_{\alpha\alpha}$, without any explicit change happening due to the variation in the mass of the lightest neutrino. On the other hand, for $\eta_{ee}$, in NO, we observe a non-trivial change as we go from $(\eta_{ee}=0.01,~m_1=0.015 \text{ eV})$ to $(\eta_{ee}=0.03,~m_1=0.005 \text{ eV})$. We discuss in detail this dependence on the $m_1$ value in section~\ref{sec:Mass_dep}.

    \item Since we only consider SNSI contribution linear in $\eta_{\alpha\alpha}$, we observe that our analytic estimates are symmetric around zero. However, note that the numerical results deviate from this behavior. This is due to higher order contributions, such as those from $\eta_{\alpha\alpha}^2$ terms.

    \item Note that, due to the complex nature of the SNSI contributions, even for the same value of $\eta_{ee}$, $\eta_{\mu\mu}$, and $\eta_{\tau\tau}$, the height and positions of the peaks and the dips can differ.
    Furthermore, the mass ordering is also observed to affect the SNSI contribution, as predicted.
    
    \item In the presence of $\eta_{ee}$, the analytic and numerical contributions agree well in both NO and IO scenarios. For the chosen values of $\eta_{ee}$ and $m_\ell$, the analytic expression matches the numerically observed behaviour quite well, with a slight variation at the position and height of peaks and dips observed for higher $\eta_{ee}$ values.
    
    \item In the presence of $\eta_{\mu \mu}$, for both the mass orderings, the numerical and analytic contributions match very well with a slight shift in the position of peaks and dips for higher $\eta_{\mu \mu}$ values.
    
    \item  In the presence of $\eta_{\tau \tau}$ as well, for both NO and IO, we observe a good match between the analytic and the numerical probabilities.
\end{itemize}

  \begin{figure}[t]
         \hspace{1.cm}
          \includegraphics[width=0.89\linewidth]{Updated_plots/plot_legend_one_row.pdf} \\[6pt]
 	 \includegraphics[width=0.33\linewidth]{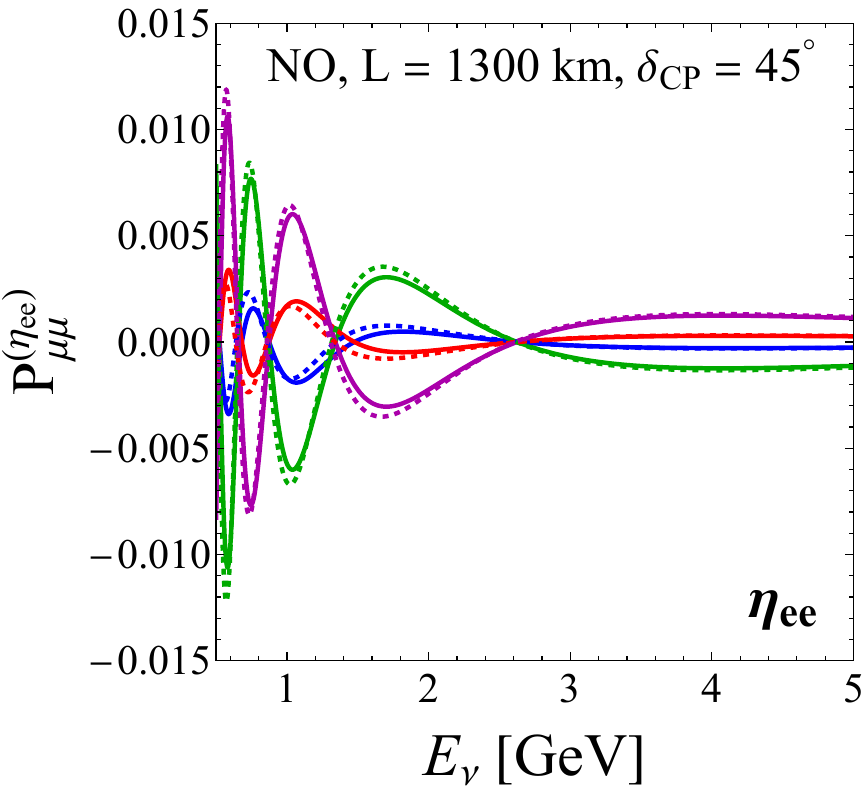} 
 	\includegraphics[width=0.31\linewidth]{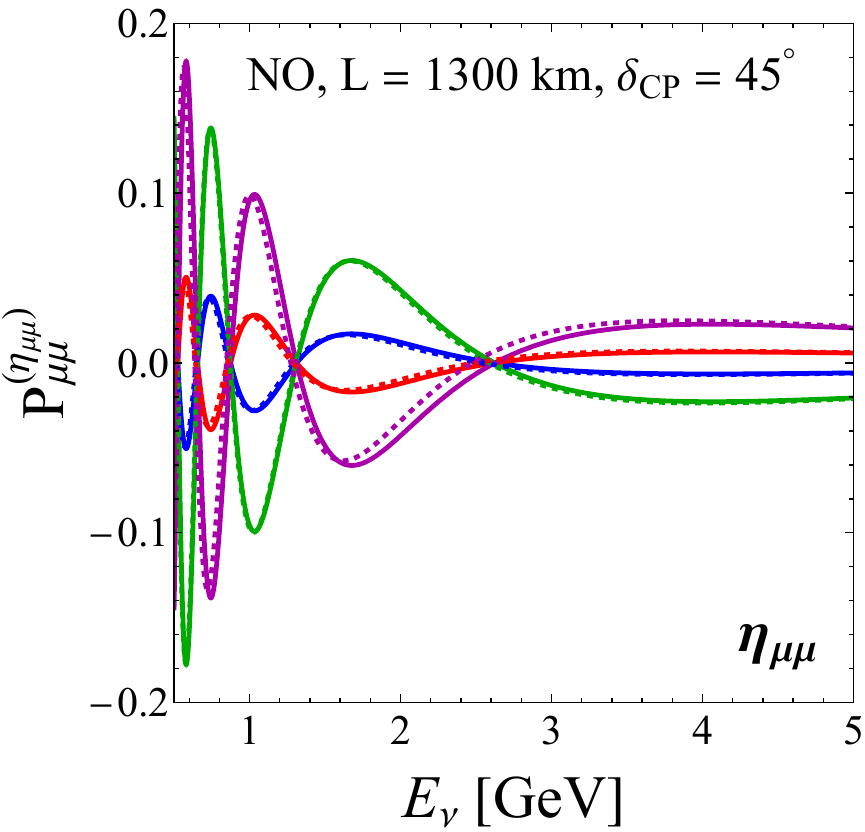} 
        \includegraphics[width=0.31\linewidth]{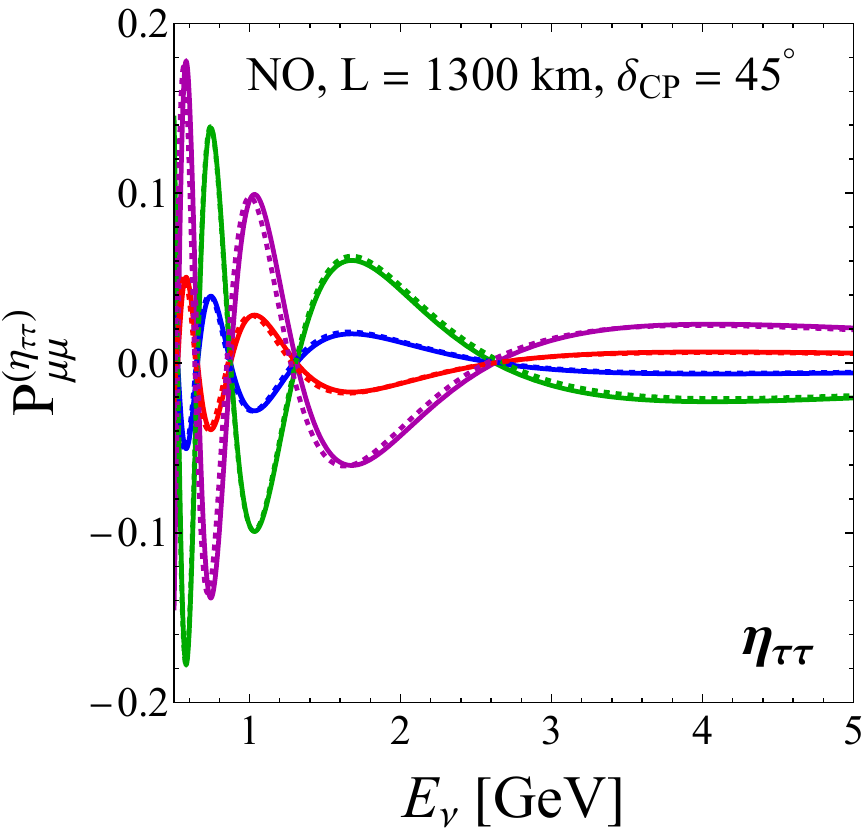} \\
       \includegraphics[width=0.33\linewidth]{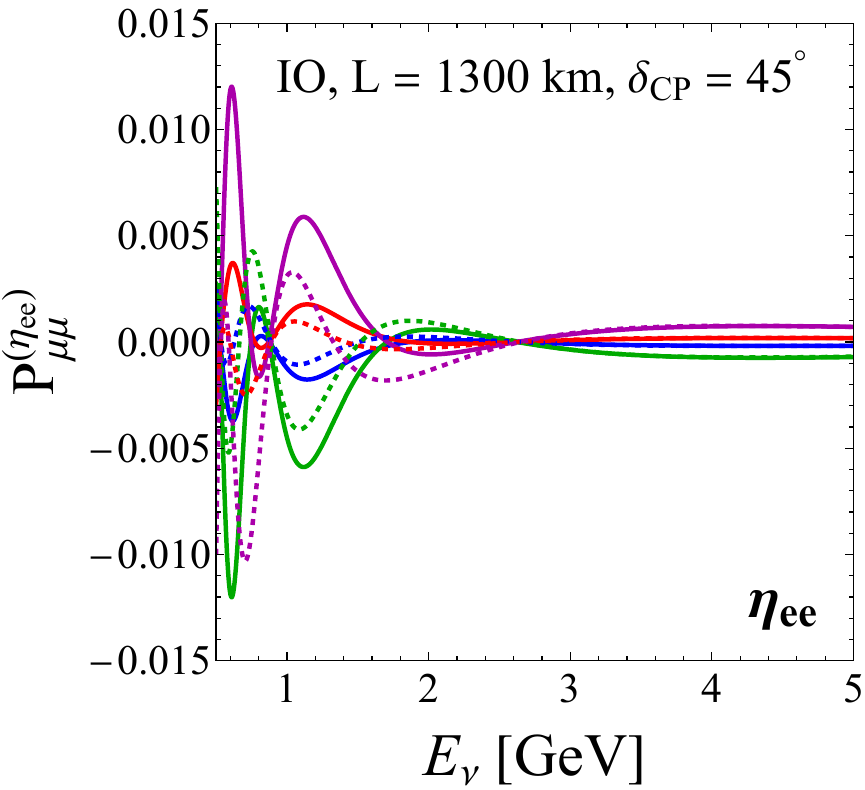} 
 	\includegraphics[width=0.31\linewidth]{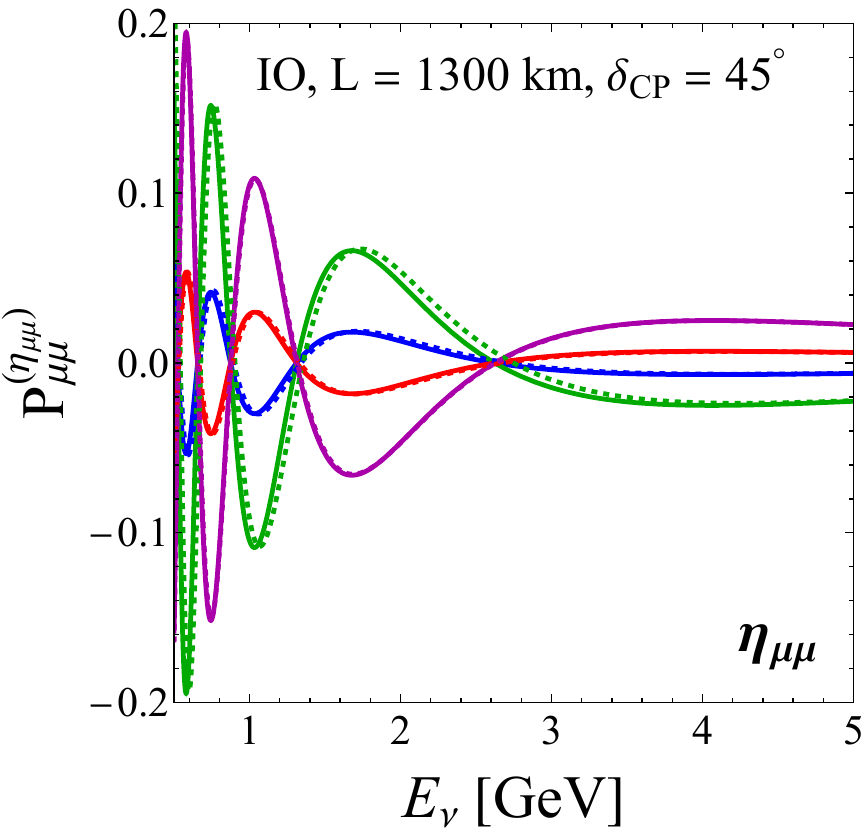} 
        \includegraphics[width=0.31\linewidth]{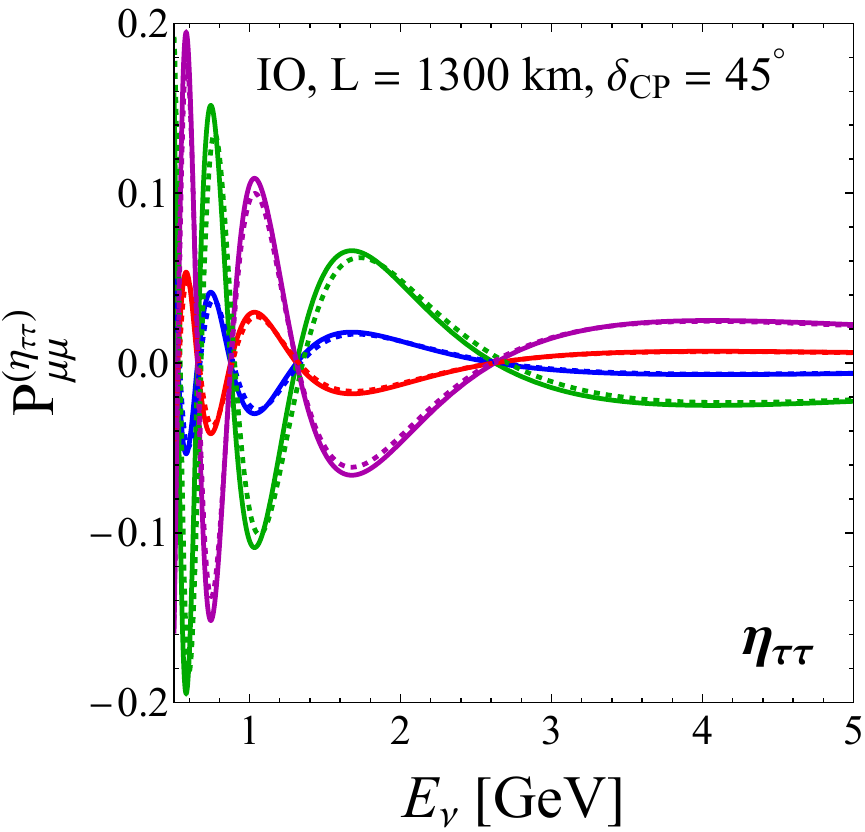} 

 	\caption{ $P_{\mu \mu}^{\eta_{\alpha\alpha}}$ vs $E_\nu$ for $\eta_{ee}$ (left), $\eta_{\mu \mu}$ (middle) and $\eta_{\tau \tau}$ (right), with  NO (top panel) and IO (bottom panel). The solid (dashed) lines correspond to analytically (numerically) calculated values. }
 	\label{fig:pmm_snsi}
 \end{figure}
 In figure \ref{fig:pmm_snsi}, we similarly present the SNSI contribution to the disappearance probability $P_{\mu \mu}$. Note that, the solid lines are obtained using the analytic expressions in equations \ref{eq:pmm_etaee}, \ref{eq:pmm_etamm} and \ref{eq:pmm_etatt}. The dashed lines are the numerically calculated probabilities. The contribution for both positive (green and blue) and negative values (red and magenta) are shown in the figure. We observe the following.

\begin{itemize}
   \item  Note that the SNSI contribution to $P_{\mu\mu}$ for $\eta_{\mu\mu}$ and $\eta_{\tau\tau}$ have similar form and amplitude, with both the SNSI contributions reaching a value of $\sim 0.2$ for $\eta_{\alpha\alpha} = 0.03$, whereas, the SNSI contribution for $\eta_{ee}$ differs, and is much smaller.

    \item Similar to the observations made for $P_{\mu e}$, the numerical results for the SNSI contribution are not fully symmetric around zero due to  $\eta_{\alpha\alpha}^2$ contributions.

    \item Interestingly, unlike the observations made for $P_{\mu e}$, for both $\eta_{\mu\mu}$ and $\eta_{\tau\tau}$, the SNSI contribution does not change significantly for the two mass ordering (NO vs. IO) scenarios.
   
    \item In case of NO, we observe a good match between the analytic and numerical SNSI contributions for both positive and negative values of $\eta_{ee}$ with small deviation at the peaks and dips.  For IO, however, we see that the analytic probabilities deviate from the numerical probabilities for both positive and negative values of $\eta_{ee}$. We also observe a change in the position of peaks and dips. However, due to the smallness of the $\eta_{ee}$ contribution to $P_{\mu\mu}$, the effect of a non-zero $\eta_{ee}$ will be difficult to observe in upcoming neutrino experiments.
    
    \item The positive values of $\eta_{\mu\mu}$ show a better match of analytic with the numerical values. However, for higher negative (positive) $\eta_{\mu\mu}$ in NO (IO) analytic contributions are slightly shifted in energy w.r.t. numerical probabilities. 
    
    \item We also observe a similar match of the analytic probabilities with the numerical one in the case of the SNSI parameter $\eta_{\tau\tau}$ for both NO and IO.
\end{itemize}

 \begin{figure}[!t]
 	\centering
        \includegraphics[width=0.73\linewidth]
        {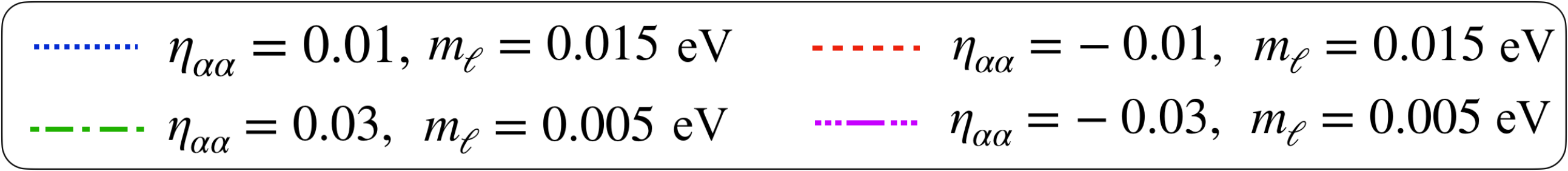} \\
        \vspace{5pt}
        \includegraphics[width=0.327\linewidth]{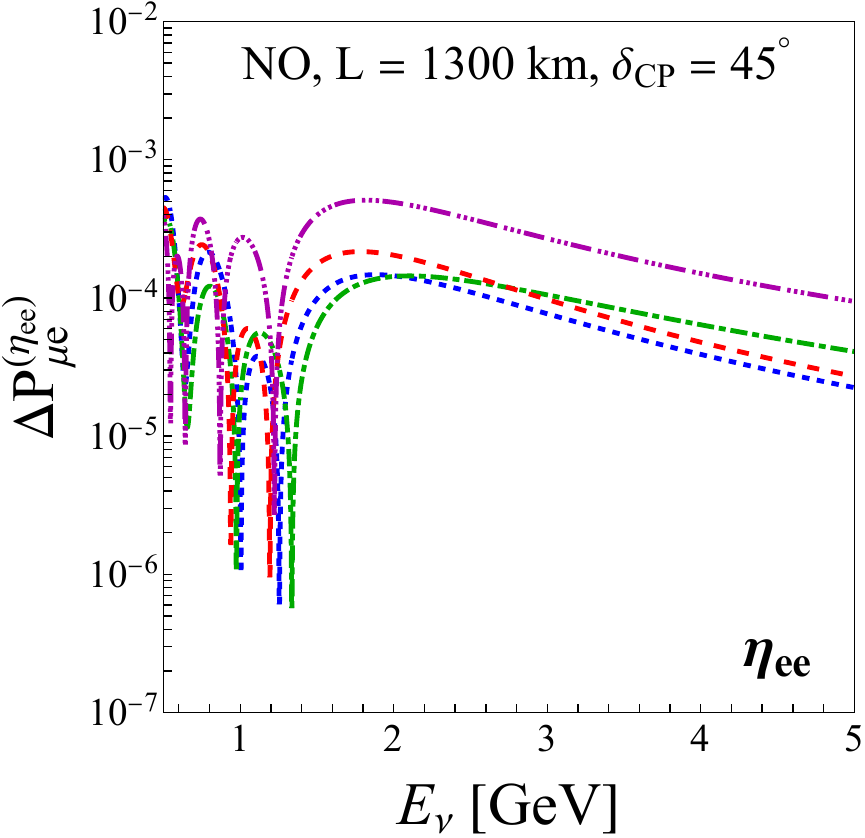} 
        \includegraphics[width=0.327\linewidth]{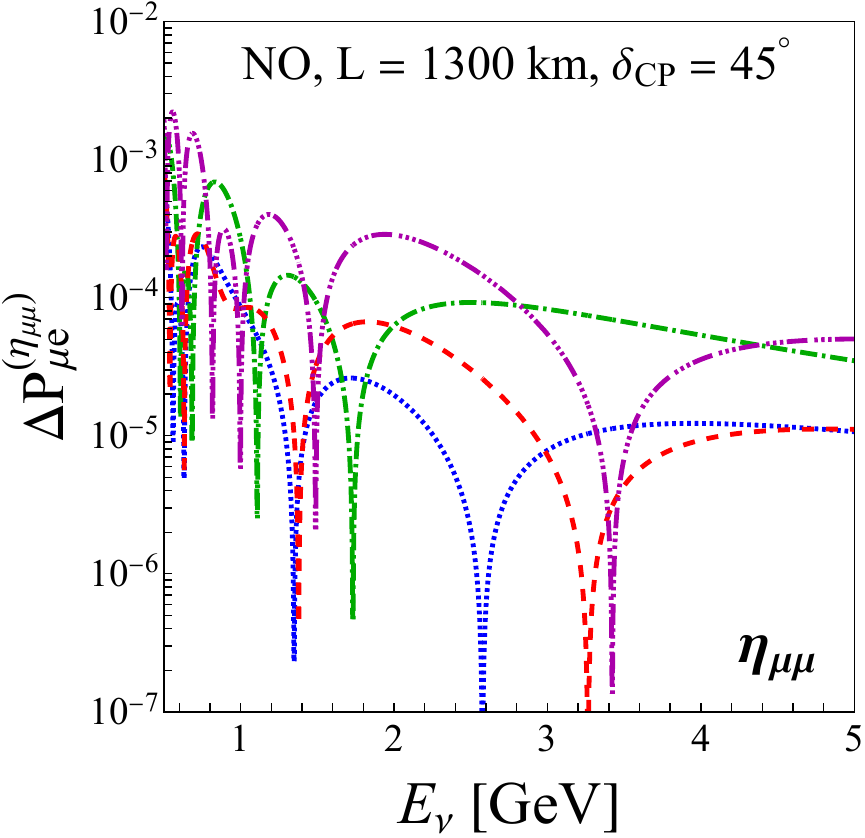} 
 	\includegraphics[width=0.327\linewidth] {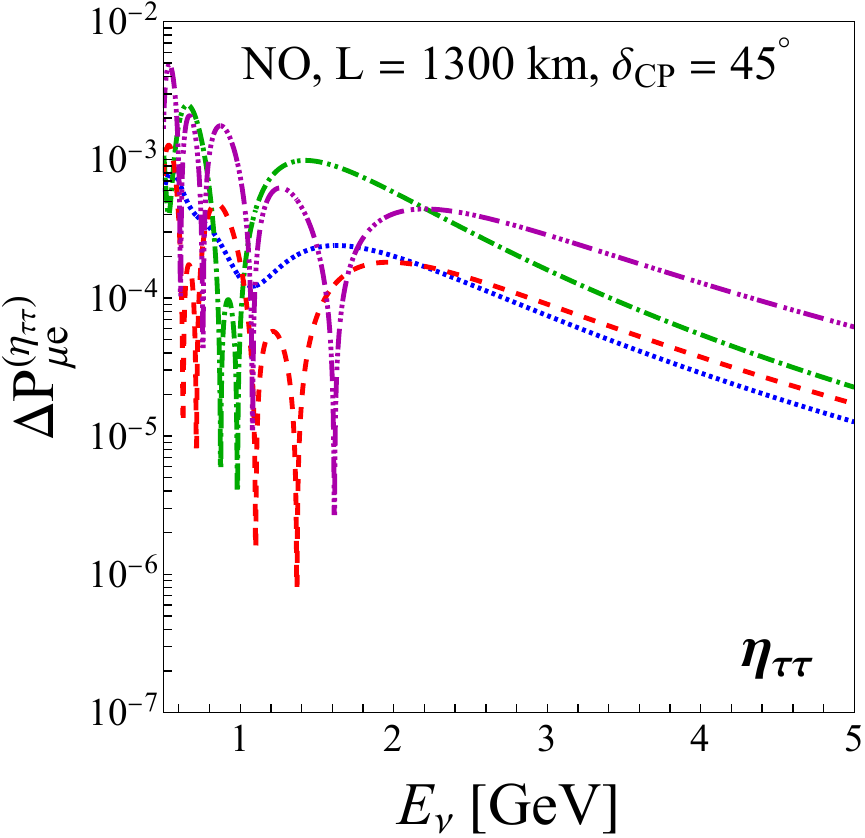}\\
        \vspace{10pt}
        \includegraphics[width=0.327\linewidth]{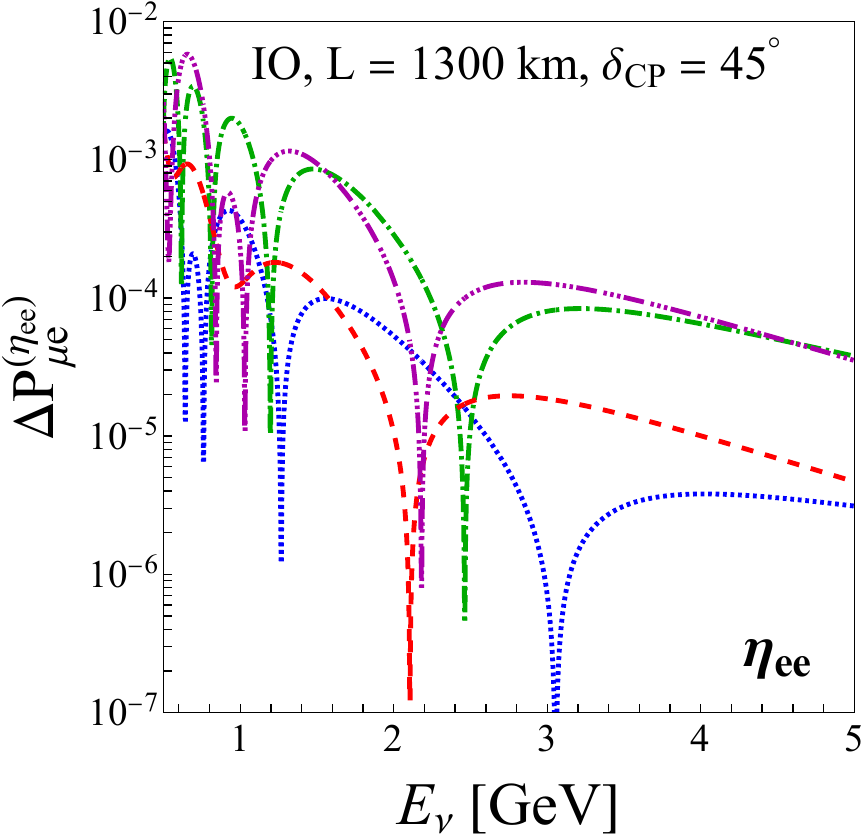}
        \includegraphics[width=0.327\linewidth]{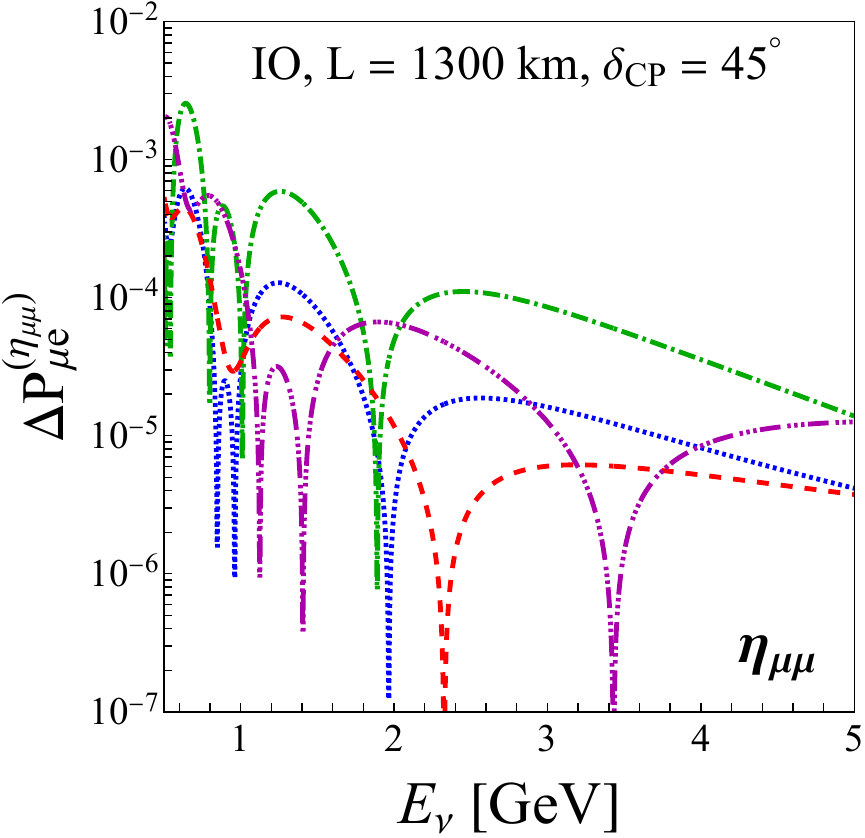}  
        \includegraphics[width=0.327\linewidth]{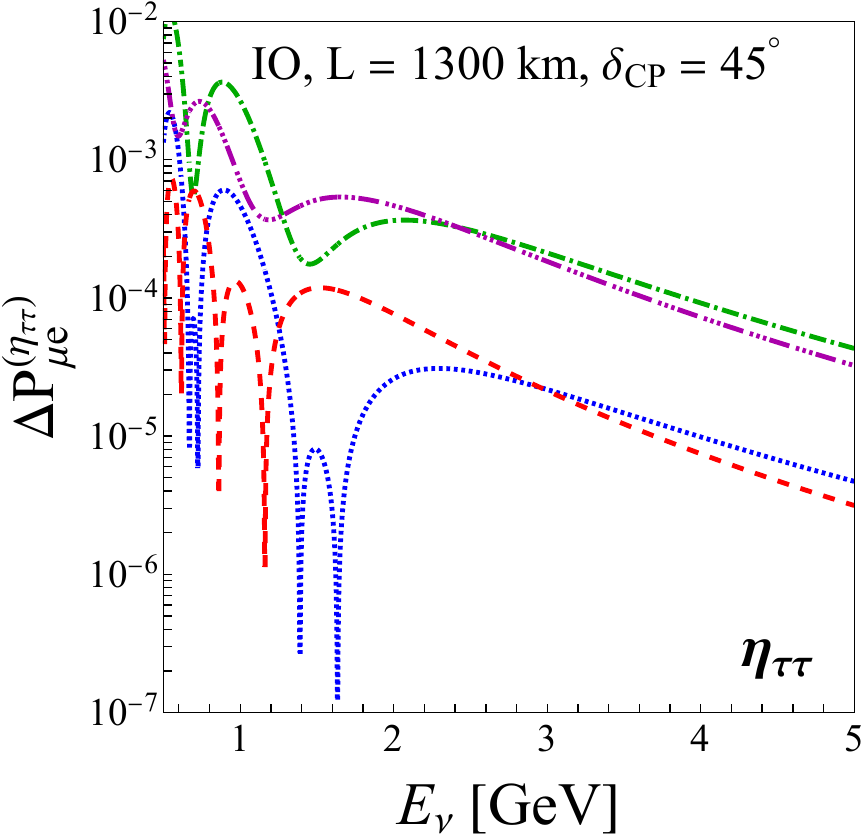}  
        \caption{Absolute error $\Delta P_{\mu e}^{(\eta_{\alpha \alpha})}$ for the analytic SNSI contributions to $P_{\mu e}$ for DUNE baseline and energy range 0.5-5 GeV. The top panel is for NO and the bottom panel is for IO. The left, middle and right panels are for $\eta_{ee}$, $\eta_{\mu \mu}$ and $\eta_{\tau \tau}$ respectively. }
 	\label{fig:accuracy_pme}
 \end{figure}

 We plot the absolute error $\Delta P_{\mu e}^{(\eta_{\alpha \alpha})}$ for the appearance channel $P_{\mu e}$ for both the neutrino mass orderings in figure \ref{fig:accuracy_pme}.  The left, middle, and right panels correspond to the absolute error in presence of the $\eta_{ee}$, $\eta_{\mu \mu}$, and $\eta_{\tau \tau}$  respectively. We use the $(\eta_{\alpha\alpha},~m_{\ell})$ combinations as specified on the top of the figure. We observe that for all chosen values of ($\eta_{\alpha \alpha}$,~$m_\ell$) the absolute error is below 1$\%$ for the entire energy range shown. In fact, in NO, the absolute error is below $0.1\%$ for most of the energy range shown. In IO, the absolute error is below $0.1\%$ for $E_\nu \gtrsim 1$ GeV.

\begin{figure}[t]
 	\centering
        \includegraphics[width=0.73\linewidth] {Updated_plots/plot_legend_accuracy.pdf} \\[1ex]
       \includegraphics[width=0.327\linewidth]{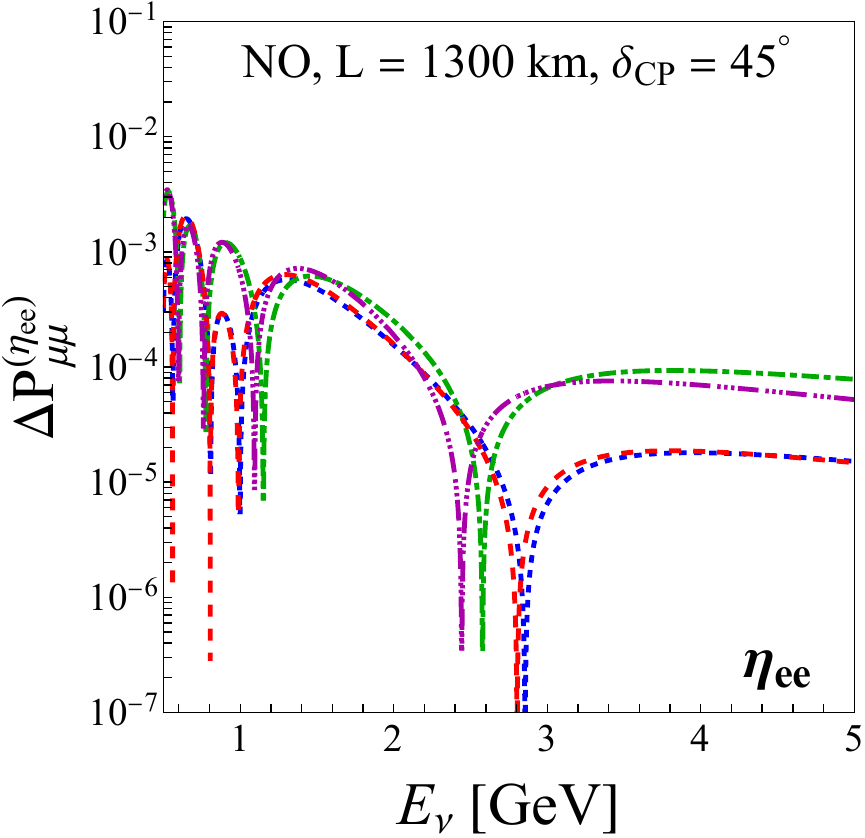} 
        \includegraphics[width=0.327\linewidth]{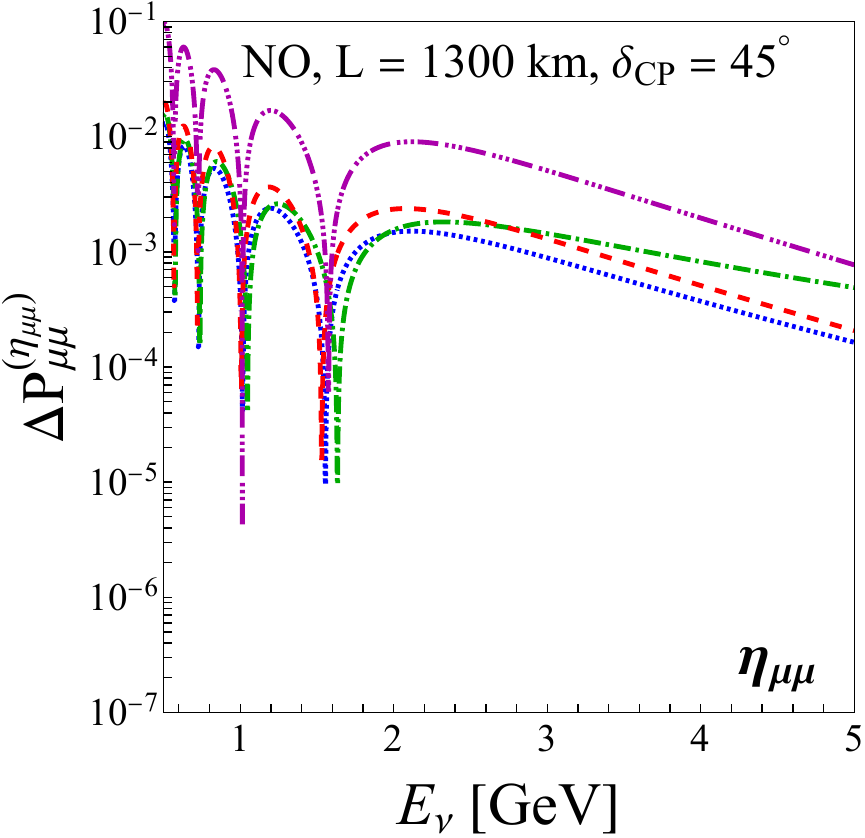} 
 	\includegraphics[width=0.327\linewidth] {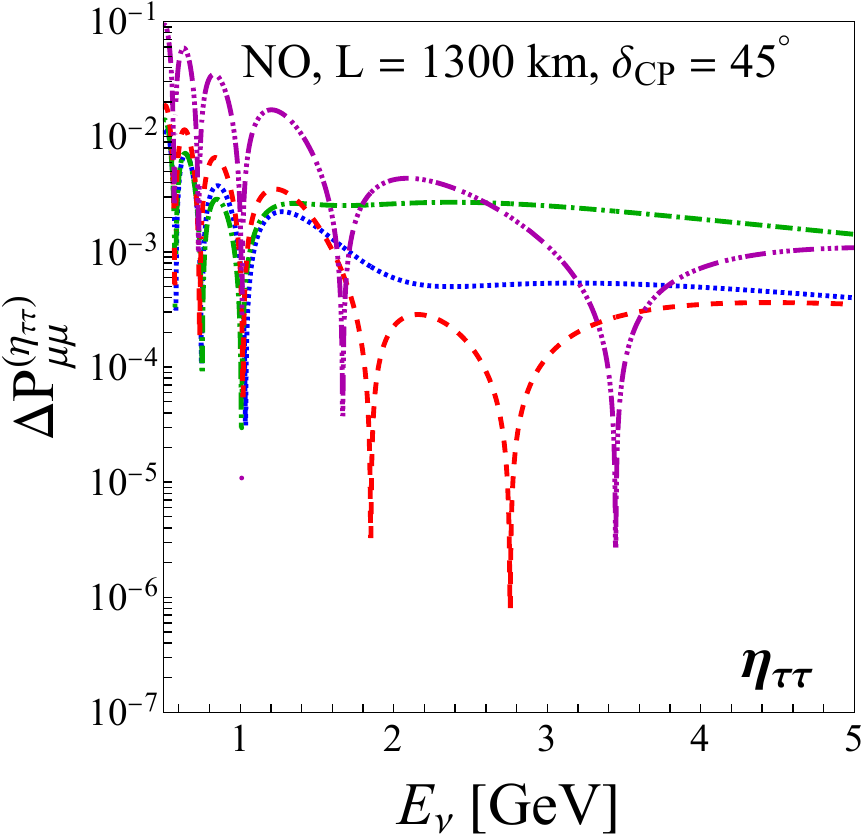}\\
        \vspace{10pt}
        \includegraphics[width=0.327\linewidth]{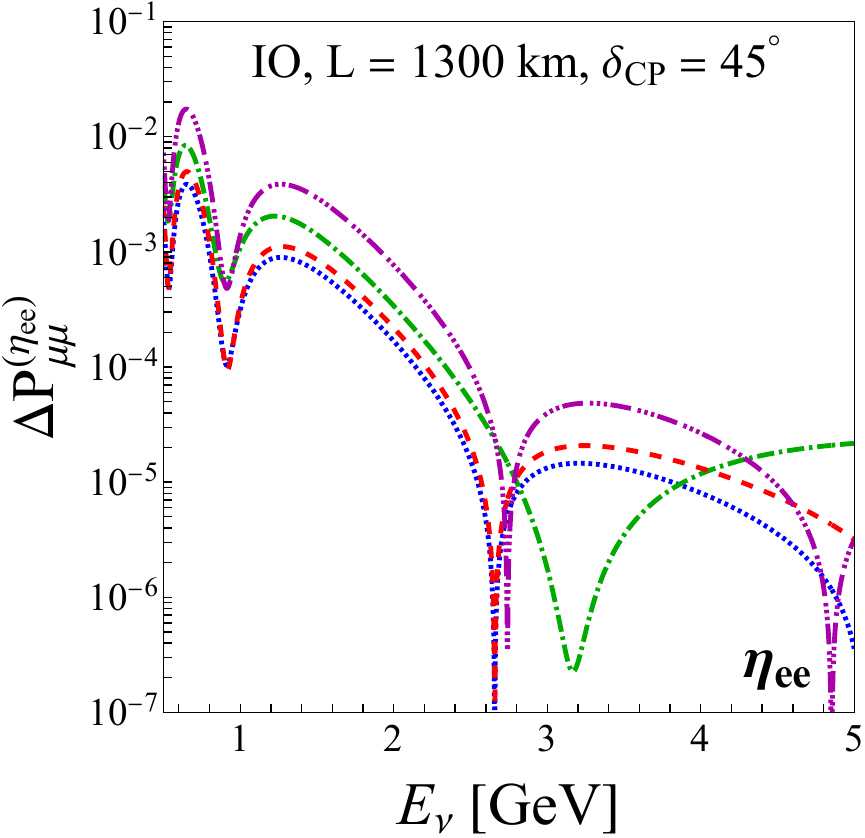}
        \includegraphics[width=0.327\linewidth]{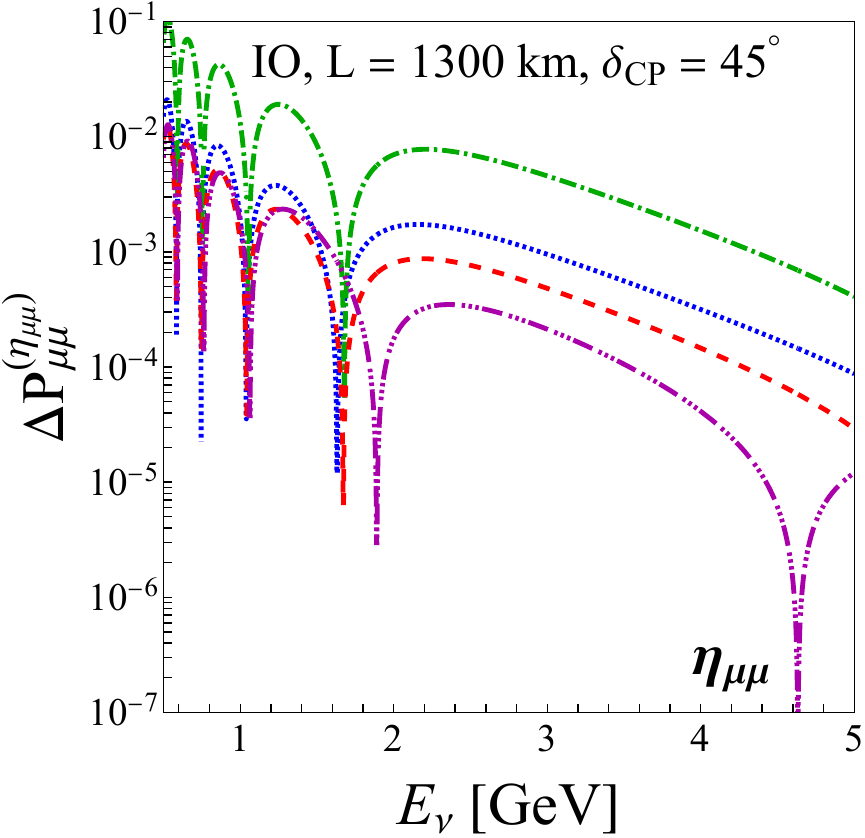}  
        \includegraphics[width=0.327\linewidth]{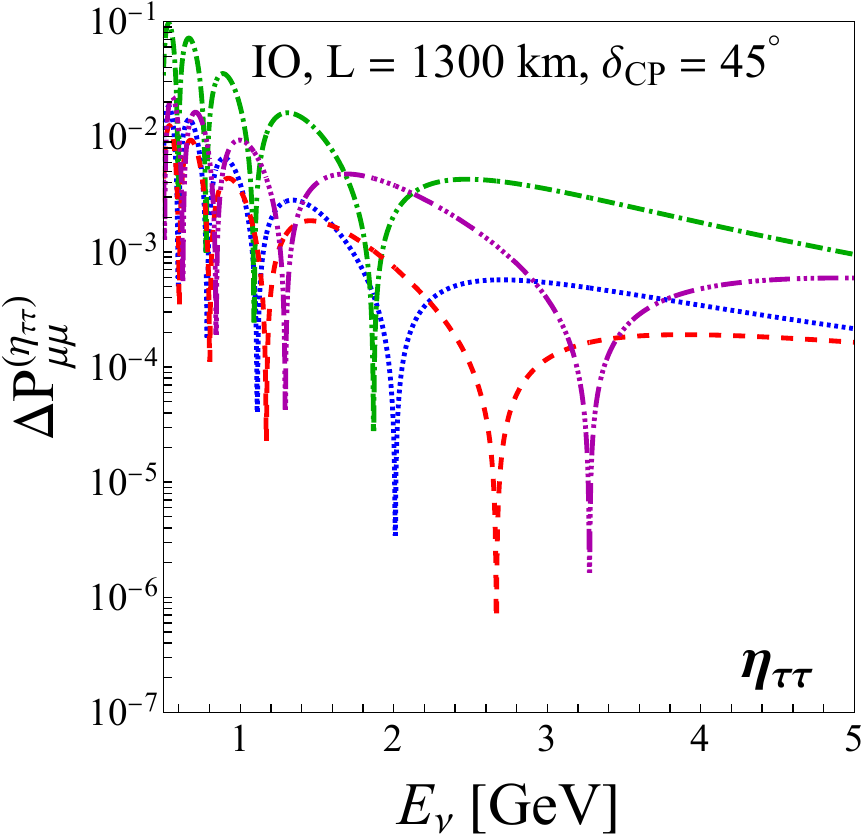}   
        \caption{$\Delta P_{\mu \mu}^{(\eta_{\alpha \alpha})}$ vs $E_\nu$ for DUNE baseline and energy range 0.5-5 GeV, for both  NO (top panel) and IO (bottom panel), for $\eta_{ee}$ (left), $\eta_{\mu \mu} (middle)$ and $\eta_{\tau \tau} $ (right) respectively. For $\eta_{\mu \mu}$ and $\eta_{\tau \tau}$, the absolute error $\Delta P_{\mu \mu}^{(\eta_{\alpha \alpha})}$ is exaggerated due to slight phase shift observed between numerically and analytically calculated SNSI contributions in figure.~\ref{fig:pmm_snsi}.} 
 	\label{fig:accuracy_pmm}
\end{figure}

In figure \ref{fig:accuracy_pmm}, we show the absolute error $\Delta P_{\mu \mu}^{(\eta_{\alpha \alpha})}$ for the disappearance channel. The top (bottom) panel is for NO (IO). The absolute error for $\eta_{ee}$, $\eta_{\mu \mu}$ and $\eta_{\tau \tau}$ contributions are shown in left, middle and right panels respectively. We use the $(\eta_{\alpha\alpha}, m_{\ell})$ combinations as specified on the top of the figure. There is mostly an excellent match between the numerical and analytic probabilities, the absolute error being below 1$\%$. However, for some cases, the errors are amplified because of the slight shift in energy that we have observed in the analytic contributions as compared to the numerically calculated results, as can be seen from figure \ref{fig:pmm_snsi}. Thus our analytic approximations for this channel will also allow us to gain physical insight into the SNSI process.

 \begin{figure}[t]
 	\centering
   	\includegraphics[width=0.4\linewidth]{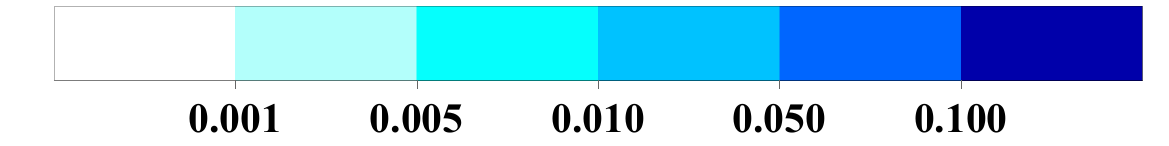} \\
        \vspace{5pt}
       \includegraphics[width=0.3\linewidth]{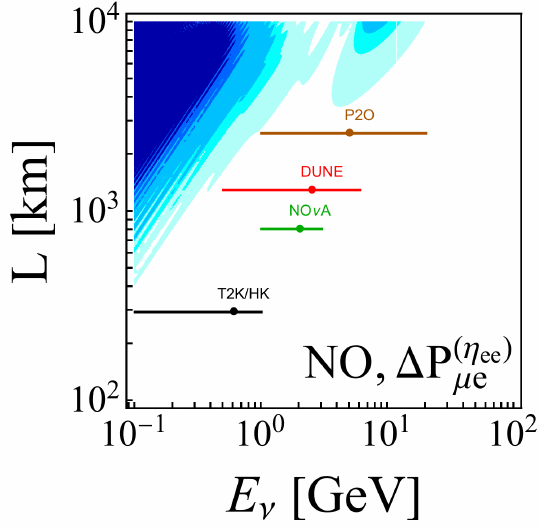} 
 	\includegraphics[width=0.3\linewidth]{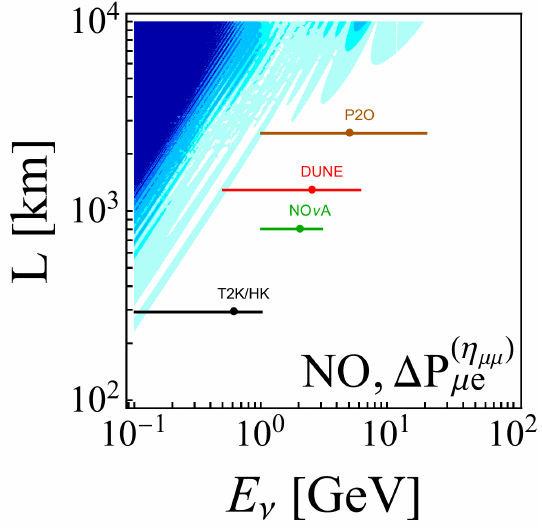} 
        \includegraphics[width=0.3\linewidth]{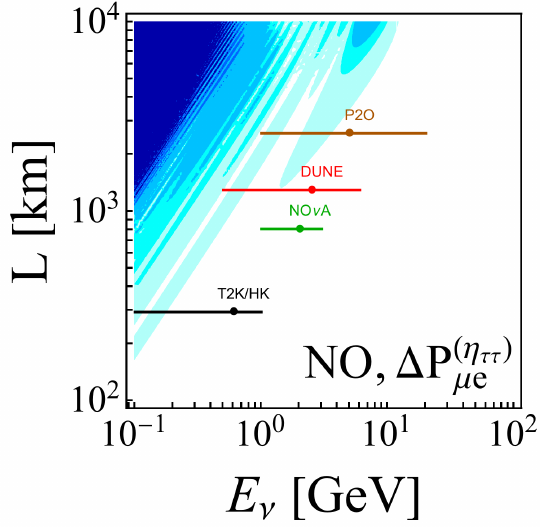}\\
        \vspace{5pt}
         \includegraphics[width=0.3\linewidth]{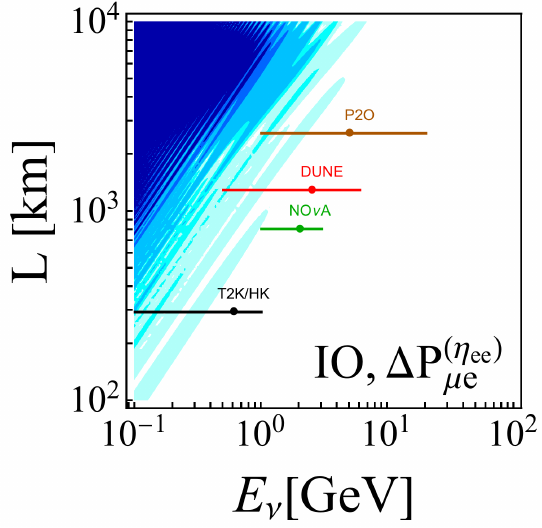} 
 	\includegraphics[width=0.3\linewidth]{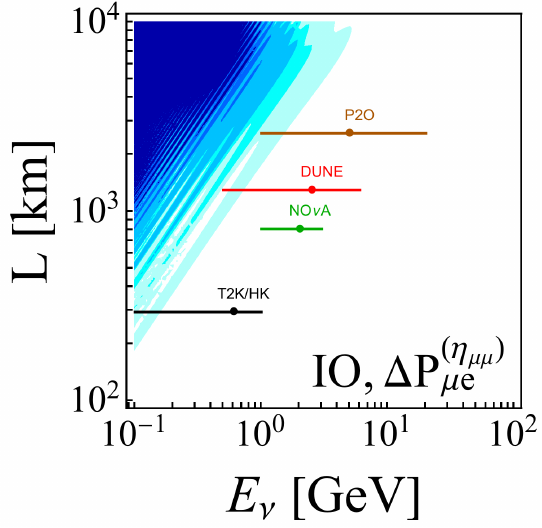} 
        \includegraphics[width=0.3\linewidth]{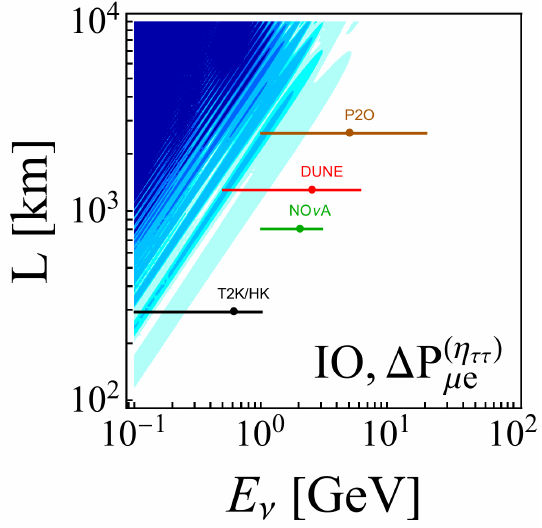}\\
        \vspace{5pt}
       \includegraphics[width=0.3\linewidth]{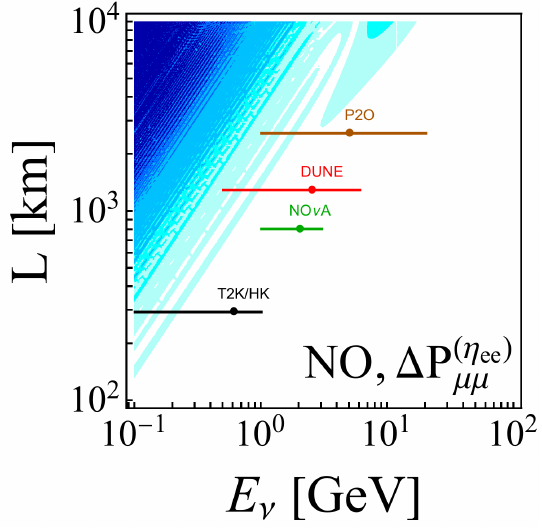} 
 	\includegraphics[width=0.3\linewidth]{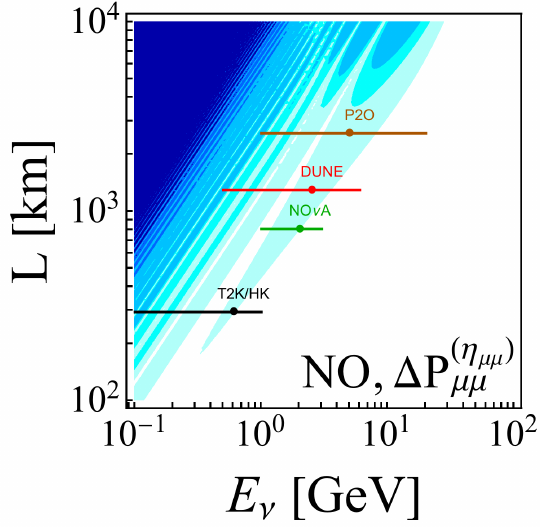} 
        \includegraphics[width=0.3\linewidth]{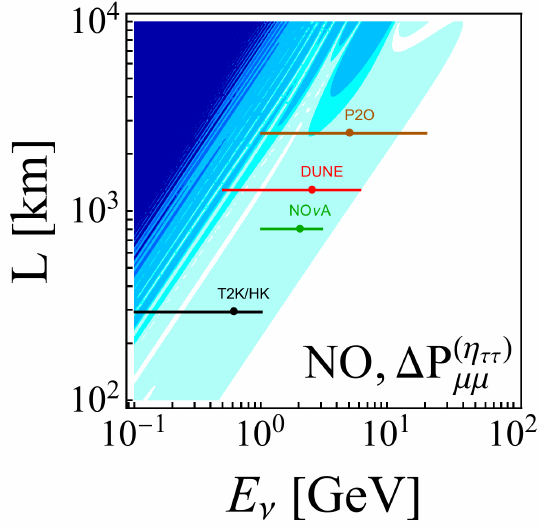} \\
        \vspace{5pt}
       \includegraphics[width=0.3\linewidth]{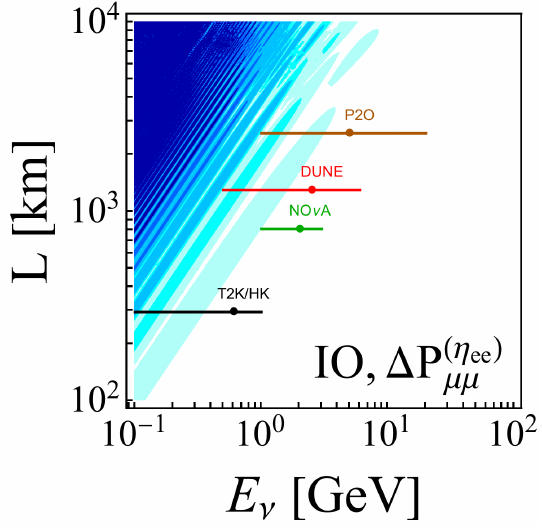} 
 	\includegraphics[width=0.3\linewidth]{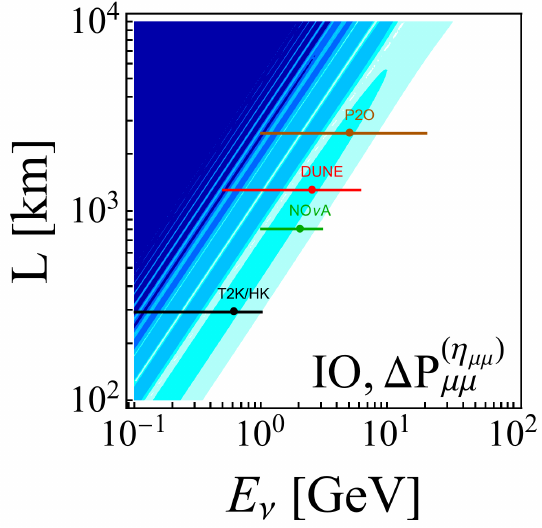} 
        \includegraphics[width=0.3\linewidth]{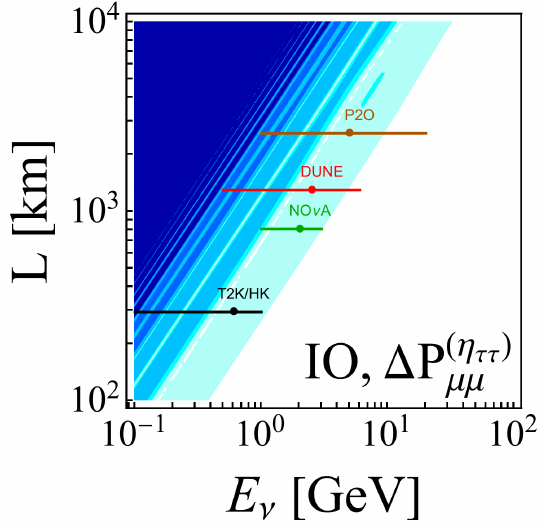} 

 	\caption{ $\Delta P_{\mu e}^{(\eta_{\alpha\alpha})}$ (top 2 rows) and $\Delta P_{\mu \mu}^{(\eta_{\alpha\alpha})}$ (bottom 2 rows) in the $(E_\nu -L)$ plane, for both NO and IO, for the SNSI parameters $\eta_{ee}$ (left), $\eta_{\mu \mu}$ (middle) and $\eta_{\tau \tau}$ (right).  Here, we take $\eta_{\alpha \alpha} = 0.03$ and $m_\ell=0.005$ eV. The colored lines in the plot represent P2O, DUNE, NO$\nu$A and T2K.}
 	\label{fig:error_snsi_2d}
 \end{figure}

We now explore the accuracy of the SNSI contribution to the neutrino probabilities over a wide $(E_\nu - L)$ range.
In the top two rows of figure \ref{fig:error_snsi_2d},  we plot $\Delta P_{\mu e}^{(\eta_{\alpha\alpha})}$, and in the bottom two rows of figure \ref{fig:error_snsi_2d}  we plot $\Delta P_{\mu \mu}^{(\eta_{\alpha\alpha})}$, on the $(E_\nu-L)$ plane. We consider $E_\nu$ in the range $(0.1-100)~\text{GeV}$ and $L$ in the range $(100 - 10000)~\text{km}$. The values of $\eta_{\alpha \alpha}$ and $m_\ell$ are taken to be $0.03$ and $0.005$ eV respectively. For both appearance and disappearance channels, the horizontal lines on the plots represent the energy range of the current and upcoming LBL experiments T2K \cite{T2K:2014xyt}, NO$\nu$A \cite{NOvA:2004blv,NOvA:2016kwd}, DUNE \cite{DUNE:2016hlj, DUNE:2015lol, DUNE:2016rla}, HK \cite{Hyper-KamiokandeProto-:2015xww,Hyper-Kamiokande:2016srs}, and P2O \cite{Akindinov:2019flp} with solid circles indicating the peak energies of those experiments. 
The regions with lighter shades correspond to higher accuracy and darker shades correspond to lower accuracy. We see that the accuracy of the expressions increases for the lower-right plane in the plots. 
We observe $\Delta P_{\mu e}$ to be mostly below $0.1\%$  and $\Delta P_{\mu\mu}$ is below $\sim5\%$ for the three $\eta_{\alpha \alpha}$ parameters at the baseline and energy range corresponding to the LBL experiments. In the case of $P_{\mu \mu}$ in presence of $\eta_{\mu \mu}$ and $\eta_{\tau \tau}$, the absolute error is magnified as there is a slight phase shift between the numerical and analytic probabilities (as observed in figure~\ref{fig:pmm_snsi}).

In the next section, we explore in detail the complex dependence of the SNSI contribution on the neutrino mass and $\delta_{CP}$ value.

\section{The Complex Behavior of the Scalar NSI Contributions} \label{sec:SNSI_mixing_param}

We explore the dependence of analytically calculated oscillation probabilities on the leptonic CP phase ($\delta_{CP}$) in section \ref{sec:dcp_sec}. We also explore the direct dependence on the absolute neutrino mass and the ordering of the masses in section \ref{sec:Mass_dep}.

\subsection{With absolute $\nu$-mass} \label{sec:Mass_dep}

In this section, we discuss the non-trivial mass dependence of the contributions from the diagonal SNSI terms. As shown in section~\ref{sec:ana_prob}, the neutrino survival and conversion probabilities in the presence of SNSI have a complex dependence on the absolute masses of the neutrino vacuum mass eigenstates.

The phenomenon of neutrino oscillations itself --- in the absence of any new physics scenarios --- does not depend on the absolute masses or the mass differences, it only depends on the mass-squared differences which regulate the neutrino oscillation frequency.
Furthermore, most of the new physics scenarios also do not lead to any direct dependence on the absolute mass scales of neutrino physics.
In the neutrino decay scenario, an indirect dependence on neutrino masses arises since the neutrino lifetime in the rest frame and the neutrino lifetime in the lab frame are related as $\tau_{\rm lab} = \tau_{\rm rest} \times (E_\nu/ m_i)$. 
However, this dependence cannot be disentangled as the rest frame lifetime $\tau_{\rm rest}$ and neutrino mass $m_i$ always appear together.
On the other hand, for SNSI the dependence on the absolute masses is very non-trivial, as any effects of new physics depend on all three neutrino masses as well as the value of the SNSI parameter $\eta_{\alpha\alpha}$.
In fact, in future neutrino experiments, it may be possible to distinguish the neutrino mass and $\eta_{\alpha\alpha}$ separately if SNSI effects are present~\cite{Medhi:2023ebi}.

From the analytic approximations for the neutrino probabilities in the presence of SNSI, we observe that the SNSI contributions to the eigenvalues as well as the  probabilities can be divided into three distinct types of terms with different mass dependences:
\begin{itemize}
    \item \textbf{Mass summation term:} mass terms of the form $m_1+m_2$, $m_1 c_{12}^2 + m_2 s_{12}^2$, and $m_1 s_{12}^2 + m_2 c_{12}^2$, since $c_{12} \sim s_{12} \sim O(1)$, the general form of this term can be understood from the general behavior of the $m_1+m_2$ term.
    \item \textbf{Mass difference term:} mass terms of the form $m_2-m_1$, the contribution from this term depends crucially on the neutrino mass ordering in the atmospheric sector (i.e., the sign of $\Delta m^2_{31}$), any contribution of this term will be heavily suppressed in the Inverted Ordering scenario.
    \item \textbf{The $m_3$ term}: mass terms which are directly proportional to the value of $m_3$. Given the current state of bounds on the sum of the neutrino masses, obtained from cosmology, this term is also expected to have a small contribution in the Inverted Ordering scenario.
\end{itemize}
For example, the SNSI contribution to conversion probability $P_{\mu e}$ for a non-zero value of $\eta_{ee}$, as given in  equation~\ref{eq:pme_etaee}, has 4 types of mass terms: $m_1+m_2$, $m_1 c_{12}^2 +m_2 s_{12}^2 $, $m_2-m_1$, and $m_3$.
However, the $m_1 c_{12}^2 +m_2 s_{12}^2 $ can be rewritten as:
\begin{equation}
    m_1 c_{12}^2 +m_2 s_{12}^2  = \frac{1}{2} \Big( m_1 + m_2 \Big) - \frac{1}{2}\cos \left(2 \theta _{12}\right) \Big( m_2 - m_1 \Big)
\end{equation}
Note that for $m_1, \, m_2 \gg 0$, the above equation can be approximated to be $m_1 c_{12}^2 +m_2 s_{12}^2  \approx \tfrac{1}{2} \Big( m_1 + m_2 \Big)$; whereas,
in the limit of $m_1 \to 0$, where the maximum possible of $(m_2-m_1)$ term can be obtained, the above equation can be approximated as $ m_1 c_{12}^2 +m_2 s_{12}^2  \approx m_2 s_{12}^2 $.

Expressing the contribution of the SNSI $\eta_{ee}$ term to the conversion probability in terms of $m_1+m_2$, $m_2-m_1$, and $m_3$, we write
\begingroup
\allowdisplaybreaks
\begin{align}\label{eq:pme_etaee_massdependence}
    P_{\mu e}^{(\eta_{ee})} = & \Big[ C_{-}^{\mu e} \left(m_2-m_1\right) +  C_{+}^{\mu e} \left( m_1 + m_2 \right) + C_{3}^{\mu e} m_3 \Big] \times  \left(\frac{S_m}{\Delta m^2_{31}}\right)\, \eta _{\text{ee}} \; ,
\end{align}
\endgroup
where the coefficients $C_{-}^{\mu e}$, $C_{+}^{\mu e}$, and $C_{3}^{\mu e}$ are complex functions of the neutrino mixing parameters, the normalized matter potential $A_e$, and the normalized length-scale $\Delta$. The coefficients can be expressed as: 
\begin{subequations}
\label{eq:pme_etaee_massdependence_detail}
\begin{align}
   C_{-}^{\mu e} = &\, 2\,  s_{13} \sin \left(2 \theta _{12}\right) \sin \left(2 \theta _{23}\right) \cos \left(\delta _{\text{CP}}+\Delta \right) \frac{\sin \left[ \left(A_e-1\right) \Delta \right]}{A_e-1} \frac{\sin \left[ A_e \Delta \right]}{A_e}  \nonumber \\
   & \; -4 s_{13}^2 \sin ^2\left(\theta _{23}\right) \cos \left(2 \theta _{12}\right) \frac{\sin \left[ \left(A_e-1\right) \Delta \right]}{A_e-1} \frac{1}{A_e-1} \Bigg( 2 \Delta  \cos \left[ \left(A_e-1\right) \Delta \right] \Bigg. \nonumber\\
   & \Bigg. \; \qquad \qquad -\left(A_e+1\right) \frac{\sin \left[ \left(A_e-1\right) \Delta \right]}{A_e-1} \Bigg) + 2 \, \alpha \, c_{23}^2 \sin ^2\left(2 \theta _{12}\right)  \frac{\sin ^2\left[ A_e \Delta \right]}{A_e^2}  \; , \\
    C_{+}^{\mu e} = & \, 4 \,s_{13}^2 s_{23}^2 \frac{\sin \left[\left(A_e-1\right) \Delta \right]}{A_e-1} \frac{1}{A_e-1} \left[ 2 \, \Delta  \cos \left[ \left(A_e-1\right)\Delta \right]- \left(A_e+1\right) \frac{ \sin \left[ \left(A_e-1\right) \Delta \right]}{A_e-1} \right]  \nonumber\\
   & \; - \alpha \, s_{13}  \sin \left(2 \theta _{12}\right) \sin \left(2 \theta _{23}\right) \cos \left(\delta _{\text{CP}}+\Delta \right)  \frac{1}{A_e-1} \times \nonumber\\
   & \qquad \Bigg( \frac{2}{A_e} \Big[ \sin \left[ \left(A_e-1\right) \Delta \right] \frac{\sin \left[ A_e \Delta \right]}{A_e}  - \Delta  \sin \left[ \left( 2 A_e-1 \right) \Delta \right] \Big] \Bigg. \nonumber\\
   & \Bigg. \hspace{30ex} + \left( A_e + 1 \right) \frac{\sin \left[ A_e \Delta \right]}{A_e} \frac{\sin \left[ \left(A_e-1\right) \Delta \right]}{A_e-1} \Bigg) \; ,  \\
    C_{3}^{\mu e} = & \, 2 \, s_{13} \frac{\sin \left[\left(A_e-1\right) \Delta  \right]}{A_e-1} \Bigg( 4 \, s_{13} \, s_{23}^2 \, \frac{\sin \left[\left(A_e-1\right) \Delta  \right]}{A_e-1} \times \Bigg. \nonumber\\
    & \Bigg. \qquad \qquad \qquad \qquad \qquad + \alpha \, \sin \left(2 \theta _{12}\right) \sin \left(2 \theta _{23}\right) \cos \left(\delta _{\text{CP}}+\Delta \right) \frac{\sin \left[ A_e \Delta \right]}{A_e} \Bigg)\; .
\end{align}
\end{subequations}
Note the $C_{3}^{\mu e}$ term, as well as the dominant contribution of the $C_{-}^{\mu e} $ term are approximately regulated by the $\sin \left[\left(A_e-1\right) \Delta \right]$ term, whereas the behavior of the $C_{+}^{\mu e} $ term is more complex, and the oscillations are not in phase with the other two terms.

\begin{figure}
    \centering
    \includegraphics[width=0.45\linewidth]{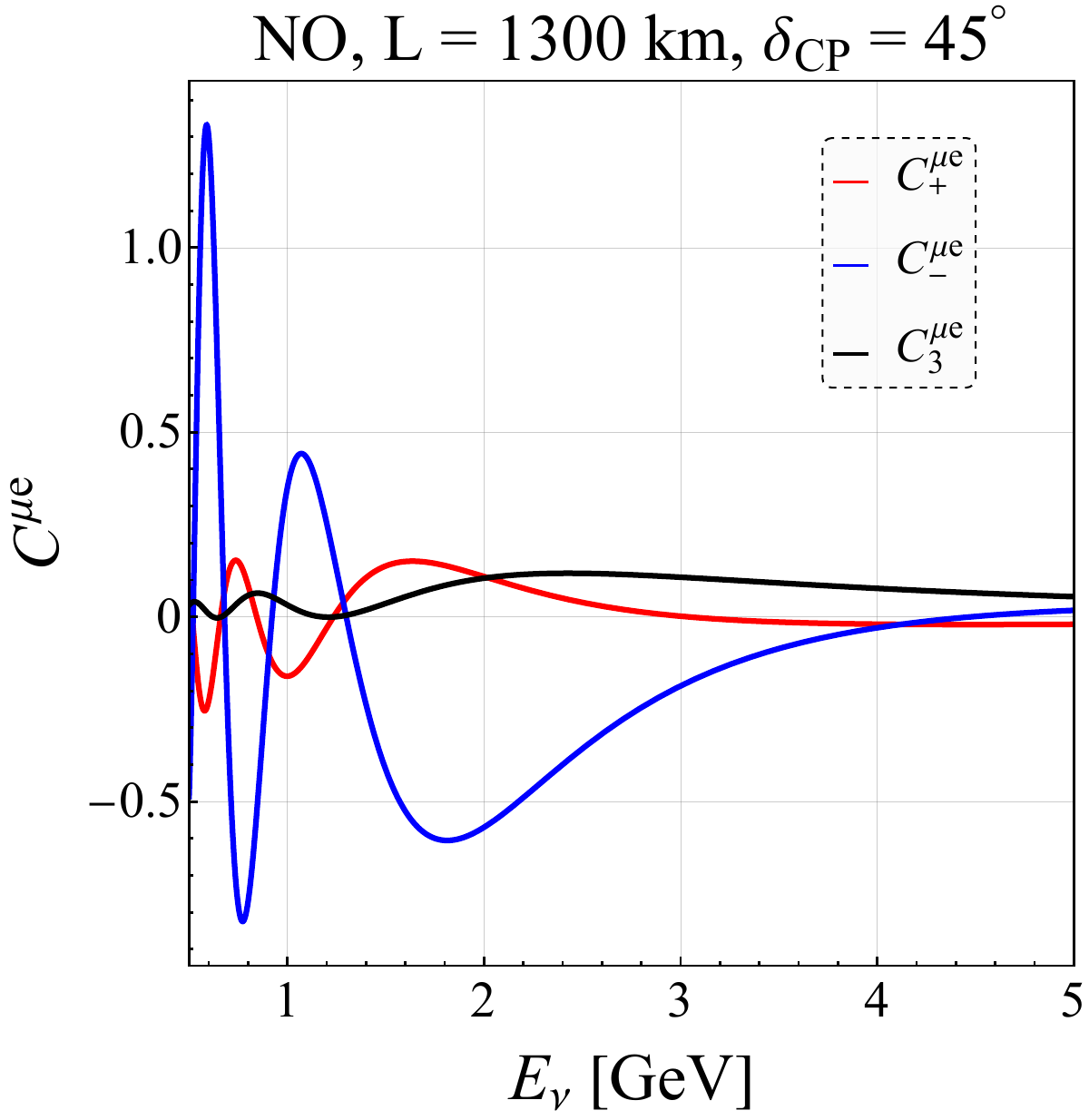}
    \hspace{10pt}
    \includegraphics[width=0.45\linewidth]{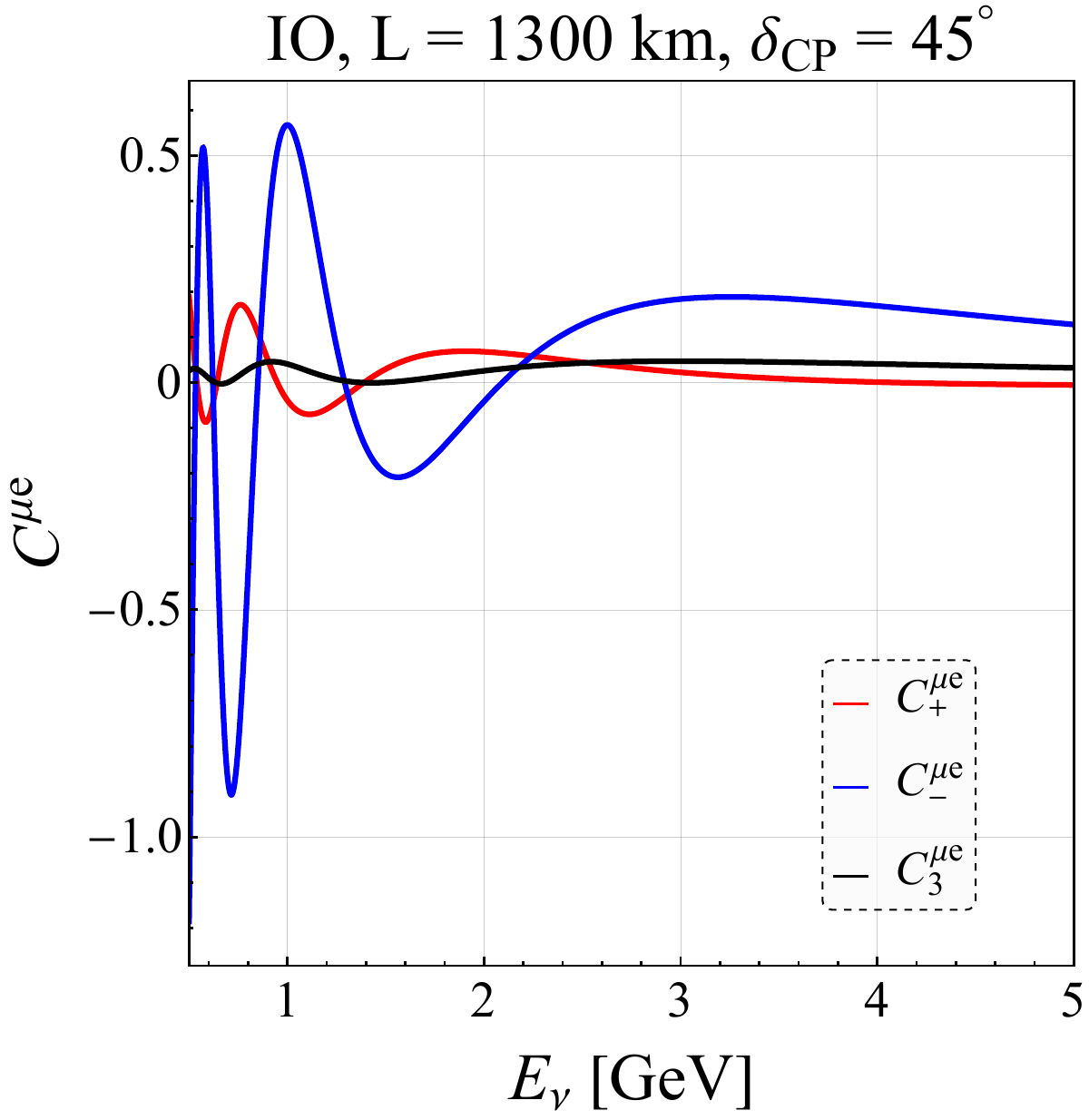}
    \caption{Variation of $C_{-}^{\mu e}$, $C_{+}^{\mu e}$ and $C_{3}^{\mu e}$ with $E
    _\nu$ for DUNE baseline, for both NO [left panel] and IO [right panel].}
    \label{fig:pme_ee_massterms}
\end{figure}

In figure \ref{fig:pme_ee_massterms}, we show the energy dependence of these three coefficients for a DUNE baseline neutrino experiment, in both the mass orderings: NO and IO. For both the mass orderings, $C_{3}^{\mu e} \gtrsim 0$, while $C_{+}^{\mu e}$ and $C_{-}^{\mu e}$ oscillates around zero. Furthermore, the amplitudes of these contributions remain approximately the same, even when the signs are reversed, with the amplitude of $C_{-}^{\mu e}$ being the largest by a factor of $\sim O(\text{few})$.

The first zero-crossing point from the right (where all three terms cross zero), approximately corresponds to $\sin \left[\left(A_e-1\right) \Delta \right] = 0$, i.e., $\left(1-A_e\right) \Delta = \pi$. Thus, we observe a small shift to higher energy for IO, as expected. Furthermore, we also note that while the amplitudes of the $C_{-}^{\mu e}$ and $C_{+}^{\mu e}$ terms are comparable at higher energies, we clearly have $|C_{-}^{\mu e}| > |C_{+}^{\mu e}|$ at lower energies. Therefore, even with slightly smaller values for $m_2-m_1$, the $C_{-}^{\mu e}$ term may have a dominant contribution in certain scenarios.
Further, we also observe that the amplitude of the $C_{3}^{\mu e}$ term is smaller by a factor of $\sim O(10)$ or more.
Thus, for the $C_{3}^{\mu e}$ coefficient to have a dominant contribution, the value of $m_3$ would need to be larger than the value of the other two mass terms by at least around $O(10)$.

In NO scenario, the mass of $\nu_3$ eigenstate ($m_3$) is larger than that of the $\nu_1$ or $\nu_2$ eigenstates, i.e., $m_3 > m_2, \, m_1$. On the other hand, for inverted ordering (IO), we have $m_3 < m_2, \, m_1$. Thus, for the same values of the SNSI parameter $\eta_{\alpha\alpha}$, the $m_3$ mass term would always have a larger contribution in NO when compared to IO. On the other hand, for the mass term of the $m_1+m_2$ form, since $m_1, \, m_2 > m_3$ in IO, it would have the dominant contribution in IO. Finally, since $m_2 - m_1$ may be expressed as
\begin{equation}
    m_2 - m_1 = \frac{m_2^2-m_1^2}{m_2+m_1} = \frac{\Delta m_{21}^2}{m_1+m_2} \;,
\end{equation}
and since the value of $\Delta m_{21}^2$ is fixed, this term would have more contributions when $m_1+m_2$ is smaller, i.e., in NO, especially in the limit of $m_1 \to 0$ eV.

Arguments regarding the maximum possible value of the three different types of mass terms can help further understand which terms will be dominant in which scenarios.
Considering a bound on the sum of the absolute mass of the neutrinos at $\sum_{\nu} m_\nu \lesssim 0.12$ eV \cite{Planck:2018vyg}, we can show, that for NO:
\begin{align}
     \left[m_1 + m_2\right]_{\max}^{\rm (NO)} \approx & \;  6.12 \times 10^{-2} \text{ eV} \; , \nonumber\\
     \quad \left[ m_2 -m_1 \right]_{\max}^{\rm (NO)} \approx &\; 8.66 \times 10^{-3} \text{ eV} \; , \nonumber\\ 
     \left[m_3\right]_{\max}^{\rm (NO)} \approx & \; 5.88 \times 10^{-2} \text{ eV} \; .
\end{align}
Thus, in NO, all three mass terms may contribute dominantly, depending on the value of the lightest neutrino mass $m_1$.
Further, note that the maximum limits for both $m_1+ m_2$ as well as $m_3$ term are obtained for the highest possible value of $\sum_{\nu} m_\nu$, whereas, the maximum limit for the $m_2 -m_1 $ is obtained for the lowest possible value of $\sum_{\nu} m_\nu$ in NO.
On the other hand, for IO, for $\sum_{\nu} m_\nu \lesssim 0.12$ eV we obtain:
\begin{align}
       \left[m_1 + m_2\right]_{\max}^{\rm (IO)} \approx & \; 1.06 \times 10^{-1} \text{ eV} \; , \nonumber\\
       \quad \left[ m_2 -m_1 \right]_{\max}^{\rm (IO)} \approx & \; 7.36 \times 10^{-4} \text{ eV} \;, \nonumber\\
       \left[m_3\right]_{\max}^{\rm (IO)} \approx & \; 1.42 \times 10^{-2} \text{ eV} \; .
\end{align}
Note that, in IO, the maximum possible value of the $m_2 -m_1$ term is smaller than that in NO, this is because only in NO we can obtain the maximum possible value of $m_2 -m_1$ in the $m_1 \to 0$ limit.
Thus, in IO, for all practical purposes, the mass difference term can be neglected.
Further, the $m_3$ term would always be smaller than the $m_1 c_{12}^2 + m_2 s_{12}^2$ term in IO. Therefore, only the mass summation term of the $m_1+m_2$ form (in this particular case, appearing as $m_1 c_{12}^2 + m_2 s_{12}^2$) needs to be considered for the SNSI contribution in IO.

We note from the analytic approximations, that the SNSI contributions to the neutrino probabilities are accompanied by a factor of $\Delta m^2_{31}$. However, we find that the effects of neutrino mass ordering are non-trivial --- much more complex than just a simple sign change of $\Delta m^2_{31}$ --- since the mass ordering also directly affects the range of absolute masses of the neutrino mass eigenstates.

Depending on the mass of the lightest neutrino as well as the mass ordering, certain mass terms will contribute more dominantly.
In such a scenario, since the coefficients of the mass terms are dependent on $\Delta$, $A_e$, and the neutrino mixing parameters, the contributions would also peak and dip at different energies for different values of the lightest neutrino mass and mass ordering choices.
In fact, the dependence of the coefficients and thus the SNSI neutrino probability contribution on the matter potential $A_e$ and the implicit dependence of the $\eta_{\alpha\alpha}$ term on matter density can, allow us to probe the absolute mass of neutrinos in the presence of SNSI~\cite{Medhi:2023ebi}. This can be achieved using various combinations of neutrino experiments, such as DUNE and T2HK.

\begin{figure}[t!]
    \centering
    \includegraphics[width=0.45\textwidth]{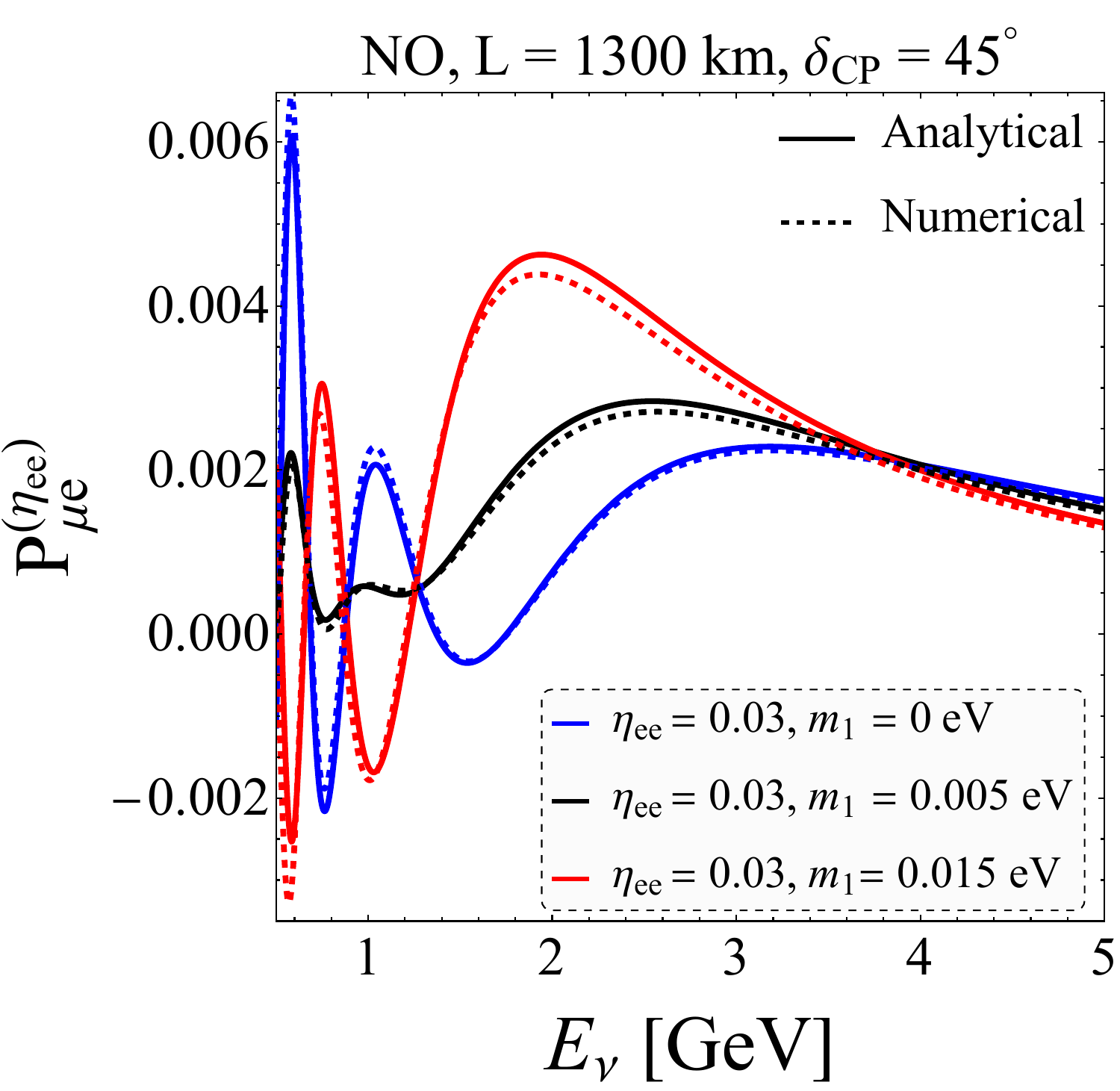} \hspace{10pt}
    \includegraphics[width=0.45\textwidth]{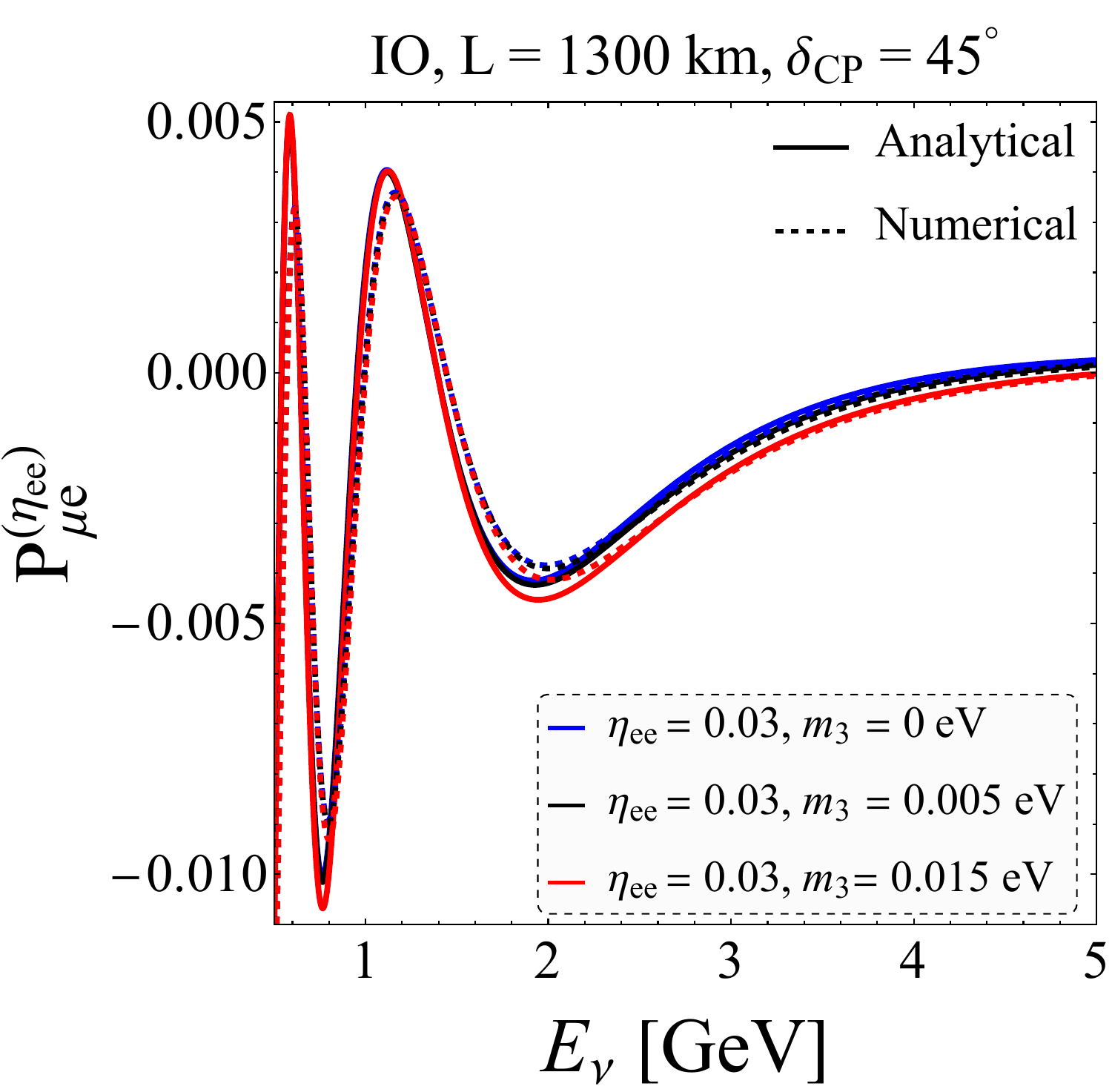}
    \caption{$P_{\mu e}^{\eta_{ee}}$  vs $E_\nu$ at different values of lightest neutrino mass $m_\ell$, for NO (left panel) and IO (right panel). Here $\eta_{ee}$ = 0.03.}
    \label{fig:pme_mass}
\end{figure}

In figure \ref{fig:pme_mass}, we illustrate the dependence of the SNSI contribution to the conversion probability $P_{\mu e}$ on the lightest neutrino mass, in both mass orderings, for a fixed value of $\eta_{ee} = 0.03$.
For different values of the lightest neutrino mass $m_\ell$ (i.e., $m_1$ in NO, and $m_3$ in IO), the SNSI contribution differs significantly for the NO and shows almost no change for IO. In fact, for NO, not only the position of the peaks and dips, but the amplitude of the SNSI contribution also differs. Further, we also observe that the amplitude of the SNSI contribution is higher for IO compared to NO.

To understand the features observed in figure~\ref{fig:pme_mass}, we outline the values of the three mass terms for each $m_{\ell}$ value, for both mass orderings, in table~\ref{tab:massdependence}.
In table~\ref{tab:massdependence}, we indicate which mass term plays the key role in the SNSI contribution to the conversion probability $P_{\mu e}$ for varying values of the lightest neutrino mass.
Note that for different values of $m_1$, in NO, different combinations of mass terms contribute dominantly for different values of $m_1$, this leads to the observed changes in $P_{\mu e}^{\eta_{ee}}$ as a function of the energy $E_\nu$. Further, the small amplitude of $P_{\mu e}^{\eta_{ee}}$ at lower energies, in NO, can be explained by the partial cancellation of the terms. In particular, since the coefficients $C_{-}^{\mu e}$ and the $C_{+}^{\mu e}$ terms are almost fully out of phase, for $m_= 0.005$ eV, the contribution from both of these terms leads to a partial cancellation and the value of $P_{\mu e}^{\eta_{ee}}$ decreases. For $m_1 = 0.015$ eV, the primary contribution in NO is from $C_{+}^{\mu e}$ and the $C_{3}^{\mu e}$ term, and thus a similar cancellation does not happen, leading to a comparatively larger amplitude of the SNSI contribution.

\begin{table}[t]
    \centering
    \renewcommand{\arraystretch}{1.25}
    \begin{tabular}{|c|c|c|c|c|}
    \hline
    \multicolumn{5}{|c|}{Absolute neutrino masses (in eV)}\\
        \hline
        \multirow{4}{*}{\rotatebox{90}{N.O.}} \quad & $m_{\ell} = m_1$ \quad & \quad $m_1+m_2$ \quad & \quad $m_2 -m_1$ \quad & \quad $m_3$ \quad \\
        \cline{2-5}
        &$0 $ & $8.66\times 10^{-3}$ &  \cellcolor{usergreen}  $8.66 \times  10^{-3} $ & \cellcolor{usergreen}   $5.06 \times 10^{-2} $ \\
        \cline{2-5}
         &$0.005$ & \cellcolor{usergreen}   $ 1.5 \times 10^{-2} $ &  \cellcolor{usergreen}  $ 5 \times  10^{-3} $ & \cellcolor{usergreen} $5.08 \times 10^{-2} $ \\
         \cline{2-5}
         &$0.015 $ & \cellcolor{usergreen}  $ 3.23 \times 10^{-2} $ &  $ 2.23 \times  10^{-3} $ & \cellcolor{usergreen}  $5.28 \times 10^{-2} $ \\
         \hline
        \hline
        \multirow{4}{*}{\rotatebox{90}{I.O.}} & \quad $m_{\ell} = m_3$ \quad & \quad $m_1+m_2$ \quad & \quad $m_2 -m_1$ \quad & \quad $m_3$ \quad \\
        \cline{2-5}
        &$0 $ & \cellcolor{usergreen} $1.02 \times 10^{-1}$  & $7.36 \times  10^{-4} $ & $ 0$ \\
        \cline{2-5}
         &$0.005$ & \cellcolor{usergreen} $1.02 \times 10^{-1}$  & $ 7.32 \times 10^{-4} $ & $5.0 \times 10^{-3} $ \\
        \cline{2-5}
         &$0.015 $ & \cellcolor{usergreen} $1.06 \times 10^{-1}$ & $ 7.06 \times 10^{-4} $ & $1.5 \times 10^{-2} $ \\
         \hline
    \end{tabular}
    \caption{The values of the three mass terms that regulate the SNSI contribution to $P_{\mu e}$ for a non-zero value of $\eta_{ee}$, for both the normal and inverted mass ordering. The green color of the box denotes that the key contribution will caused due to that particular mass term. Multiple green boxes for the same $m_1$ value in NO further highlight the complex nature of the SNSI contribution, with multiple competing mass terms modifying the probability.}
    \label{tab:massdependence}
\end{table}

From table~\ref{tab:massdependence} we further observe that for IO the dominant contribution is always from the $m_1 c_{12}^2 + m_2 s_{12}^2$ term. In fact, the variation in the value of $m_3$ does not change the value of $m_1 c_{12}^2 + m_2 s_{12}^2$ significantly. Thus, we expect $P_{\mu e}^{\eta_{ee}}$ to show negligible changes with varying values of $m_3$, this matches the observed behavior in figure~\ref{fig:pme_mass}.

As observed in section~\ref{sec:applicability}, for both $\eta_{\mu \mu}$ and $\eta_{\tau \tau}$, the SNSI contribution is significant in the survival channel $P_{\mu\mu}$. Therefore, we also briefly discuss the mass dependence of the $P_{\mu\mu}$ SNSI contribution for a non-zero $\eta_{\mu \mu}$ and $\eta_{\tau \tau}$. For both $P_{\mu\mu}^{\eta_{\mu\mu}}$ and $P_{\mu\mu}^{\eta_{\tau\tau}}$, the mass terms are of the form $m_1 s_{12}^2 + m_2 c_{12}^2$ and $m_3$. The $m_1 s_{12}^2 + m_2 c_{12}^2$ term can be further decomposed into
\begin{equation}
    m_1 s_{12}^2 + m_2 c_{12}^2 = \frac{1}{2} \left(m_1+m_2\right) +\frac{1}{2} \cos (2\theta_{12}) \left(m_2 - m_1\right) \; .
\end{equation}
Therefore, in NO, the $\left(m_2 - m_1\right)$ term will contribute to the SNSI contributions. However, since the coefficients of the $\left(m_1+m_2\right)$ term, $\left(m_2 - m_1\right)$ term, and the $m_3$ term are comparable (see equation~\ref{eq:pmm_etamm} and \ref{eq:pmm_etatt}), that means that due to the subdominant nature of the $\left(m_2 - m_1\right)$ contribution, this term may be ignored. for example, for $\eta_{\mu \mu}$, this leads to the SNSI contribution having the approximate form of:
\begin{align}\label{eq:pmm_etamm_approx}
    P_{\mu \mu}^{(\eta_{\mu \mu})} \approx & \; 2 \sin ^2\left(2 \theta _{23}\right) \sin [\Delta] \Bigg[ \frac{1}{2} \biggl( 2 \Delta c_{23}^2 \cos [\Delta ]- \cos \left(2 \theta _{23}\right) \sin [\Delta ]  \biggr) \Big( m_1+m_2\Big)  \Big. \nonumber\\
    & \, \Big. \qquad \qquad -\biggl(2\, \Delta s_{23}^2 \cos [\Delta] + \cos \left(2 \theta _{23}\right) \sin [\Delta ] \biggr) m_3 \Bigg]  \left(\frac{S_m}{\Delta m^2_{31}}\right) \eta _{\mu \mu } \; .
\end{align}
Note that for the range of allowed $\theta_{23}$ values obtained from~\cite{Esteban:2020cvm}, the value of $\cos \left(2 \theta _{23}\right)$ will be small and thus the energy dependence will be primarily regulated by a  $\sin [\Delta] \cos[\Delta] $ term, for both the mass terms. Thus, the position of the peaks and the dips of the SNSI contribution to $P_{\mu \mu}$, for both $\eta_{\mu \mu}$ and $\eta_{\tau\tau}$, are not expected to change as we vary the mass of the lightest neutrino $m_\ell$. However, the heights of the peaks and the dips will decrease for larger values of $m_\ell$, due to cancellation between the two mass terms.
Thus, the leading order behavior shows that it would be difficult to disentangle the neutrino mass with SNSI parameter value, for $\eta_{\mu \mu}$ and $\eta_{\tau\tau}$, if we only consider the survival probability $P_{\mu \mu}$.

Thus, here we show that our analytic estimates for the neutrino probabilities also enable us to explain the non-trivial dependence of the SNSI contribution on the absolute mass of neutrinos, for both mass ordering scenarios.
Any future search for SNSI effects can be aided with such analytic approximations for the neutrino survival and conversion probabilities.

\subsection{ With leptonic CP phase $\delta_{CP}$}\label{sec:dcp_sec}


We have discussed in section \ref{sec:ana_prob}, the SNSI contributions to the oscillation probabilities can be expressed as factors of $(m_2 - m_1)$, $(m_2 + m_1)$, $(m_1 s_{12}^2 + m_2 c_{12}^2)$, $(m_1 c_{12}^2 + m_2 s_{12}^2)$, and $m_3$. The SNSI probabilities also have a complex dependence on the neutrino mixing parameters and neutrino energy. In this section, we explore the dependence of the SNSI contributions on leptonic CP phase $\delta_{CP}$.

To explore the $\delta_{CP}$ dependence of $P_{\mu e}^{(\eta_{ee})}$, we can split it into $\delta_{CP}$ dependent and independent parts as shown in the equation below, 
\begin{equation}\label{eq:delCP_ee_terms}
    P_{\mu e}^{(\eta_{ee})} = \underbrace{(\mathbb{X}_0)_{\mu e}}_{\text{independent of $\delta_{CP}$}} + \underbrace{(\mathbb{X}_1)_{\mu e}^{\delta_{CP}}}_{\text{$\delta_{CP}$  dependent term}}.
\end{equation}

In section \ref{sec:ana_prob}, we have expressed the SNSI contributions as third, fourth and fifth order terms of the bookkeeping parameter $\lambda$. Interestingly, for all three SNSI parameters in the appearance channel ($P_{\mu e}^{\eta_{\alpha \alpha}}$), $\delta_{CP}$ is present only in the third and fifth order correction terms. The fourth order correction term is always $\delta_{CP}$ independent. Therefore, from equation \ref{eq:pme_etaee}, $(\mathbb{X}_0)_{\mu e}$ and $(\mathbb{X}_1)_{\mu e}^{\delta_{CP}}$ can be written as,
\begin{subequations}
\begin{align}
    (\mathbb{X}_1)_{\mu e}^{\delta_{CP}}= & \left[P_{\mu e}^{(\eta_{ee})} \right]^{(3)} + \left[P_{\mu e}^{(\eta_{ee})} \right]^{(5)} \\
     (\mathbb{X}_0)_{\mu e} = & \left[P_{\mu e}^{(\eta_{ee})} \right]^{(4)}
\end{align}
\label{eq:pme_etaee_dcp}
\end{subequations}

\begin{figure}[h!]
    \centering
    \includegraphics[width=0.45\linewidth]{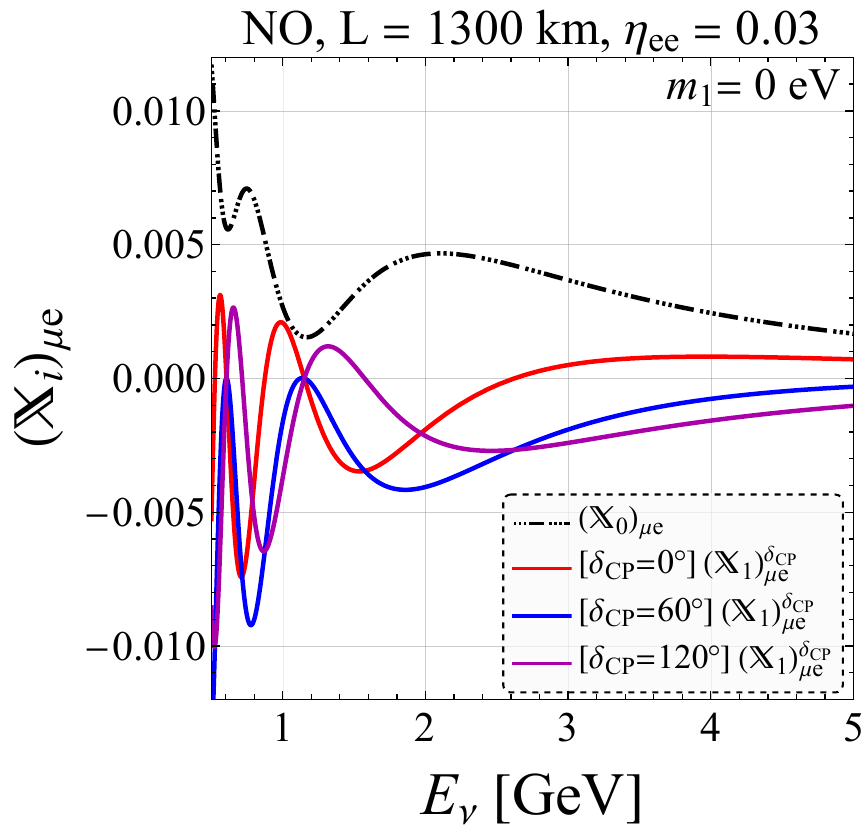} \hspace{10pt}
    \includegraphics[width=0.45\linewidth]{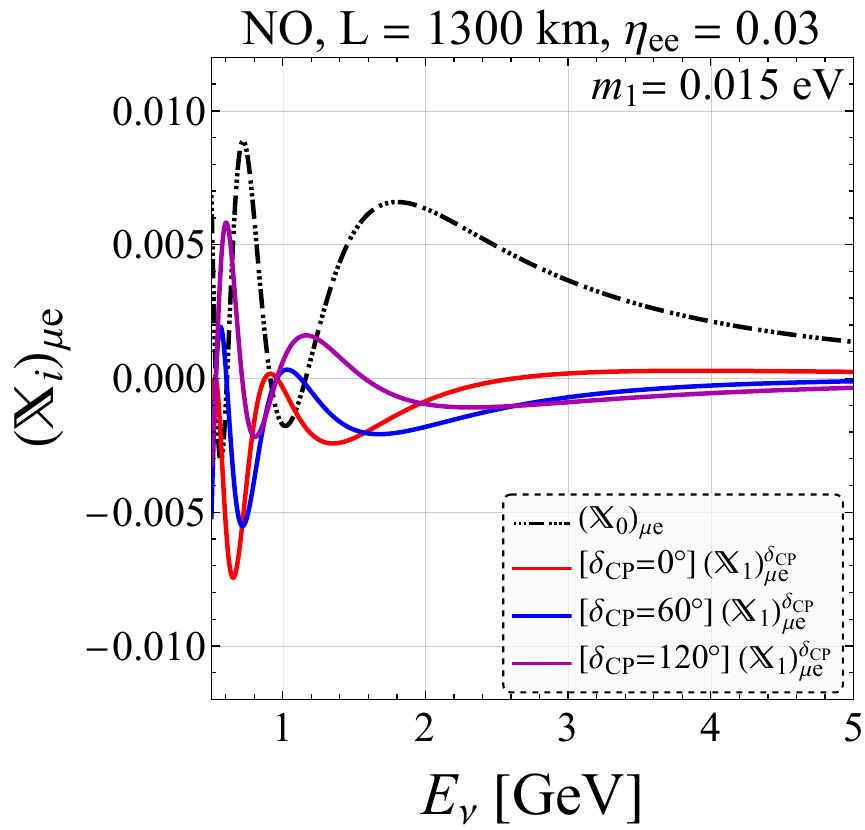} \\ \vspace{5pt}
    \includegraphics[width=0.45\linewidth]{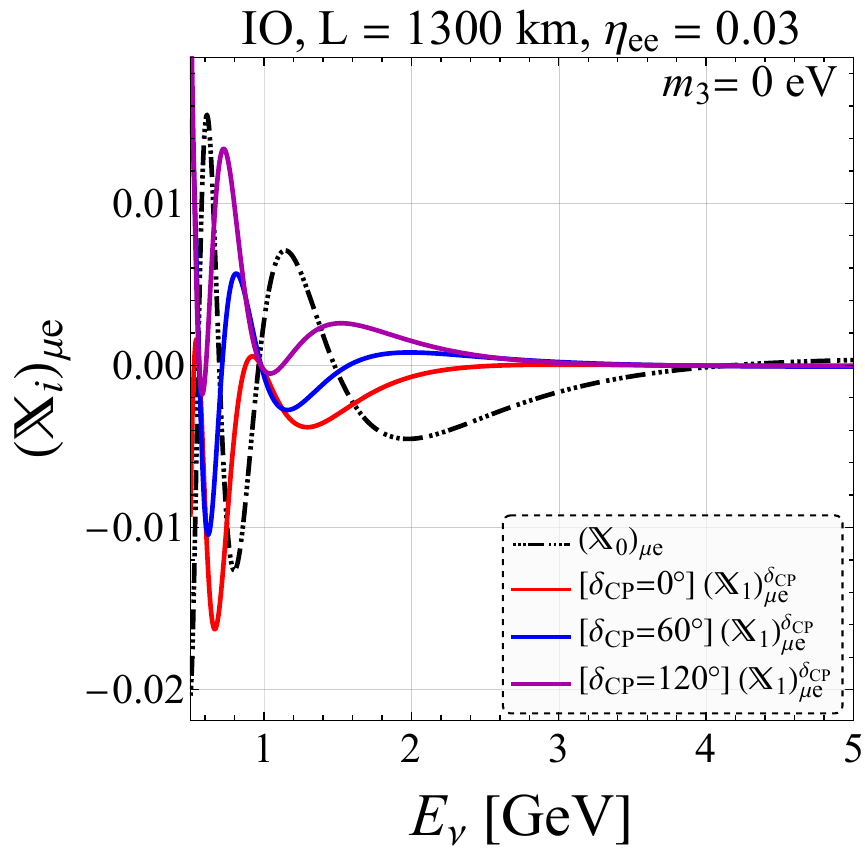} \hspace{10pt}
    \includegraphics[width=0.45\linewidth]{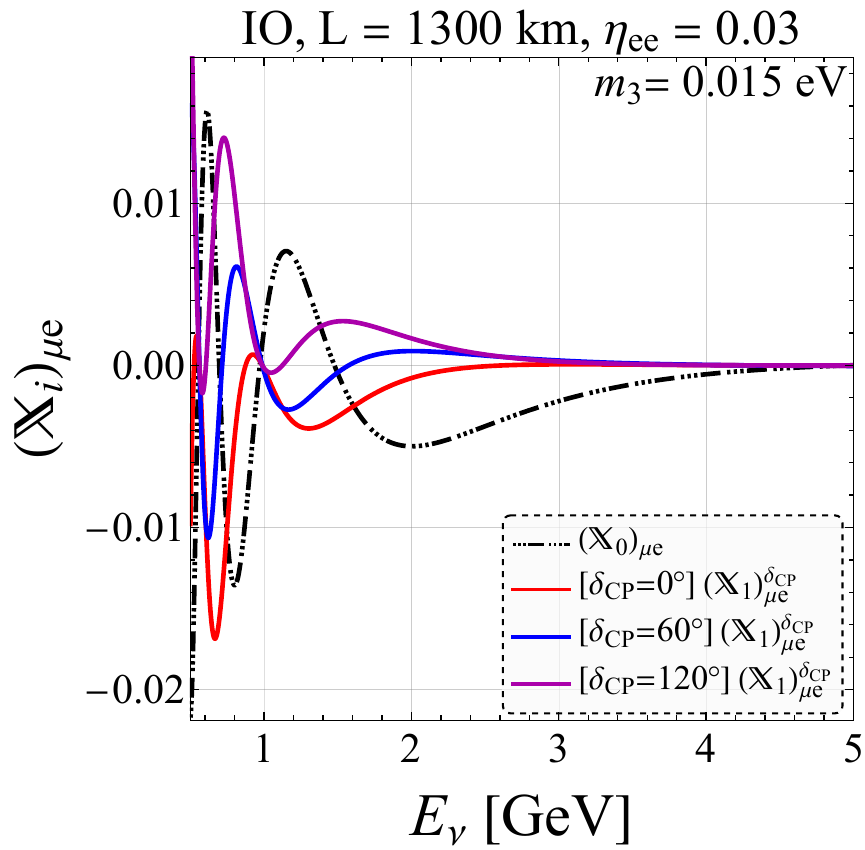}
    \caption{$(\mathbb{X}_i)_{\mu e}$ [$(\mathbb{X}_0)_{\mu e}$ and $(\mathbb{X}_1)_{\mu e}^{\delta_{CP}}$] vs $E_\nu$ for DUNE baseline for $\eta_{ee} = 0.03$, for NO (top panel) and IO (bottom panel).  Here,  $m_\ell$ = 0 eV [left panel] and 0.015 eV [right panel]. Red, blue and the magenta lines corresponds to $(\mathbb{X}_1)_{\mu e}^{\delta_{CP}}$ for  $\delta_{CP}$ values, $0^\circ$, $60^\circ$ and $120^\circ$ respectively. }
    \label{fig:pme_dcp_no}
\end{figure}

\begin{figure}[h!]
    \centering
    \includegraphics[width=0.45\linewidth]{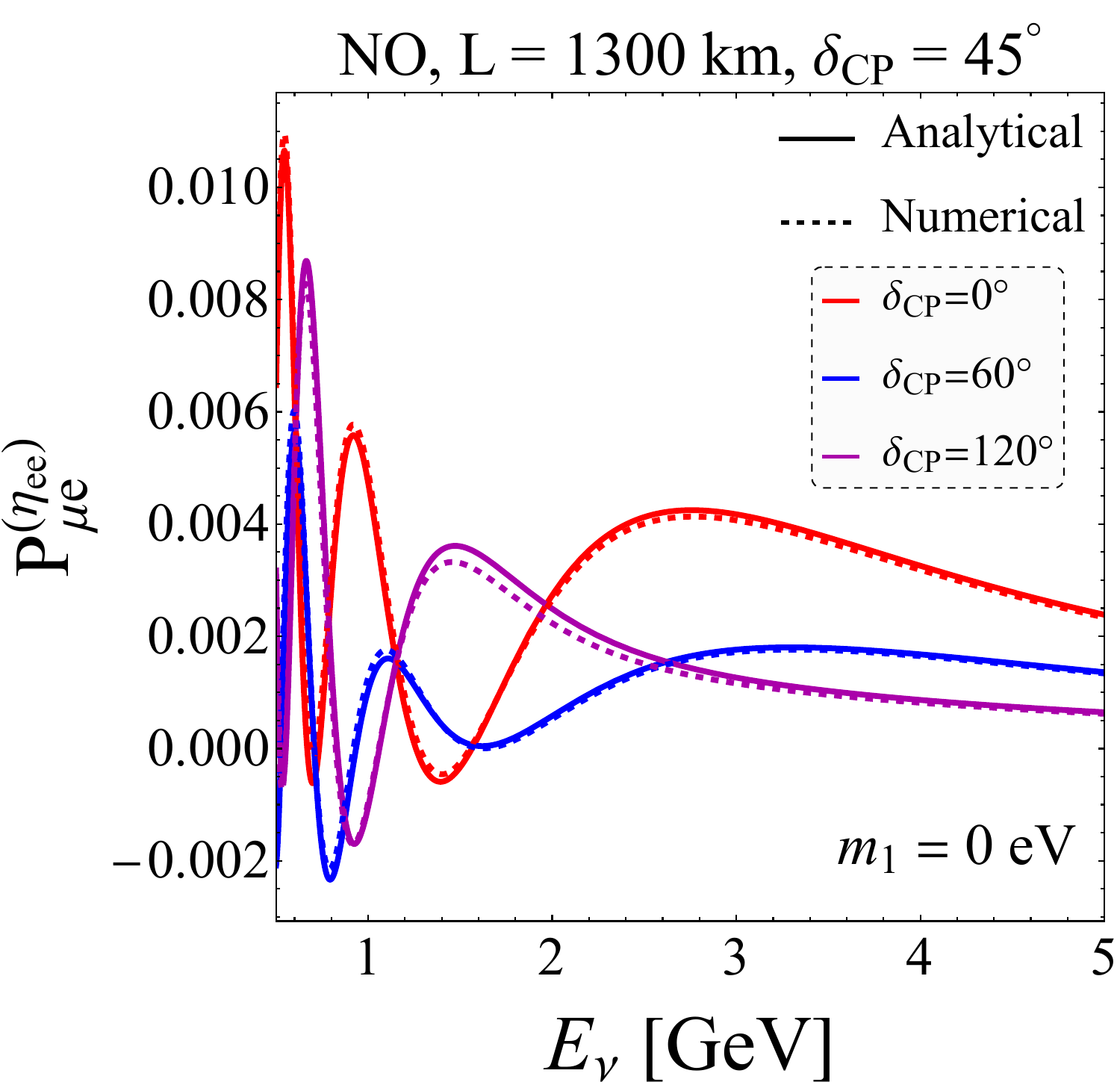} \hspace{10pt}
    \includegraphics[width=0.45\linewidth]{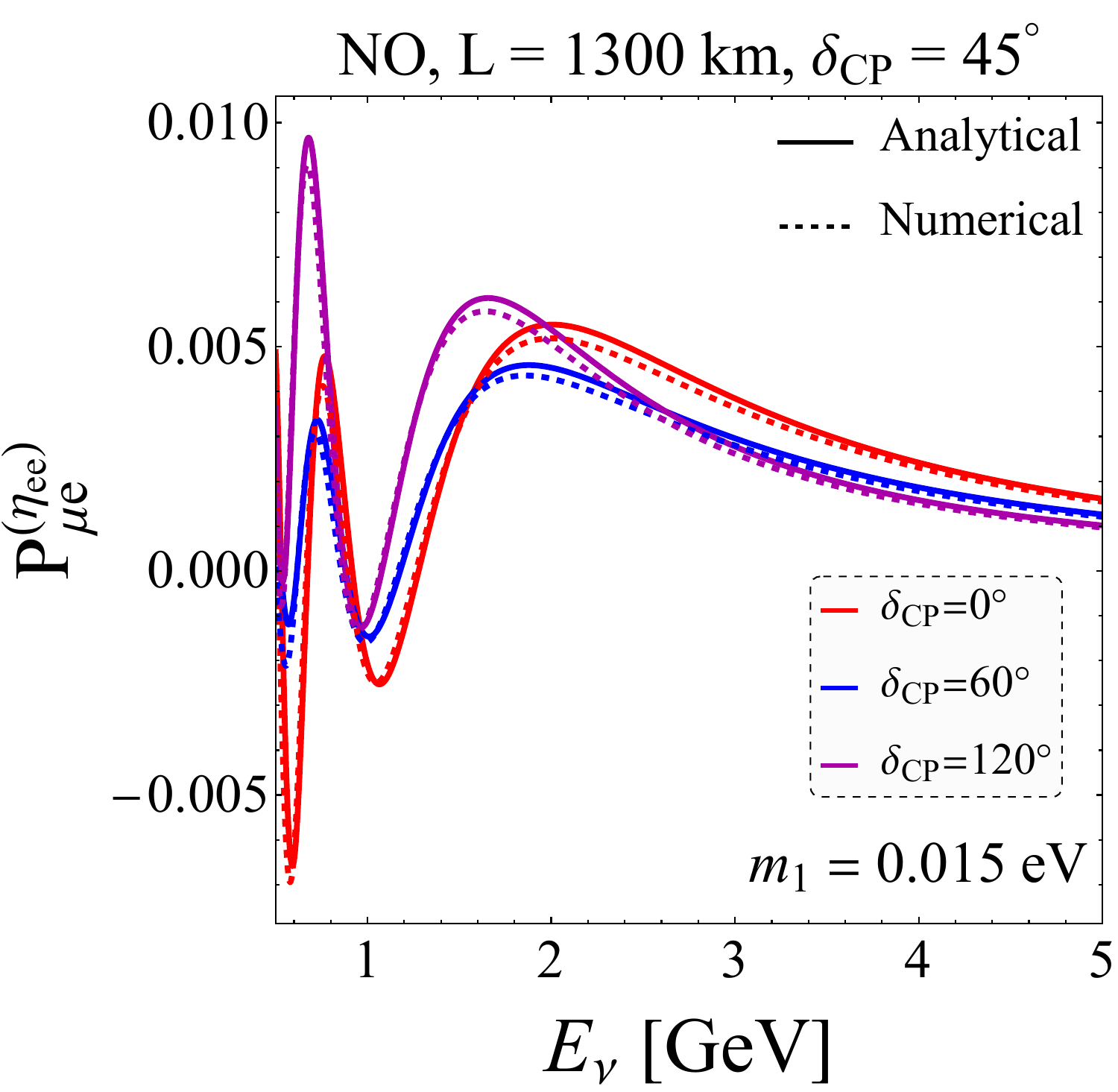} \\ \vspace{5pt}
    \includegraphics[width=0.45\linewidth]{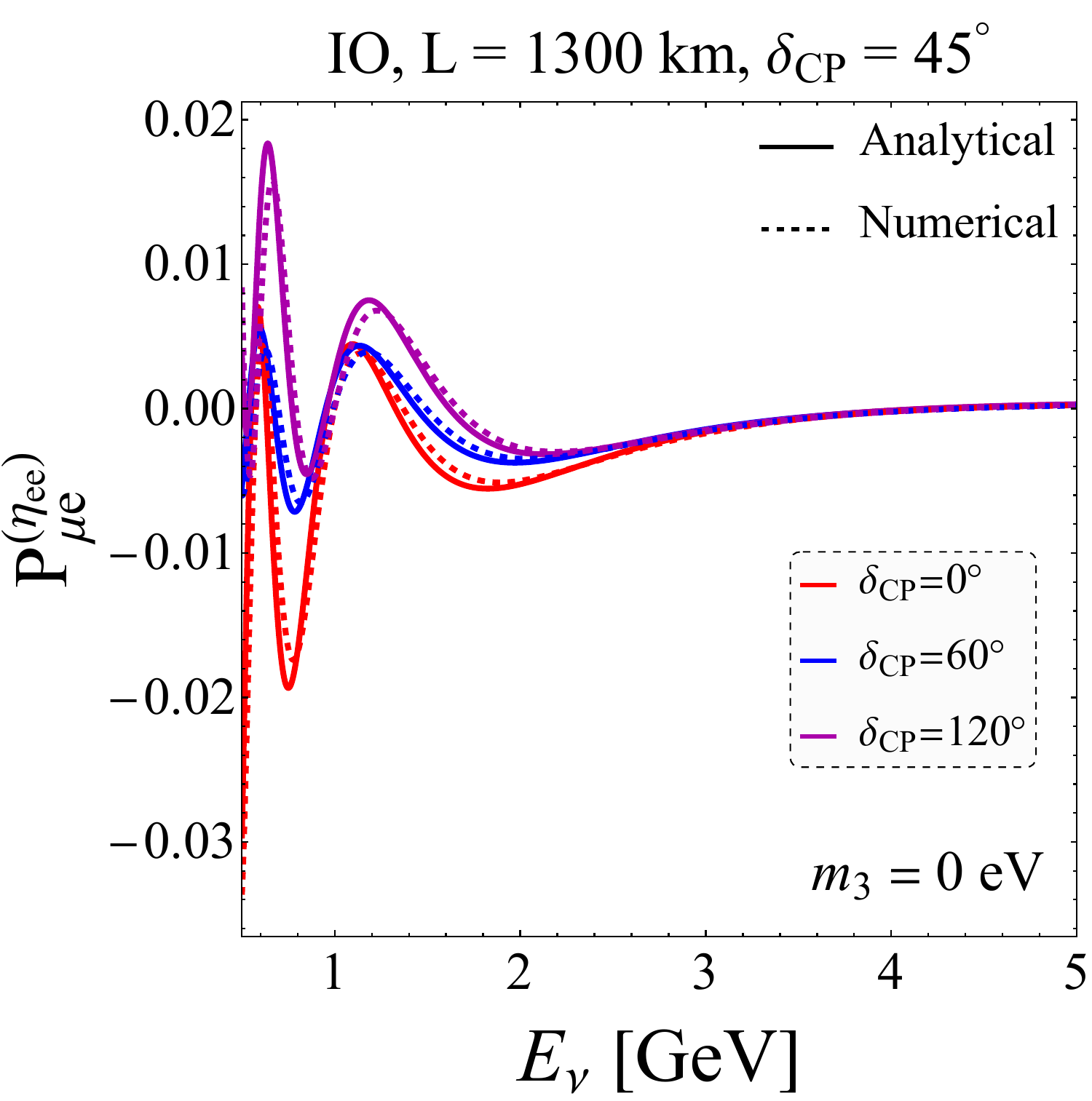} \hspace{10pt}
    \includegraphics[width=0.45\linewidth]{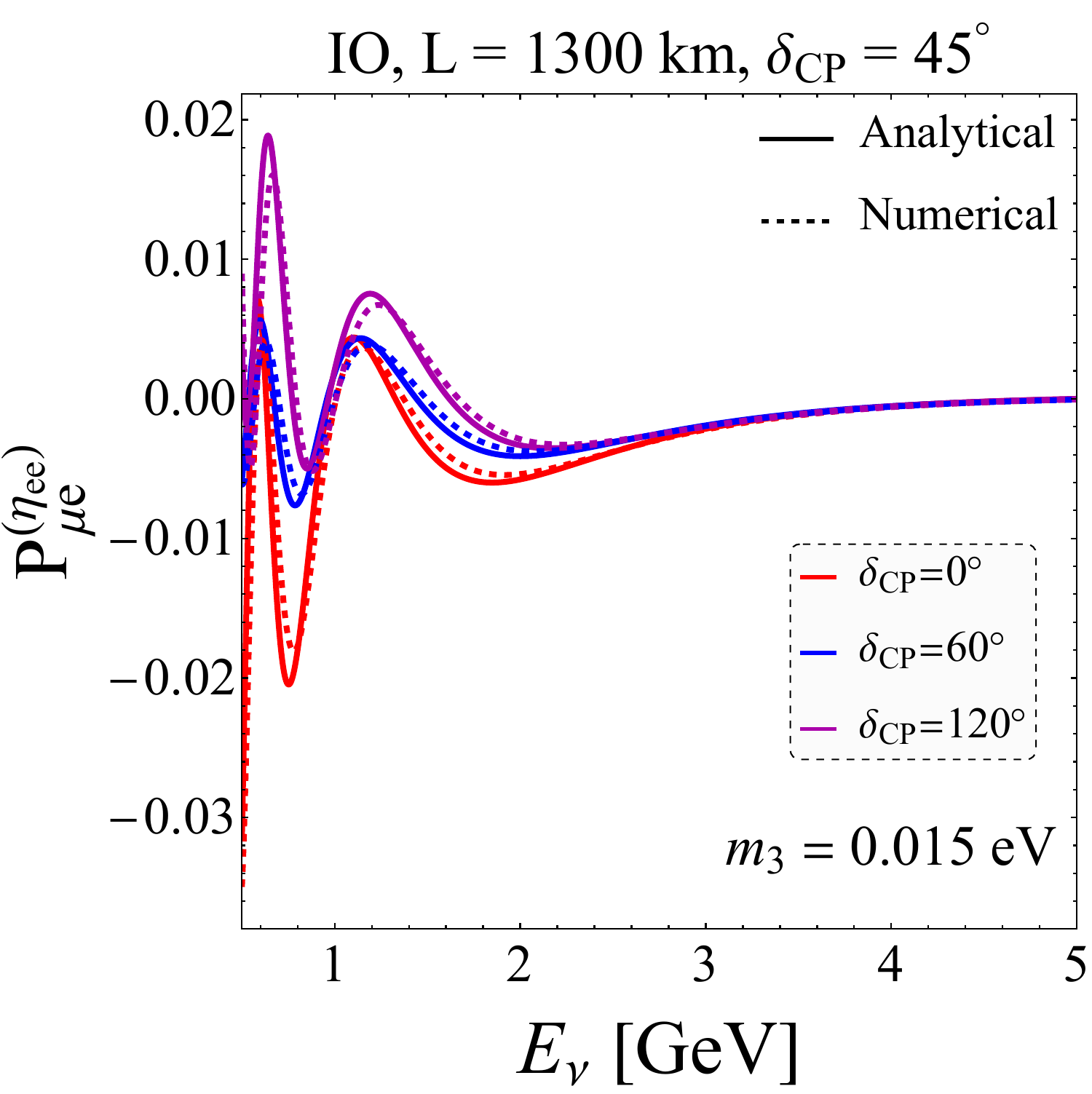}
    \caption{$P_{\mu e}^{(\eta_{ee})}$ vs  $E_{\nu}$ for DUNE baseline for
$\eta_{ee} = 0.03$, for NO (top panel) and IO (bottom panel). Here,  $m_\ell$ = 0 eV [left panel] and 0.015 eV [right panel]. Red, blue and magenta lines correspond to $\delta_{CP}$ values, 0$^\circ$, 60$^\circ$ and 120$^\circ$ respectively. Solid (dashed) lines are for analytic (numerically calculated) contributions. }
\label{fig:pme_dcp_ana_num}
\end{figure}

In figure~\ref{fig:pme_dcp_no}, we show the behavior of the $(\mathbb{X}_1)_{\mu e}^{\delta_{CP}}$ and $(\mathbb{X}_0)_{\mu e}$ terms from equation~\ref{eq:pme_etaee_dcp}. The top panels correspond to NO, and the bottom panels to IO. Two different values of $m_\ell$ are considered: 0 eV in the left panel and 0.015 eV in the right panel, with the parameter $\eta_{ee}$ fixed at 0.03 for all the plots.
Due to the complex dependence of $(\mathbb{X}_1)_{\mu e}^{\delta_{CP}}$ on $A_e$ and $\Delta$, we observe a shift in the position of peaks and the centre of the oscillatory pattern of $(\mathbb{X}_1)_{\mu e}^{\delta_{CP}}$ depending on the value of $\delta_{CP}$. Further, the position of peaks and dips for $(\mathbb{X}_0)_{\mu e}$ and  $(\mathbb{X}_1)_{\mu e}^{\delta_{CP}}$ are also different. The total SNSI contribution, $P_{\mu e}^{(\eta_{ee})}$ will therefore have peaks and dips depending on whether $(\mathbb{X}_0)_{\mu e}$ or $(\mathbb{X}_1)_{\mu e}^{\delta_{CP}}$ will contribute more. 
In figure \ref{fig:pme_dcp_ana_num}, we show the behaviour of $P_{\mu e}^{(\eta_{ee})}$ for three different $\delta_{CP}$ values, plotting both the numerical result as well as the analytic approximation.

For NO scenario, when $m_1 = 0$ eV, the $(m_2 - m_1)$ and the $m_3$ terms will have dominant contributions as discussed in section \ref{sec:Mass_dep} and further elaborated upon in table~\ref{tab:massdependence}. At leading order, the coefficient of the $(m_2 - m_1)$ term has $\delta_{CP}$ dependence through $\cos(\delta_{CP} + \Delta)$. It also has small perturbative terms with no $\delta_{CP}$ dependence. In the $m_3$ term, $\delta_{CP}$ dependence occurs at subleading order. On the other hand, when $m_1$ is 0.015 eV, the dominant contribution will be from the $(m_1 + m_2)$ and the $m_3$ terms. The coefficient of the $(m_1 + m_2)$ term i.e. $C^{\mu e}_+$ has both  $\delta_{CP}$ dependent and independent terms, whereas in the coefficient of the $m_3$ term, $\delta_{CP}$ dependence occurs at subleading order. 
Thus, the overall effect of the $(m_1 + m_2)$ and the $m_3$ terms lead to an increase in the $\delta_{CP}$ independent contribution, as can be seen from equation~\ref{eq:pme_etaee_massdependence_detail}.
Therefore, we expect the $\delta_{CP}$ dependence to be slightly less prominent for larger values of $m_1$. This can be observed from the top panels of figure~\ref{fig:pme_dcp_no}, where the $\delta_{CP}$ dependent term is larger for the $m_1=0$ eV scenario. The complex and competing effects of the interplay between the $\delta_{CP}$ dependent and independent terms can also be seen from the top panels of figure \ref{fig:pme_dcp_ana_num}. However, for $m_1 = 0.015$ eV (top right panel of figure \ref{fig:pme_dcp_ana_num}), different values of $\delta_{CP}$ have less effect in $P_{\mu e}^{(\eta_{ee})}$.

In IO, for all values of $m_3$, the contribution from the $(m_1 + m_2)$ term will be the dominant one and $(m_2 - m_1)$ and  $m_3$ will contribute subdominantly. As shown in equation~\ref{eq:pme_etaee_massdependence_detail}, $C_{+}^{\mu e}$ has $\delta_{CP}$ dependent and independent terms. Furthermore, from the bottom panels of figure~\ref{fig:pme_dcp_no}, we observe that the $\delta_{CP}$ dependent and independent terms have with roughly the same amplitude. Further, when we change $m_3$, the value of $(m_1 + m_2)$ doesn't change significantly, as evident from table \ref{tab:massdependence}. Thus, for IO, we expect that the $\delta_{CP}$ dependent term of $P_{\mu e}^{(\eta_{ee})}$ will not vary as a function of mass. In the bottom panel of figure~\ref{fig:pme_dcp_ana_num} comparing the cases $m_3 =0$ eV (left) and 0.015 eV (right), we observe nominal changes in $\delta_{CP}$ dependence of the SNSI contribution. 

Note that, for $\eta_{\mu \mu}$ and  $\eta_{\tau \tau}$, the relevant channel will be the disappearance channel $P_{\mu\mu}$. Leading order contribution of $P_{\mu \mu}^{(\eta_{\mu \mu})}$ and $P_{\mu \mu}^{(\eta_{\tau \tau})}$ do not have $\delta_{CP}$ dependence. Therefore, $\delta_{CP}$ will not impact these SNSI contributions significantly.

\section{Summary and Conclusions} \label{sec:summary}

Scalar NSI (SNSI) interactions, involving neutrino interaction with standard model fermions mediated via a BSM scalar, is expected to appear as a sub-dominant effect that can impact the $\nu$-oscillations. It brings a unique association of $\nu$--oscillation probabilities with the absolute neutrino masses.
In this work, we have calculated the compact analytic expressions of the neutrino oscillation probabilities $P_{\mu e}$ and $P_{\mu \mu}$, in the presence of the diagonal SNSI elements $\eta_{ee}$, $\eta_{\mu\mu}$, and $\eta_{\tau\tau}$.
Comparing our analytic expressions with numerically calculated oscillation probabilities, we find that for current and upcoming long-baseline experiments the discrepancy remains below 0.1\% for $P_{\mu e}$, and $1\%$ for $P_{\mu \mu}$ in most of their energy ranges.
These expressions can be implemented in long-baseline neutrino experiments. We obtain the analytic expressions by a perturbative expansion in the small parameters $\alpha$, $s_{13}$ and the diagonal SNSI parameters $\eta_{\alpha \alpha}$. We have considered only one element at a time while calculating the expressions for the diagonal elements. The composite probability, in the presence of multiple $\eta_{\alpha \alpha}$ parameters, may be obtained by summing the individual SNSI probability contributions. This is because we only consider SNSI contribution up to linear order in $\eta_{\alpha \alpha}$.
The expressions presented in this study have revealed several interesting features in the oscillation probabilities in the presence of SNSI. Our findings will provide valuable insights into how SNSI influences neutrino behaviour in the long-baseline and atmospheric neutrino sectors, as long as the constant density approximation holds. 

 We show that the derived probability expressions have a direct dependence on the neutrino masses with terms of the form: $(m_1 + m_2)$, $(m_1 c_{12}^2 + m_2 s_{12}^2)$, $(m_1 s_{12}^2 + m_2 c_{12}^2)$, $(m_2 -m_1)$, and $m_3$. These contributory mass terms behave differently with NO and IO scenarios, which indicates that SNSI contributions have distinct impacts based on the mass ordering. Particularly in the NO scenario, we observe that the $(m_2 -m_1)$ term has a noticeable effect compared to the other mass terms only when the lightest neutrino mass is small. However, it has a negligible impact in the IO scenario. The primary contribution in NO can arise from the $(m_2 -m_1)$ term, the mass summation term of the form $(m_1 + m_2)$, and the $m_3$ term. On the other hand, in IO, the dominant contribution always comes from the mass summation terms. The $\eta_{\alpha\alpha}$ contributions depend on the normalized matter potential $A_e$, and the normalized length scale $\Delta$. This, in turn, results in shifts in peaks and dips at different energies. Note that both the position and amplitude of the SNSI contributions on the probabilities explicitly depend on the lightest neutrino mass and the neutrino mass ordering. 

We also explore the complex dependence of the SNSI contributions on the probabilities with the leptonic CP phase $\delta_{CP}$. We note the interplay of the $\delta_{CP}$--dependent terms with the factors with neutrino masses which implies that the imprint of $\delta_{CP}$ would also be impacted by the value of lightest neutrino mass and the neutrino mass ordering. For example, in the appearance channel ($P_{\mu e}$) in the presence of $\eta_{ee}$, the $\delta_{CP}$ dependence is expected to be more prominent with decreasing value of $m_1$ for NO. On the other hand, the value of $m_3$ has a nominal impact on the $\delta_{CP}$ dependence for IO.
 
Considering the precision of the upcoming neutrino experiments, the analytic expressions prove useful for gaining qualitative insight into the oscillation probabilities in the presence of SNSI. These expressions can serve as a useful tool for understanding the underlying physics and for guiding experimental efforts in probing SNSI effects.

\section*{Acknowledgments}

DB acknowledges the DST INSPIRE Fellowship (DST/INSPIRE Fellowship/2022/IF220161) for providing financial support. 
DSC acknowledges support from the Department of Atomic Energy (DAE), Government of India, under Project Identification No. RTI4002. AM acknowledges the support of the project titled ``Indian Institutions - Fermilab Collaboration in Neutrino Physics" of IIT Guwahati funded by DST, Government of India. AS acknowledges the CSIR SRF fellowship (09/0796(12409)/2021-EMR-I) received from CSIR-HRDG. MMD acknowledges the Science and Engineering Research Board (SERB), DST for  the grant CRG/2021/002961. The authors also acknowledge the XXV$^{\text{th}}$ DAE-BRNS HEP Symposium, held at IISER Mohali and the Workshop in High Energy Physics Phenomenology (WHEPP-XVII), held at IIT Gandhinagar.


\bibliographystyle{JHEP}
\bibliography{scalar_NSI_analytical}

\end{document}